%% file: main.tex
\documentclass{llncs}

\makeatletter
\let\proof\@undefined
\let\endproof\@undefined
\makeatother

\usepackage[a4paper,margin=2.5cm]{geometry}

\usepackage{url}
\usepackage[latin1]{inputenc}
\usepackage{PNota}
\usepackage{dsfont}
\usepackage{multirow}
\usepackage{xspace}
\usepackage[figuresleft]{rotating}

\usepackage{color}

\usepackage{pgf}
\usepackage{tikz}
\usetikzlibrary{myautomata}
\usetikzlibrary{extshapes}
\usepgflibrary{arrows}
\usetikzlibrary{shapes}
\tikzstyle{mstate}=[state,minimum size=5mm]
\tikzstyle{wstate}=[lstate]

\makeatletter
\tikzoption{initial angle}{\tikzaddafternodepathoption{\def\tikz@initial@angle{#1}}}
\makeatother

\usepackage{mathrsfs}

\usepackage{amssymb}
\usepackage{amsfonts}
\usepackage{amsthm}
\usepackage{stmaryrd}
\usepackage{eurosym}
\usepackage{color}

\usepackage{epsfig}

\usepackage{enumitem}

\usepackage{subfig}

\usepackage{listings}

\lstdefinelanguage{IPN}[]{Java}{morekeywords={in,out,subnet,UnitaryNet,CompositeNet,transition,synchronization,place,state},basicstyle=\small}
\newcommand{\nat}%
{\ensuremath{\mathds{N}}}
\newcommand{\zrel}%
{\ensuremath{\mathds{Z}}}
\newcommand{\rea}%
{\ensuremath{\mathds{R}}}
\newcommand{\bool}%
{\ensuremath{\mathds{B}}}

\newcommand{\tr}{\xrightarrow}

\newcommand{\pre}[1]{\ifmmode^{\bullet}\mspace{-2mu}#1\else$^{\bullet}\!\!#1$\fi}
\newcommand{\post}[1]{\ifmmode#1^{\bullet}\else$#1^{\bullet}$\fi}

\newcommand{\lab}{\ensuremath{\lambda}}
\newcommand{\lang}{\ensuremath{\mathcal{L}}}

\def\CD{\ensuremath{\mathcal{D}}}
\def\CQ{\ensuremath{\mathcal{Q}}}
\def\CF{\ensuremath{\mathcal{F}}}
\def\CT{\ensuremath{\mathcal{T}}}
\def\CA{\ensuremath{\mathcal{A}}}
\DeclareMathOperator{\Acc}{Acc}

\DeclareMathOperator{\ReachF}{ReachF}
\DeclareMathOperator{\SuccF}{SuccF}
\DeclareMathOperator{\FReach}{FReach}
\DeclareMathOperator{\FSucc}{FSucc}

\newcommand{\AP}{\ensuremath{\mathrm{AP}}}
\newcommand{\FV}{\ensuremath{\mathrm{FV}}}
\newcommand{\SF}{\ensuremath{\mathrm{SF}}}
\newcommand{\KS}{KS\xspace}
\newcommand{\kst}{\ensuremath{\Gamma}}
\newcommand{\resteq}[1]{\ensuremath{\stackrel{#1}{=}}}
\newcommand{\restneq}[1]{\ensuremath{\stackrel{#1}{\neq}}}

\DeclareMathOperator{\F}{\textbf{\textsf{F}}}
\DeclareMathOperator{\G}{\textbf{\textsf{G}}}
\DeclareMathOperator{\U}{\textbf{\textsf{U}}}
\DeclareMathOperator{\X}{\textbf{\textsf{X}}}
\DeclareMathOperator{\sop}{{\widehat{\otimes}}}

\newtheorem{thm}{Theorem}
\newtheorem{cor}{Corollary}

\def\clap#1{\hbox to 0pt{\hss#1\hss}}

\title{Combining Explicit and Symbolic Approaches for Better On-the-Fly LTL Model Checking}

\author{Alexandre Duret-Lutz\inst{1} \and Kais Klai\inst{2} \and Denis Poitrenaud\inst{3} \and Yann Thierry-Mieg\inst{3}}

\institute{LRDE, EPITA, Kremlin-Bicêtre, France.\and
           LIPN, Universit\'e Paris-Nord, Villetaneuse, France.\and
           LIP6/MoVe, Universit\'e Pierre \& Marie Curie, Paris, France.}

\begin{document}

\maketitle
 \input{abstract}

\input{intro}
%\input{figs}
%\input{related}
\input{preliminaries}
\input{sop} %DSOG
\input{slap}%SLOG
% \input{implement}
\input{perfs}
\input{conclusion}

\bibliographystyle{abbrv}
\bibliography{obsgraph}

\end{document}

%% file: abstract.tex
\begin{abstract}
  We present two new hybrid techniques that replace the synchronized
  product used in the automata-theoretic approach for LTL model
  checking.  The proposed products are explicit graphs of aggregates
  (symbolic sets of states) that can be interpreted as Büchi
  automata. These hybrid approaches allow on the one hand to use
  classical emptiness-check algorithms and build the graph on-the-fly,
  and on the other hand, to have a compact encoding of the state space
  thanks to the symbolic representation of the aggregates.  The
  \emph{Symbolic Observation Product} assumes a globally stuttering
  property (e.g., LTL$\setminus \X$) to aggregate states.  The
  \emph{Self-Loop Aggregation Product} does not require the
  property to be globally stuttering (i.e., it can tackle full LTL),
  but dynamically detects and exploits a form of stuttering where
  possible.  Our experiments show that these two variants, while
  incomparable with each other, can outperform other existing
  approaches.
\end{abstract}

% LocalWords:  LTL Büchi

%% file: intro.tex
\section{Introduction}
Model checking for Linear-time Temporal Logic (LTL) is usually based
on converting the property into a Büchi automaton, composing the
automaton and the model (given as a Kripke structure), and finally
checking the language emptiness of the composed
system~\cite{vardi.96.banff}.  This verification process suffers from
a well known state explosion problem.%~\cite{valmari.98.lpn}.
Among the various techniques that have been suggested as improvement,
we can distinguish two large families: explicit and symbolic
approaches.

\textbf{Explicit model checking approaches} explore an explicit
representation of the product graph. % of the model and property.
A common optimization builds the graph on-the-fly as required by
the emptiness check algorithm: the construction stops as soon as a
counterexample is found~\cite{courcoubetis.90.cav}. \par
Another source of optimization is to take advantage of stuttering
equivalence between paths in the Kripke structure when verifying a
stuttering-invariant property~\cite{etessami.99.cav}: this has been
done either by ignoring some paths in the Kripke
structure~\cite{kaivolav.92.concur}, or by representing the property
using a \textit{testing automaton}~\cite{hansen.02.fmics}.  To our
knowledge, all these solutions require dedicated algorithms to check
the emptiness of the product graph.

\textbf{Symbolic model checking} tackles the state-explosion problem
by representing the product automaton symbolically, usually by means
of decision diagrams (a concise way to represent large sets or
relations).  Various symbolic algorithms exist to verify LTL using
fixpoint computations (see~\cite{FislerFVY01,SRKB02} for comparisons
and~\cite{KestenPR98} for the clarity of the presentation).  As-is,
these approaches do not mix well with stuttering invariant reductions
or on-the-fly emptiness checks.

However explicit and symbolic approaches are not exclusive, some
combinations have already been
studied~\cite{BiereCZ99,HaIlKa04,sebastiani.05.cav,KlaiP08} to get the best of
both worlds.  They are referred to as~\textbf{hybrid approaches}.

Most of these approaches consist in replacing the \KS{} by an explicit
graph where each node contains sets of states of the \KS (called
aggregates throughout this paper), that is an abstraction of the \KS{}
preserving properties of the original \KS.  In~\cite{BiereCZ99} for
instance, each aggregate contains states that share their atomic
proposition values, and the successor aggregates contain direct
successors of the previous aggregate, thus preserving LTL but not
branching temporal properties. In~\cite{HaIlKa04} this idea is taken
one step further in the context of stuttering invariant properties,
and each aggregate now contains sets of consecutive states that share
their atomic proposition values. In both of these approaches, an
explicit product with the formula automaton is built and checked for
emptiness, allowing to stop early (on-the-fly) if a witness trace is
found.

The approach of~\cite{sebastiani.05.cav} is a bit different, as it builds one
aggregate for each state of the Büchi automata (usually few in
number), and uses a partitioned symbolic transition relation to check
for emptiness of the product, thus resorting to a symbolic
emptiness-check (based on a symbolic SCC hull computation).

The hybrid approaches we define in this paper are based on explicit
graphs of aggregates (symbolic sets of states) that can be interpreted
as Büchi automata.  With this combination, we can use classical
emptiness-check algorithms and build the graph on-the-fly, moreover
the symbolic representation of aggregates gives us a compact encoding
of the state space along with efficient fixpoint algorithms.

The first technique we present extends the Symbolic Observation Graph
(SOG) technique~\cite{HaIlKa04,KlaiP08} (which itself can be seen as a
specialization of the work of Biere et~al.~\cite{BiereCZ99} for
stuttering-invariant properties).  Given a property, only a subset of
atomic propositions of the system need to be observed.  The SOG
approach aggregates consecutive states of the Kripke structure that
share the same values for the observed atomic propositions.  The SOG
is an aggregated Kripke structure that is stuttering equivalent to the
original Kripke structure.  We combine this principle with an idea
presented by Kokkarinen~et~al.~\cite{kokkarinen.97.cav} in the context
of partial order reductions: as we progress in the Büchi automaton,
the number of atomic propositions to observe diminishes and allows
further aggregation.  We call this new graph a \textit{Symbolic
Observation Product} (SOP), because it replaces the product between
the Kripke structure and the Büchi automaton in the explicit approach.

The second technique we present also defines an aggregation graph
which is a product: the \textit{Self-Loop Aggregation Product}
(SLAP). It uses a different aggregation criterion based on the study
of the self-loops around the current state of the Büchi automaton.
Roughly speaking, consecutive states of the system are aggregated
when they are compatible with the labels of self-loops. Unlike the
previous approach, SLAP is not limited to stuttering-invariant
properties. It dynamically allows to stutter according to a boolean
formula computed as the disjunction of the labels of self-loops of
the automata.

This paper is organized as follows.  Section~\ref{sec:prelim}
introduces our notations, presents the basic automata-theoretic
approach and compares it to the (existing) SOG approach.
Sections~\ref{sec:sop} and~\ref{sec:slap} define our two new hybrid
approaches: SOP and SLAP.  We explain how we implemented these
approaches and evaluate them in Section~\ref{sec:perf}.

% LocalWords:  Kripke LTL Büchi SCC fixpoint SOG Biere al Kokkarinen

%% file: preliminaries.tex
\section{Preliminaries}
\label{sec:prelim}
\subsection{Boolean Formulas}
Let $\AP$ be a set of (atomic) propositions, and let
$\bool=\{\bot,\top\}$ represent Boolean values.  We denote
$\bool(\AP)$ the set of all Boolean formulas over $\AP$, i.e.,
formulas built inductively from the propositions $\AP$, $\bool$,
and the connectives $\land$, $\lor$, and $\lnot$.  If $\AP' \subseteq
\AP$, then we have $\bool(\AP')\subseteq\bool(\AP)$ by construction.
For any formula $f$, we will note $FV(f)$ (for Free Variables) the set
of propositions that occurs in $f$, e.g., $FV(b\lor \lnot
a)=\{a,b\}$.

An assignment is a function $\rho: \AP\to \bool$ that assigns a truth
value to each proposition.  We denote $\bool^\AP$ the set of all
assignments of $\AP$.  Given a formula $f\in\bool(\AP)$ and an
assignment $\rho\in \bool^\AP$, we denote $\rho(f)$ the evaluation of
$f$ under $\rho$.\footnote{This can be defined straightforwardly as
  $\rho(f\land g)=\rho(f)\land\rho(g)$, $\rho(\lnot f)=\lnot \rho(f)$,
  etc.}  In particular, we will write $\rho\models f$ iff $\rho$ is a
satisfying assignment for $f$, i.e., $\rho\models f \iff
\rho(f)=\top$. The set $\bool^\star(\AP) = \{
f\in\bool(\AP)\mid\exists\rho\in\bool^\AP, \rho \models f\}$ contains
all satisfiable formulas.

We will use assignments to label the states of the model we want to
verify, and the propositional functions will be used as labels in the
automaton representing the property to check.  The intuition is that a
behavior of the model (a sequence of assignments) will match the
property if we can find a sequence of formulas in the automaton
that are satisfied by the sequence of assignments.

We will write $\rho \stackrel{E}{=} \rho'$ iff $\rho_{|E}=\rho'_{|E}$,
where $\rho_{|E}$ denotes the restriction of the function $\rho$ to
the domain $E$.  This means that assignments $\rho$ and $\rho'$
match on the propositions $E$.

It is sometimes convenient to interpret an assignment $\rho$ as a
formula that is only true for this assignment.  For instance the
assignment $\{a\mapsto\top,b\mapsto\top,c\mapsto\bot\}$ can be
interpreted as the formula $a\land b\land \lnot c$.  So we may use an
assignment where a formula is expected, as if we were abusively
assuming that $\bool^\AP\subset\bool(\AP)$.

\input{figs}

\subsection{TGBA}
A \emph{Transition-based Generalized Büchi Automaton} (TGBA) is a
Büchi automaton in which generalized acceptance conditions are
expressed in term of transitions that must be visited infinitely
often.  The reason we use these automata is that they allow a
more compact representation of properties than traditional Büchi
automata (even generalized Büchi automata)~\cite{duret.04.mascots}
without making the emptiness check harder~\cite{couvreur.05.spin}.
\begin{definition}[TGBA]\label{def:tgba}
  A Transition-based Generalized Büchi Automata is a tuple
  $A = \tuple{ \AP, \CQ, \CF, \delta, q^0 }$ where
\begin{itemize}[topsep=0pt]
\item $\AP$ is a finite set of atomic propositions,
\item $\CQ$ is a finite set of states,
\item $\CF \neq \emptyset$ is a finite and non-empty set of acceptance conditions,
\item $\delta \subseteq \CQ \times \bool^\star(\AP) \times 2^\CF
  \times \CQ$ is a transition relation. We will commonly denote
  $q_1\tr{f, ac}q_2$ an element $(q_1,f,ac,q_2)\in\delta$,
\item $q^0 \in \CQ$ is the initial state.
\end{itemize}
\end{definition}

An execution (or a run) of $A$ is an infinite sequence of transitions
$\pi=(s_1,f_1,ac_1,d_1)\cdots\break(s_i,f_i,ac_i,d_i)\cdots\in\delta^\omega$
with $s_1=q^0$ and $\forall i, d_i=s_{i+1}$.  We shall simply denote
it as $\pi=s_1\tr{f_1,ac_1}s_2\tr{f_2,ac_2}s_3\cdots$.  Such an
execution is \textit{accepting} iff it visits each acceptance
condition infinitely often, i.e., if $\forall a\in\CF,\,\forall
i>0,\,\exists j\ge i,\,a\in ac_j$.  We denote $\Acc(A)\subseteq
\delta^\omega$ the set of accepting executions of $A$.

A behavior of the model is an infinite sequence of assignments:
$\rho_1\rho_2\rho_3\cdots\in (\bool^\AP)^\omega$, while an execution
of the automaton $A$ is an infinite sequence of transitions labeled by
Boolean formulas.  The language of $A$, denoted $\lang(A)$, is the
set of behaviors compatible with an accepting execution of $A$: $
\lang(A) = \{ \rho_1\rho_2\cdots \in (\bool^\AP)^\omega \mid \exists
s_1\tr{f_1,ac_1}s_2\tr{f_2,ac_2}\cdots \in \Acc(A) \text{~and~}
\forall i\ge 1, \rho_i\models f_i\} $

The non-emptiness constraint on $\CF$ was introduced into
definition~\ref{def:tgba} to avoid considering $\CF=\emptyset$ as a
separate case. If no acceptance conditions exist, one can be
artificially added to some edges, ensuring that every cycle of the
TGBA bears one on at least an edge. Simply adding this artificial
acceptance condition to all edges might seriously hurt subsequent
verification performance, as some emptiness-check algorithms are sensitive
to the position of acceptance conditions.

Fig.~\ref{fig:tgba} represents a TGBA for the LTL formula $a \U
b$. The black dot on the self-loop $q_1 \tr{\top,\{\bc\}} q_1$ denotes
an acceptance conditions from $\CF = \{\bc\}$. The labels on edges
($a\bar b,b,\top$) represent the Boolean expressions over $\AP =
\{a,b\}$. There are many other TGBA in Fig.~\ref{fig:examples}, that
represent product constructions of this TGBA and the Kripke Structure
of Fig.~\ref{fig:kripke}.

A language $\lang(A)$ is \emph{stuttering-invariant} if any letter of
a word can be repeated without affecting its membership to the
language.  In other words, $\lang(A)$ is stuttering-invariant iff for
any finite sequence $u\in(\bool^\AP)^*$, any assignment
$\rho\in\bool^\AP$, and any infinite sequence $v\in(\bool^\AP)^\omega$ we
have $u\rho v\in\lang(A) \iff u\rho\rho v\in\lang(A)$.

Two sequences $w_1$ and $w_2$ are \emph{stuttering equivalent} iff
they are equal after removing all repeated letters.  Two languages
$\lang(A)$ and $\lang(B)$ are \emph{stuttering equivalent} iff any
word of $\lang(A)$ is stuttering equivalent to a word of $\lang(B)$
and vice versa.

LTL$\setminus\X$ is the set of LTL formulas that do not use the $\X$
(next-time) operator.  It is known that formulas in LTL$\setminus\X$
describe stuttering-invariant properties (i.e., the language of the
corresponding TGBA is stuttering-invariant), and that any
stuttering-invariant property can be expressed in
LTL$\setminus\X$~\cite{peled.97.ipl}. The TGBA of Fig.~\ref{def:tgba}
corresponds to the LTL formula $a \U b$ and consequently has a
stuttering-invariant language.

\subsection{Kripke Structure}
For the sake of generality, we use \emph{Kripke Structures} (\KS for short) as a
framework, since the formalism is well adapted to state-based
semantics.
\begin{definition}[Kripke structure]
  A \emph{Kripke structure} is a $4$-tuple
  $\CT = \tuple{\AP, \kst, \lab, \Delta, s_0}$ where:
\begin{itemize}[topsep=0pt]
\item $\AP$ is a finite set of atomic propositions,
\item $\kst$ is a finite set of \emph{states},
\item $\lab: \kst \to \bool^{\AP}$ is a state labeling function,
\item $\Delta \subseteq \kst \times \kst$ is a \emph{transition
    relation}. We will commonly denote $s_1\tr{}s_2$ the element
  $(s_1,s_2)\in\Delta$.
\item $s_0 \in \kst$ is the \emph{initial state}.
\end{itemize}
\end{definition}

Fig.~\ref{fig:kripke} represents a Kripke structure over $\AP =
\{a,b,c\}$. The state graph of a system is typically represented by a
\KS whose labeling function gives the truth values of the atomic
propositions for a given state of the system.  The SOG
construction of Fig.~\ref{fig:sog} also represents a \KS; it is an
aggregated abstraction built from Fig.~\ref{fig:kripke} by observing
only labels $a$ and $b$.

We now define a synchronized product for a TGBA and a \KS, such that
the language of the resulting TGBA is the intersection of the languages
of the two automata.

\begin{definition}[Synchronized product of a TGBA and a Kripke structure]
\label{def:tgbaprod}
  Let $\CA = \tuple{\AP', \CQ, \CF, \delta, q^0}$ be a TGBA and $\CT =
  \tuple{\AP, \kst,\lab,\Delta,s_0}$ be a Kripke structure over over $\AP \supseteq \AP'$.

  The \emph{synchronized product} of $\CA$ and $\CT$ is the TGBA denoted
  by $\CA\otimes \CT=\tuple{\AP, \CQ^{}_\otimes, \CF,
    \delta^{}_\otimes, q^0_\otimes}$ defined as:
  \begin{itemize}[topsep=0pt]
  \item $\CQ^{}_\otimes=\CQ\times\kst$,
  \item $\delta^{}_\otimes\subseteq \CQ_\otimes
    \times \bool^\star(\AP) \times 2^{\CF} \times \CQ_\otimes$
    where
    \[\delta_\otimes=\left\{(q_1, s_1)\tr{f, ac}(q_2, s_2) \left|
    \begin{aligned}
      &s_1\tr{}s_2 \in \Delta,\,\lab(s_1) = f\text{~and}\\
      &\exists g \in \bool^\star(\AP)\text{~s.t.~}
      q_1\tr{g, ac}q_2 \in \delta\text{~and~}\lab(s_1) \models g\\
    \end{aligned}
    \right.\right\}\]
  \item $q^0_\otimes=(q^0,s_0)$.
  \end{itemize}
\end{definition}

Fig.~\ref{fig:product} represents such a product of the TGBA $a \U
b$ of Fig.\ref{fig:tgba} and the Kripke structure of
Fig.~\ref{fig:kripke}. State $(s_0,q_0)$ is the initial state of the
product.  Since $\lab(s_0) = a\bar b c$ we have $\lab(s_0) \models
a\bar b$, successors $\{s_1,s_4\}$ of $s_0$ in the \KS{} will be
synchronized through the edge $q_0 \tr{a\bar b, \emptyset} q_0$ of the
TGBA with $q_0$. In state $(q_0,s_4)$ the product can progress through
the $q_0 \tr{b,\emptyset} q_1$ edge of the TGBA, since $\lab(s_4) =
ab\bar c\models b$. Successor $s_5$ of $s_4$ in the \KS{} is thus
synchronized with $q_1$. The TGBA state $q_1$ now only requires states
to verify $\top$ to validate the acceptance condition $\bc$, so any
cycle in the $\KS$ from $s_5$ will be accepted by the product. The
resulting edge of the product bears the acceptance conditions
contributed by the TGBA edge, and the atomic proposition Boolean
formula label that comes from the \KS. The size of the product in both
nodes and edges is bounded by the product of the sizes of the TGBA and
the \KS.

The emptiness-check on a TGBA verifies if there exist a cycle that
pass through an accepting edge (with the black dot). All of the TGBA
product constructions in Fig.~\ref{fig:examples} agree in having a
non-empty language, since language emptiness is the property these
abstractions (SOP, SLAP) guarantee to preserve. These specialized
synchronized products that are the main contribution of this paper
will be discussed as they are defined.

\subsection{Symbolic Observation Graph (SOG)}

A symbolic observation graph over $\AP'$ is an abstraction of a \KS{}
over $\AP$ ($\AP' \subseteq \AP$) built to allow preservation of
stuttering-invariant properties~\cite{HaIlKa04,KlaiP08}.

It uses a symbolic data structure to represent sets of states of the
\KS{} that have been aggregated.  The SOG is not a quotient graph,
since the predicate we use to aggregate states is not
an equivalence relation. Hence its worst case size in number of
states (aggregates) is bounded by $2^\kst$, while the number of
successors of each aggregate is bounded by $2^{\AP'} -1$.

However, in practice, particularly when the set of observed
propositions $\AP'$ is small, which the case of a typical $LTL$
formula, the SOG is much smaller than the underlying \KS. Since the
states in each aggregate are stored symbolically, the size of these
aggregates is not necessarily the dominating factor in the overall
complexity.

\textbf{Notations} For a set of states $a \subseteq \kst$ and a
Boolean formula $f \in \bool(\AP)$, let us denote by $\SuccF(a,
f)=\{s' \in \Gamma \mid \exists s \in a, s \rightarrow s' \in \Delta
\land \lambda(s') \models f\}$, i.e., the set of the
\textbf{Succ}essors states of $a$ \textbf{F}iltered to keep only those
satisfying $f$.

Furthermore, we denote by $\ReachF(a, f)$ the least subset of
$\kst$ satisfying:
\begin{itemize}[topsep=0pt]
  \item $a \subseteq \ReachF(a, f)$
  \item $\SuccF(\ReachF(a, f), f) \subseteq \ReachF(a, f)$
\end{itemize}

\begin{definition}[Homogeneous aggregate]\label{def:homogeneous}
  Let $a \in 2^\Gamma \setminus \{\emptyset\}$ be a set of states.  We
  say that $a$ is \emph{a homogeneous aggregate} with respect to a
  given subset of atomic propositions $\AP' \subseteq \AP$ iff
  $\forall s, s' \in a, \lambda(s) \resteq{\AP'} \lambda(s')$.
  Furthermore, for a homogeneous aggregate $a$ w.r.t. $\AP' \subseteq
  \AP$, we write $\lambda_{\AP'}(a) = \lambda(s)_{|\AP'}$ for some
  state $s \in a$.
\end{definition}
A homogeneous aggregate $a$ w.r.t. $\AP'$ is then a set of states that
share the same values for atomic propositions in $AP'$. The associated
label is the label of one of its states. Obviously, a homogeneous
aggregate $a$ w.r.t. $\AP'$ is homogeneous w.r.t. any $\AP'' \subseteq
\AP'$.
\begin{definition}[Symbolic Observation Graph]\label{def:sog}
  Let $\CT = \tuple{\AP, \kst, \lab, \Delta, s_0}$ be a KS. A
  \emph{symbolic observation graph} over $\AP' \subseteq \AP$ of $\CT$
  is the \KS over $\AP'$ defined as $\mathcal{G}_{\AP'} = \tuple{\AP',
    S', \lab', \Delta', a_0}$ satisfying :
  \begin{enumerate}[topsep=0pt]
        \item \label{sog:item1}$\displaystyle S' = \Gamma' \cup \bool^{\AP'}\text{~with~}
          \Gamma' = \left\{ a \in 2^\Gamma \setminus \{\emptyset\} \left|
              \begin{aligned}
                &a\text{~is homogeneous w.r.t.~}\AP'\\
                &a=\ReachF(a, \lambda_{\AP'}(a))
              \end{aligned}
            \right.\right\}$\\
          Elements of $\Gamma'$ are called \emph{aggregates} and
          elements of $\bool^{\AP'}$ are \emph{divergent states}.
        \item $\forall a \in S'$, $\lambda'(a) =
          \begin{cases}
            \lambda_{\AP'}(a) & \text{if~} a\in \Gamma'\\
            a & \text{if~} a\in\bool^{\AP'}\\
          \end{cases}$
        \item $\displaystyle
          \begin{aligned}[t]
            \Delta' =\,& \{ a\tr{} a' \in \Gamma'\times\Gamma' \mid
            a'=\ReachF(\SuccF(a,\lambda'(a'))\setminus a,\lambda'(a'))\} \\
            \cup\,& \{ a\tr{} l \in \Gamma'\times \bool^{\AP'} \mid
            a\text{~contains a cycle and~}l=\lambda'(a)\} \\
            \cup\,& \{ l\tr{} l \mid l\in \bool^{\AP'}\}
          \end{aligned}$\\
         % (Note: $\setminus a$ ensures that $\lambda'(a)\ne \lambda'(a')$)
        \item $a_0 = \ReachF(\{s_0\},\lambda(s_0)_{|\AP'})$.
  \end{enumerate}
\end{definition}
Following point~\ref{sog:item1} of the above Definition, the nodes of
a SOG are of two kinds: (1) homogenous aggregates $a$ satisfying
$a=\ReachF(a, \lambda_{\AP'}(a))$, i.e., if a state $s\in a$ then each
successor $s'$ of $s$ belongs to $a$ as soon as $\lambda(s') \resteq{\AP'}
\lambda(s)$, and, (2), divergent states, labeled with atomic
propositions of $\AP'$. The transition relation can be informally
explained as follows: three kind of edges can connect the nodes of a
SOG. If $a$ and $a'$ are two aggregates of $\Gamma'$ then $a\tr{} a'$
iff $\lambda'(a)\ne \lambda'(a')$ and each state $s'\in \Gamma$
satisfying $\lambda(s')_{|\AP'}=\lambda'(a')$ and $s\tr{}s'$ is in $a'$. Given
$a\in\Gamma'$ and $l$ a divergent state then $a\tr{} l$ iff $a$
contains a cycle and is labeled with $l$. Finally each divergent state
has a self-loop.

Fig.~\ref{fig:sog} represents the SOG built over the
\KS{} of Fig.~\ref{fig:kripke} by disregarding the value of $c$. We
can see in this product one divergent states labeled $a \bar b$ that
represents the presence of a cycle in the states of its predecessor
aggregate $\{s_0,s_1,s_2,s_3\}$. States are aggregated as long as they
agree on the value of the subset of observed atomic propositions. The
SOG is still a \KS{}, that allows to check any stuttering-invariant
property over the alphabet $\{a,b\}$. For instance, its product with
the TGBA of $a \U b$ produces the TGBA of
Fig.~\ref{fig:sogproduct}. Both this abstraction and its product are
smaller than their equivalents based on the plain Kripke structure of
Fig.~\ref{fig:kripke}.

\begin{thm}[\cite{KlaiP08}]\label{th:sog}
  Given a Kripke Structure $\CT$ defined on $\AP$, then the SOG
  $\mathcal{G}_{\AP'}$ of $\CT$ built over $\AP'\subseteq\AP$ preserves any
  stuttering-invariant property $\CA$ on $\AP'$.  In other words:
  $\lang(\CA\otimes \CT)\ne\emptyset\iff \lang(\CA\otimes
  \mathcal{G}_{\AP'})\ne\emptyset$.
\end{thm}

% LocalWords:  iff TGBA Büchi topsep pt LTL Kripke YANN SOG

%% file: figs.tex
\begin{figure}[tbp]
  \centering
\subfloat[TGBA $\CA$ for $a\U b$]{\label{fig:tgba}
\begin{tikzpicture}[automaton,thick]
  \node[state,initial] (S0) at (0,0) {$q_0$};
  \node[state] (S1) at (2,0) {$q_1$};
  \path[->] (S0) edge[loop above] node[above]{$a \bar b$} (S0)
                 edge node[above]{$b$} (S1)
            (S1) edge[loop above] node[below=-5pt]{\bc}
                                  node[above=2pt]{$\top$} (S1);
  \node[below=-2pt] at (S0.-90) {$\vphantom{a\bar b}$};
\end{tikzpicture}
}
\hfill
\subfloat[Kripke structure \CT]{\label{fig:kripke}
\begin{tikzpicture}[automaton,thick,scale=1.1]
  \node[state,initial,initial angle=135] (S0) at (1.5,0) {$s_0$};
  \node[state] (S1) at (1.5,-1.5) {$s_1$};
  \node[state] (S2) at (0,-1.5)   {$s_2$};
  \node[state] (S3) at (0,0)      {$s_3$};
  \node[state] (S4) at (3,0)      {$s_4$};
  \node[state] (S5) at (4.5,0)    {$s_5$};
  \node[state] (S6) at (4.5,-1.5) {$s_6$};
  \node[state] (S7) at (3,-1.5)   {$s_7$};
  \draw[->] (S0) -- (S1);
  \draw[->] (S1) -- (S2);
  \draw[->] (S2) -- (S3);
  \draw[->] (S3) -- (S0);
  \draw[->] (S0) -- (S4);
  \draw[->] (S4) -- (S5);
  \draw[->] (S5) -- (S6);
  \draw[->] (S6) -- (S7);
  \draw[->] (S7) -- (S4);

  \node[above=-2pt] at (S0.90)  {$a\bar b c$};
  \node[below=-2pt] at (S1.270) {$a\bar b \bar c$};
  \node[below=-2pt] at (S2.270) {$a\bar b c$};
  \node[above=-2pt] at (S3.90)  {$a\bar b \bar c$};
  \node[above=-2pt] at (S4.90)  {$ab \bar c$};
  \node[above=-2pt] at (S5.90)  {$a\bar b c$};
  \node[below=-2pt] at (S6.270) {$\bar a b\vphantom{\bar b} \bar c$};
  \node[below=-2pt] at (S7.270) {$\bar a b\vphantom{\bar b} c$};
\end{tikzpicture}
}
\hfill
\subfloat[Aggregated Kripke structure SOG \rlap{$\widehat\CT_{\{a,b\}}$}]{\label{fig:sog}
\begin{tikzpicture}[automaton,thick,yscale=0.8]
  \node[wstate,initial] (A0) at (0,0) {
    $\left\{\begin{smallmatrix}s_0 &s_1\\s_2&s_3\\\end{smallmatrix}\right\}$
  };
  \node[wstate]   (Adiv) at (-1.2,-2) {$a\bar b$};
  \node[wstate]   (A1) at (0,-2) {$\{s_4\}$};
  \node[wstate]   (A2) at (1.9,0) {$\{s_5\}$};
  \node[wstate]   (A3) at (1.9,-2) {$\{s_6s_7\}$};

  \path[->] (A0) edge (A1)
            (A0) edge (Adiv)
            (Adiv) edge[loop left] (Adiv)
            (A1) edge (A2)
            (A2) edge (A3)
            (A3) edge (A1);

  \node[above=-2pt] at (A0.90)  {$a\bar b$};
  \node[below=-2pt] at (Adiv.270) {$a\bar b$};
  \node[below=-2pt] at (A1.270) {$ab\vphantom{a\bar b}$};
  \node[above=-2pt] at (A2.90) {$a\bar b$};
  \node[below=-2pt] at (A3.270)  {$\bar a b\vphantom{\bar b}$};
\end{tikzpicture}
}

\subfloat[TGBA of product $\CA\otimes\CT$]{\label{fig:product}
\begin{tikzpicture}[automaton,thick,scale=1.1]
  \node[state,initial,initial angle=135] (S0) at (1.5,0) {$q_0,s_0$};
  \node[state] (S1) at (1.5,-1.5) {\small $q_0,s_1$};
  \node[state] (S2) at (0,-1.5)   {\small $q_0,s_2$};
  \node[state] (S3) at (0,0)      {\small $q_0,s_3$};
  \node[state] (S4) at (3,0)      {\small $q_0,s_4$};
  \node[state] (S5) at (4.5,0)    {\small $q_1,s_5$};
  \node[state] (S6) at (4.5,-1.5) {\small $q_1,s_6$};
  \node[state] (S7) at (6,-1.5)   {\small $q_1,s_7$};
  \node[state] (S4bis) at (6,0)   {\small $q_1,s_4$};

  \draw[->] (S0) -- node[right]{$a\bar b c$} (S1);
  \draw[->] (S1) -- node[above]{$a\bar b \bar c$} (S2);
  \draw[->] (S2) -- node[left]{$a\bar b c$} (S3);
  \draw[->] (S3) -- node[above]{$a\bar b \bar c$} (S0);
  \draw[->] (S0) -- node[above]{$a\bar b c$} (S4);
  \draw[->] (S4) -- node[above]{$ab\vphantom{\bar b}\bar c$} (S5);
  \draw[->] (S5) -- node[pos=.3]{\bc} node[left=1mm]{$a\bar b c$} (S6);
  \draw[->] (S6) -- node[pos=.3]{\bc} node[above]{$\bar a b\vphantom{\bar b}\bar c$} (S7);
  \draw[->] (S7) -- node[pos=.3]{\bc} node[right=1mm]{$\bar a b\vphantom{\bar b} c$} (S4bis);
  \draw[->] (S4bis) -- node[pos=.3]{\bc} node[above]{$ab\vphantom{\bar b}\bar c$} (S5);
\end{tikzpicture}
}
\hfill
\subfloat[TGBA of the product $\CA\otimes\widehat\CT_{\{a,b\}}$]{\label{fig:sogproduct}
\begin{tikzpicture}[automaton,thick,xscale=1.1,yscale=0.9]
  \node[wstate,initial] (A0) at (0,0) {
    $q_0,\left\{\begin{smallmatrix}s_0 &s_1\\s_2&s_3\\\end{smallmatrix}\right\}$
  };
  \node[wstate]   (A1) at (-0,-2) {$q_0,a\bar b$};
  \node[wstate]   (A2) at (2.2,0) {$q_0,\{s_4\}$};
  \node[wstate]   (A3) at (4.4,0) {$q_1,\{s_5\}$};
  \node[wstate]   (A4) at (4.4,-2) {$q_1,\{s_6s_7\}$};
  \node[wstate]   (A5) at (2.2,-2) {$q_1,\{s_4\}$};

  \path[->] (A0) edge node[left]{$a\bar b$} (A1)
                 edge node[above]{$a\bar b$} (A2)
            (A1) edge[loop right] node[above,pos=0.2]{$a\bar b$} (A1)
            (A2) edge node[above]{$ab$} (A3)
            (A3) edge node{\bc} node[right=1mm]{$a\bar b$} (A4)
            (A4) edge node[pos=.3]{\bc} node[above]{$\bar a b$}(A5)
            (A5) edge node{\bc} node[above left]{$ab$} (A3);

  %\node[below=-2pt] at (A1.-90) {$\vphantom{a\bar b}$};
\end{tikzpicture}
}

\hspace*\fill
\subfloat[TGBA of the SOP $\CA\sop\CT$]{\label{fig:sop}
\begin{tikzpicture}[automaton,thick,xscale=1.25]
  \node[wstate,initial] (A0) at (0,0) {
    $q_0,\left\{\begin{smallmatrix}s_0 &s_1\\s_2&s_3\\\end{smallmatrix}\right\}$
  };
  \node[wstate]   (A1) at (0,-2) {$q_0,a\bar b$};
  \node[wstate]   (A2) at (2,0) {$q_0,\{s_4\}$};
  \node[wstate]   (A3) at (4,0) {
    $q_1,\left\{\begin{smallmatrix}s_4&s_5\\s_6&s_7\\\end{smallmatrix}\right\}$
  };
  \node[wstate]   (A4) at (4,-2) {$q_1,\top$};

  \path[->] (A0) edge node[left]{$a\bar b$} (A1)
                 edge node[above]{$a\bar b$} (A2)
            (A1) edge[loop right] node[right]{$a\bar b$} (A1)
            (A2) edge node[above]{$ab$} (A3)
            (A3) edge node{\bc} node[right]{$\top$} (A4)
            (A4) edge[loop left] node[left=-6pt]{\bc}  node[left]{$\top$} (A4);
\end{tikzpicture}
}
\hfill
\subfloat[TGBA of the SLAP $\CA\boxtimes\CT$]{\label{fig:slap}
\begin{tikzpicture}[automaton,thick]
  \node[wstate,initial] (A0) at (1.2,0) {$q_0,\left\{\begin{smallmatrix}s_0 &s_1\\s_2&s_3\\s_4&\\\end{smallmatrix}\right\}$};
  \node[wstate]   (A1) at (2.4,-2) {$q_1,\{s_5\}$};
  \node[wstate]   (A2) at (0,-2) {$q_1,\left\{\begin{smallmatrix}s_4&s_5\\s_6&s_7\\\end{smallmatrix}\right\}$};

  \path[->] (A0.-40) edge node[above right]{$\top$} (A1)
            (A1) edge node[pos=.4]{\bc} node[above]{$\top$} (A2)
            (A2) edge[loop above] node[above=-6pt]{\bc} node[above]{$\top$} (A2);
\end{tikzpicture}
}
\hspace*\fill
\caption{\label{fig:examples}Examples}

\end{figure}
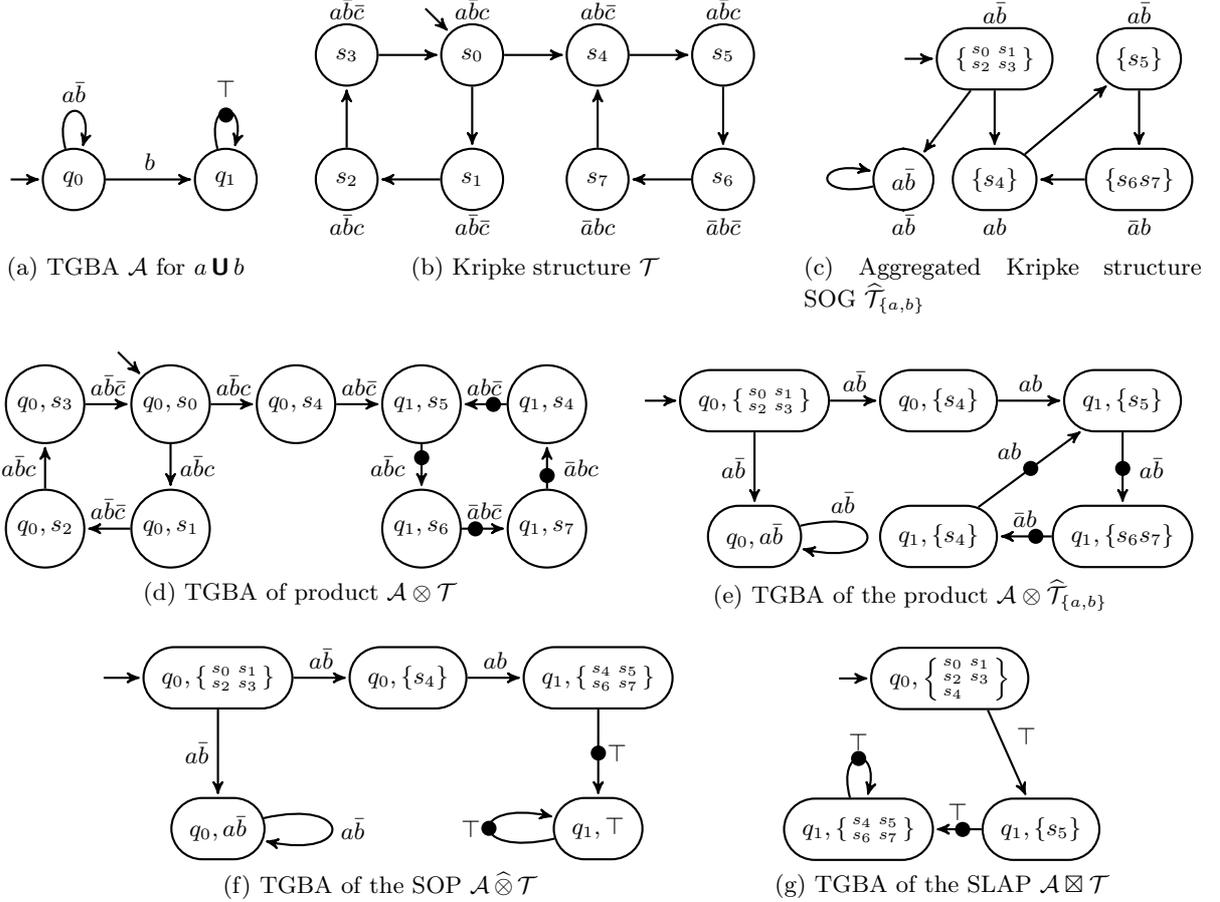

%%% Local Variables:
%%% mode: latex
%%% TeX-master: t
%%% End:

%% file: sop.tex
\section{Symbolic Observation Product (SOP)}
\label{sec:sop}
A SOP is a dynamic extension of the SOG~\cite{HaIlKa04,KlaiP08}.  Both
approaches focus exclusively on stuttering-invariant properties.  The
SOP is a hybrid synchronized product between the TGBA of a
stuttering-invariant property and a \KS.  In this product, the size of
the observed alphabet $\AP'$ decreases as the construction progresses
(an idea also presented by Kokkarinen~et~al.~\cite{kokkarinen.97.cav} in
the context of partial order reductions), therefore allowing more
aggregations.

\subsection{Definition}

Given a TGBA $A = \tuple{ \AP, \CQ, \CF, \delta, q^0 }$, let us define
the alphabet $\FV(q)$ of a state $q\in \CQ$ as the union of the atomic
propositions which can be observed from $q$: i.e., $\FV(q) =
\bigcup_{q_1\tr{f, ac}q_2 \in \delta^\star(q)} \FV(f) $ where
$\delta^\star(q)$ designates the set of transitions reachable from a
state $q$. For instance, $\FV(q^0)=\AP$. It is clear that for any
$q_1\tr{f, ac}q_2 \in \delta$, we have $\FV(q_1) \supseteq
\FV(q_2)$. The set of observed atomic propositions in a given state
and its future reduces or at worse stays stable as we advance through
the automaton.
\begin{definition}[SOP of a TGBA and a \KS]\label{def:sop}
  Given a TGBA $\CA = \langle \AP', \CQ, \CF, \delta, q^0 \rangle$ and
  a Kripke structure $\CT = \langle \AP, \Gamma, \lambda, \Delta, s_0
  \rangle$ over $\AP \supseteq \AP'$, the \emph{Symbolic Observation
    Product} of $\CA$ and $\CT$ is the TGBA denoted $\CA \sop \CT =
  \langle \AP', \CQ^{}_{\sop}, \CF, \delta^{}_{\sop}, q^0_{\sop} \rangle$
  where:
\begin{itemize}[topsep=0pt]
\item $\CQ^{}_{\sop}= \CQ' \cup \CD'$ where states of the automaton are
  synchronized with aggregates in $\CQ'$ and with divergent states in $\CD'$:\\
  $\displaystyle\begin{aligned}[t]
    \CQ' &= \left\{ (q,a) \in \CQ\times(2^\Gamma\setminus\{\emptyset\}) \left|
    \begin{aligned}
      & a\text{~is homogeneous w.r.t.~}\FV(q) \\
      & a=\ReachF(a, \lambda_{\FV(q)}(a)) \\
    \end{aligned}\right.\right\} \\
    \CD' &= \{(q, l) \mid q \in \CQ \text{~and~} l \in \bool^{\FV(q)}\} \\
  \end{aligned}$
  \item $\displaystyle
    \begin{aligned}[t]
      \delta^{}_{\sop} = &\left\{
      (q_1, a_1)\tr{l, ac}(q_2, a_2) \left|
      \begin{aligned}
        & (q_1, a_1) \in \CQ', (q_2, a_2) \in \CQ', l = \lambda_{\FV(q_1)}(a_1),\\
        & \exists f\in\bool(\AP) \text{~s.t.~} q_1\tr{f, ac}q_2 \in \delta,\, \text{~and~} l\models f,\\
        & \exists l'\in\bool^{\FV(q_2)} \text{~s.t.~}
           a_2 = \ReachF(\SuccF(a_1, l')\setminus a_1, l')\\
      \end{aligned}
      \right.
      \right\}\\
      \cup&\left\{ (q_1, a)\tr{l_1, ac}(q_2, l_2) \left|
      \begin{aligned}
         & (q_1, a)\in \CQ', (q_2, l_2)\in \CD', l_1 = \lambda_{\FV(q_1)}(a),\\
         & a \text{~contains a cycle}, l_2={l_1}_{|\FV(q_2)},\\
         & \exists f\in \bool(\AP) \text{~s.t.~} q_1\tr{f, ac}q_2 \in \delta, \text{~and~} l_1 \models f\\
      \end{aligned}
      \right.\right\}\\
      \cup&\,\left\{ (q_1, l_1)\tr{l_1, ac}(q_2, l_2) \left|
      \begin{aligned}
        & (q_1, l_1) \in \CD', (q_2, l_2) \in \CD', l_2 = {l_1}_{|\FV(q_2)},\\
        & \exists f \in\bool(\AP) \text{~s.t.~} q_1\tr{f, ac}q_2 \in \delta, \text{~and~} l_1 \models f\\
      \end{aligned}
      \right.\right\} \\
    \end{aligned}$
  \item $q^0_{\sop} = (q_0, \ReachF(\{s_0\}, \lambda(s_0)_{|\FV(q_0)}))$
\end{itemize}
\end{definition}
Let us explain the intuition behind the three terms of the transition
relation.  The first rule strongly resembles the SOG aggregation rule,
except that the aggregate built from the successors of states in $a_1$
that bear the appropriate label only observes the atomic propositions
in $\FV(q_2)$. This rule is the main ingredient that allows to observe
less atomic propositions as the product progresses, and hence be more
efficient. The next two rules define the cycle detection routine
similar to the SOG, but taking into account the reduction of
the set of atomic propositions to be observed.

Fig.~\ref{fig:sop} represents a SOP built from our example \KS\, and the
TGBA of $a \U b$. Because in state $q_1$ of the formula the observed
alphabet is empty, the SOP aggregates the states of the cycle
$\{s_4,s_5,s_6,s_7\}$ visible on the right of the product that uses the
SOG (Fig.~\ref{fig:sogproduct}). This cycle is then identified by the
cycle detection rules, and visible as a divergent state in the SOP.

\subsection{Proof of correctness}

Our ultimate goal is to establish that, given a \KS and a TGBA, the
emptiness of the language of the corresponding SOP is equivalent to
the emptiness of the language of the original synchronized product
(see Theorem~\ref{th:sop}). This result is progressively demonstrated
in the following. We proceed by construction i.e., if there exists an
accepting run of the SOP then we build an accepting run of the
original product and vice versa. In order to ease the proof of the
first direction, we define a new synchronized product having a
language stuttering equivalent to that of the original synchronized
product (Lemma~\ref{lem:acc:stut}). Then, the desired accepting run is
built in this new product (Lemma~\ref{lem:sop:onestep} and
Lemma~\ref{lem:sop:stut}) and not in the original synchronized
product. Hence, the desired result follows immediately
(Corollary~\ref{lem:sop:left}). Conversely, the proof of the second
direction (each accepting run of the original product corresponds to
an accepting run of the SOP) is not based on such a product but is
nevertheless facilitated by two intermediate lemmas
(Lemma~\ref{lem:sop:finitepath} and Lemma~\ref{lem:sop:right}).

\begin{definition}[Stuttering synchronized product of a TGBA and a
  Kripke structure]
\label{def:tgbastutprod}
  Let $\CA = \tuple{\AP, \CQ, \CF, \delta, q^0}$ be a stuttering-invariant TGBA and $\CT =
  \tuple{\AP, \kst,\lab,\Delta,s_0}$ be a Kripke structure over the same
  atomic proposition set \AP.

  The \emph{stuttering synchronized product} of $A$ and $\CT$ is the
  TGBA denoted by $\CA\tilde\otimes \CT=\tuple{\AP, \CQ^{}_{\tilde\otimes},
    \CF, \delta^{}_{\tilde\otimes}, q^0_{\tilde\otimes}}$ defined as:
  \begin{itemize}[topsep=0pt]
  \item $\CQ^{}_{\tilde\otimes}=\CQ\times\kst$,
  \item $\delta^{}_{\tilde\otimes}\subseteq \CQ_\times
    \times \bool^\star(\AP) \times 2^{\CF} \times \CQ_{\tilde\otimes}$
    where
    \[
    \begin{aligned}
      \delta_{\tilde\otimes}= &
       \left\{(q_1, s_1)\tr{f, ac}(q_2, s_2) \left|
           \begin{aligned}
             &s_1\tr{}s_2 \in \Delta,\,\lab(s_1) = f\text{~and}\\
             &\exists g \in \bool^\star(\AP)\text{~s.t.~}
             q_1\tr{g, ac}q_2 \in \delta\text{~and~}\lab(s_1) \models g\\
           \end{aligned}\right.\right\}\\
           \cup&
           \left\{(q_1, s_1)\tr{f, \emptyset}(q_1, s_2) \left|
           \begin{aligned}
             &s_1\tr{}s_2 \in \Delta,\,\lab(s_1) = f\text{~and}\\
             &\lambda(s_1)\resteq{\FV(q_1)}\lambda(s_2)\\
           \end{aligned}\right.\right\}\\
    \end{aligned}
    \]
  \item $q^0_{\tilde\otimes}=(q^0,s_0)$.
  \end{itemize}
\end{definition}

\begin{lemma}\label{lem:acc:stut}
  Let $\CA$ and $\CT$ be defined as in Definition~\ref{def:tgbastutprod}.
  We have
  $\Acc(\CA\tilde\otimes \CT)\ne\emptyset \iff \Acc(\CA\otimes
  \CT)\ne\emptyset$.
\end{lemma}
\begin{proof}
  By definition, the product $\CA\tilde\otimes \CT$ contains all the
  transitions of $\CA\otimes \CT$, and adds only stuttering
  transitions of the form $(q_i,s_i)\tr{f_i,ac_i}(q_{i+1},s_{i+1})$
  such that $q_i=q_{i+1}$, $f_i\resteq{\FV{q_i}}f_{i+1}$, and
  $ac_i=\emptyset$.  Hence, the language $\lang(\CA\tilde\otimes
  \CT)$ is stuttering equivalent to the language $\lang(\CA\otimes
  \CT)$.  Therefore $\lang(\CA\tilde\otimes \CT)\ne\emptyset \iff
  \lang(\CA\otimes \CT)\ne\emptyset$ and the lemma follows.
\end{proof}

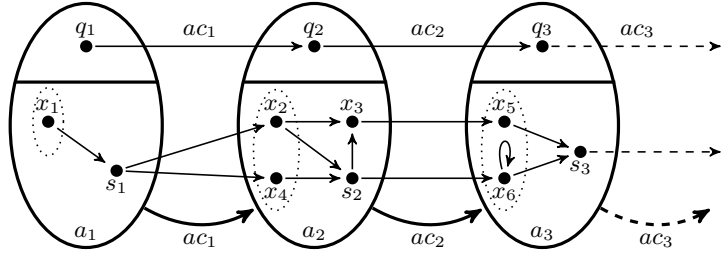
\begin{figure}[tbp]
  \centering
  \begin{tikzpicture}[automaton]
    \tikzstyle{bigell}=[ellipse,minimum width=20mm, minimum height=32.36mm]
    \begin{scope}[very thick]
    \node[draw,bigell] (S1) at (0,0) {};
    \draw (S1.31) -- (S1.149);
    \node[draw,bigell] (S2) at (3,0) {};
    \draw (S2.31) -- (S2.149);
    \node[draw,bigell] (S3) at (6,0) {};
    \draw (S3.31) -- (S3.149);
    \node[overlay,bigell] (S4) at (9,0) {};
    \path[->] (S1.-55) edge[bend right] node[below]{$ac_1$} (S2.-125)
              (S2.-55) edge[bend right] node[below]{$ac_2$} (S3.-125)
              (S3.-55) edge[bend right,dashed] node[below]{$ac_3$} (S4.-125);
    \end{scope}

    \filldraw
    (S1) +(0,10.5mm) circle(2pt) node(q1){} node[above]{\small $q_1$}
    (S2) +(0,10.5mm) circle(2pt) node(q2){} node[above]{\small $q_2$}
    (S3) +(0,10.5mm) circle(2pt) node(q3){} node[above]{\small $q_3$}
    (S1) +(-5mm,0.5mm) circle(2pt) node(x1){} node[above]{\small $x_1$}
    (S1) +(4mm,-6mm) circle(2pt) node(ss1){} node[below]{\small $s_1$}
    (S2) +(-5mm,0.5mm) circle(2pt) node(x2){} node[above]{\small $x_2$}
    (S2) +(5mm,0.5mm) circle(2pt) node(x3){} node[above]{\small $x_3$}
    (S2) +(-5mm,-7mm) circle(2pt) node(x4){} node[below]{\small $x_4$}
    (S2) +(5mm,-7mm) circle(2pt) node(ss2){} node[below]{\small $s_2$}
    (S3) +(-5mm,0.5mm) circle(2pt) node(x5){} node[above]{\small $x_5$}
    (S3) +(-5mm,-7mm) circle(2pt) node(x6){} node[below]{\small $x_6$}
    (S3) +(5mm,-3.5mm) circle(2pt) node(ss3){} node[below]{\small $s_3$};

    \node[draw,ellipse,dotted,minimum width=4mm,minimum height=9mm] (In1) at (x1) {};
    \node[draw,ellipse,dotted,minimum width=6mm,minimum height=16mm] (In2) at (barycentric cs:x2=1,x4=1) {};
    \node[draw,ellipse,dotted,minimum width=6mm,minimum height=16mm] (In3) at (barycentric cs:x5=1,x6=1) {};

    \path[overlay] (S4) +(-5mm,10.5mm) circle(2pt) node(q4){}
                   (S4) +(-5mm,-3.5mm) circle(2pt) node(ss4){};

    \path (S1.-90) node[above] {$a_1$};
    \path (S2.-90) node[above] {$a_2$};
    \path (S3.-90) node[above] {$a_3$};

    \path[->] (q1) edge node[above]{$ac_1$} (q2)
              (q2) edge node[above]{$ac_2$} (q3)
              (q3) edge[dashed] node[above]{$ac_3$} (q4)
              (x1) edge (ss1)
              (x2) edge (x3)
              (x4) edge (ss2)
              (x2) edge (ss2)
              (ss2) edge (x3)
              (x5) edge (ss3)
              (x6) edge (ss3)
              (x6) edge[loop above] (x6)
              (ss1) edge (x4)
              (ss1) edge (x2)
              (x3) edge (x5)
              (ss2) edge (x6)
              (ss3) edge[dashed] (ss4)
              ;

  \end{tikzpicture}
  \caption{\label{fig:sopex} A prefix
    $(q_1,a_1)\fireseq{ac_1}(q_2,a_2)\fireseq{ac_2}(q_2,a_2)$ of a run
    of some SOP $\CA\sop\CT$ (with \emph{different} $\CA$ and $\CT$
    from Fig.~\ref{fig:examples}) is shown using big ellipses and
    bended arrows.  The straight lines also shows the underlying
    connections between the states $\{q_1,q_2,q_3,\ldots\}$ of the
    automaton $\CA$ and between the states
    $\{s_1,s_2,\ldots,x_1,x_2\ldots\}$ of the Kripke structure $\CT$
    that have been aggregated as $a_1,a_2,a_3\ldots$ The acceptance
    conditions have been depicted as $ac_i$ and the labels of the
    transitions have been omitted for clarity.  The dotted ellipses
    show the set of input states ($In(a_1)$, $In(a_2)$, $In(a_3)$) as
    used in the proof of Lemma~\ref{lem:sop:onestep}.}
\end{figure}

\begin{lemma}\label{lem:sop:onestep}
  Let $\CA$ and $\CT$ be defined as in Definition~\ref{def:tgbastutprod}.
  Let $(q_1,a_1)\fireseq{ac}(q_2,a_2)\in\delta_{\sop}$ be a transition
  of a SOP $\CA\sop\CT$ such that $(q_2,a_2)\in \CQ'$.  For any state
  $s_2\in a_2$ there exists at least one (possibly indirect) ancestor
  $s_1\in a_1$ such that
  $(q_1,s_1)\fireseq{ac}(q_2,t_1)\fireseq{}(q_2,t_2)\fireseq{}\cdots
  (q_2,t_n)\fireseq{}(q_2,s_2)$ is a sequence of the stuttering product
  $\CA\tilde\otimes\CT$ with $\forall i,\,t_i\in a_2$.
\end{lemma}

\noindent
For example, consider transition $(q_1,a_1)\fireseq{ac_1}(q_2,a_2)$ on
Fig.~\ref{fig:sopex}, and some state in $a_2$, say $s_2$.  Then
$s_1\in a_1$ is an indirect ancestor of $s_2$ s.t.
$(q_1,s_1)\fireseq{ac_1}(q_2,x_2)\fireseq{}(q_2,s_2)$.

\begin{proof}
  Let us define
  the set of input states of the aggregate $a_2$ as
  $In(a_2)=\{ s'\in a_2\mid \exists s\in a_1, s\fireseq{}s' \in \Delta
  \}$.  This set cannot be empty since $(q_1,a_1)\fireseq{ac}(q_2,a_2)$.

  Consider a state $s_2\in a_2$.  By construction of $a_2$,
  $s_2$ is reachable from some state in $t_1\in In(a_2)$, so there
  exists a path $t_1\fireseq{}t_2\fireseq{}\cdots\fireseq{}s_2$ in the
  Kripke structure.  Furthermore, all these states
  $t_1,t_2,\ldots,s_2$ are homogeneous w.r.t. $\FV(q_2)$, and the
  property is stuttering invariant, so there exists a path
  $(q_2,t_1)\fireseq{}(q_2,t_2)\fireseq{}\cdots\fireseq{}(q_2,s_2)$ in
  the stuttering product $\CA\tilde\otimes\CT$.

  Moreover, since $t_1\in In(a_2)$, there exists a state $s_1$ in
  $a_1$ such that $(q_1,s_1)\fireseq{}(q_2,t_1)$.

  Consequently, the path
  $(q_1,s_1)\fireseq{ac}(q_2,t_1)\fireseq{}(q_2,t_2)\fireseq{}\cdots\fireseq{}(q_2,s_2)$
  satisfies the lemma.
\end{proof}

\begin{lemma}\label{lem:sop:stut}
  Let $\CA$ and $\CT$ be defined as in Definition~\ref{def:tgbastutprod}.
  If there exists $\sigma\in\Acc(\CA\sop\CT)$ an infinite run accepted
  by the SOP, then there exists an accepting run
  $\pi\in\Acc(\CA\tilde\otimes\CT)$ in the stuttering synchronized
  product.
\end{lemma}
\begin{proof}
  There are two cases to consider:
  \begin{enumerate}
  \item Either $\sigma$ contains some divergent states, and then, by
    definition of $\delta_{\sop}$, $\sigma$ must necessarily have a
    finite prefix made of states from $\CQ'$ followed by an infinite
    suffix made of states of $\CD'$ and the last state of that prefix
    has an aggregate that contains a cycle.

    Let us denote
    $\sigma=(q_1,a_1)\fireseq{ac_1}(q_2,a_2)\fireseq{ac_2}\cdots(q_k,a_k)
    \fireseq{ac_k}(q_{k+1},l_{1})\fireseq{ac_{k+1}}(q_{k+2},l_{2})\cdots$
    such an accepting run of $\CA\sop\CT$.

    Let $s_k,s_{k+1},\ldots,s_{k+n-1}$ be a cycle of $a_k$.  Applying
    Lemma~\ref{lem:sop:onestep} from $a_k$ to $a_1$, we can build a
    (possibly larger) sequence
    $\pi_p=(q_1,s_1)\tr{ac_1}\cdots(q_k,s_k)$ of transitions of
    $\CA\tilde\otimes\CT$.  Since $s_1\in a_1$, i.e., it belongs to
    the initial aggregate, it is accessible from $s_0$ by definition
    of $q^0_{\sop}$.  Therefore $\pi_p$ can be prefixed by a sequence
    starting from $(q^0,s_0)$; let $\pi_i$ be this complete
    prefix, going from $(q^0,s_0)$ to $(q_k,s_k)$.

    Let us now complete $\pi_i$ to an infinite sequence.  Because all
    the states $s_k,s_{k+1},\ldots,s_{k+n-1}$ are homogeneous
    w.r.t. $\FV(q_k)$, they are also homogeneous w.r.t. $\FV(q_{k+i})$
    for any $i\ge 0$.  By definition of the stuttering synchronized
    product,
    $\pi_s=(q_k,s_k)\tr{ac_k}\cdots(q_{k+n},s_{k+n})\tr{ac_{k+n}}
    \cdots (q_{k+i},s_{k+(i \mod n)})\tr{ac_{k+i}} \cdots$ is a path
    of $\CA\tilde\otimes\CT$.

    Consequently, the infinite sequence $\pi_i\pi_s$ starts from the
    initial state, and visits the same acceptance conditions of
    $\sigma$, thus $\pi_i\pi_s\in \Acc(\CA\tilde\otimes\CT)$.

  \item Or $\sigma$ traverses only states from $\CQ'$.

    Let us denote
    $\sigma=(q_1,a_1)\fireseq{ac_1}(q_2,a_2)\fireseq{ac_2}(q_3,a_3)\fireseq{ac_3}\cdots$
    such an accepting run of $\CA\sop\CT$.
    Let us build an infinite tree in which all nodes (except the root)
    are states of $\CA\tilde\otimes\CT$.  Let us call $\top$ the root,
    at depth $0$.  The set of nodes at depth $n>0$ is exactly the
    finite set of pairs $\{(q_n,s)\mid s\in a_n\}\subseteq
    \CQ\times\Gamma$.

    The parent of any node at level $1$ is $\top$.  For any $i>0$, the
    parent of a node $(q_{i+1},s')$ with $s'\in a_{i+1}$ is the node
    $(q_i,s)$ for any state $s\in a_i$ such that $(q_{i},s)$ is a
    (possibly indirect) ancestor of $(q_{i+1},s')$ such that we
    observe $ac_i$ on the path between these two states.  We know such
    a state $(q_i,s)$ exists because of Lemma~\ref{lem:sop:onestep}.
    As a consequence of this parenting relation, every edge in this
    tree, except those leaving the root, correspond to a path between
    two states of $\CA\tilde\otimes\CT$.

    Because the set of nodes at depth $n>0$ is finite, this infinite
    tree has finite branching.  By König's Lemma it therefore contains
    an infinite branch.  By following this branch and ignoring the
    first edge, we can construct a path of $\CA\tilde\otimes\CT$ that
    starts in $(q_1,s_1)$ for some $s_1\in a_1$, and that visits at
    least all the acceptance conditions $ac_i$ of $\sigma$ in the same
    order (and maybe more).  To prove that this accepting path we have
    constructed actually occurs in a run of $\CA\tilde\otimes\CT$, it
    remains to show that $(q_1,s_1)$ is a state that is accessible
    from the initial state of $\CA\tilde\otimes\CT$.

    Obviously $q_1=q^0$ because $(q_1,a_1)=q^0_{\sop}$ is the initial
    state of $\CA\sop\CT$.  Furthermore we have $s_1\in a_1$, so by
    definition of $q^0_{\sop}$, $(q^0,s_1)$ must be reachable from (or
    equal to) $(q^0,s_0)$ in $\CA\tilde\otimes\CT$.
  \end{enumerate}
\end{proof}

\begin{cor}\label{lem:sop:left}
  If there exists $\sigma\in\Acc(\CA\sop\CT)$ an infinite run accepted
  by the SOP, then there exists an accepting run
  $\pi\in\Acc(\CA\otimes\CT)$ in the synchronized product.
\end{cor}
\begin{proof}
  Follows from Lemma~\ref{lem:sop:stut} and Lemma~\ref{lem:acc:stut}.
\end{proof}

\begin{lemma}\label{lem:sop:finitepath}
  Let $\CA$ and $\CT$ be defined as in Definition~\ref{def:tgbastutprod}.
  For a given $n$ and a finite path
  $\pi_n=(q_0,s_0)\tr{f_0,ac_0}(q_1,s_1)\cdots\tr{f_{n-1},ac_{n-1}}(q_n,s_n)$
  of $\CA\otimes\CT$, there exists a finite path
  $\sigma_m=(q_{\varphi(0)},a_0)\tr{ac_{\varphi(0)}}(q_{\varphi(1)},a_1)
\cdots\linebreak[1]\tr{ac_{\varphi(m-1)}}(q_{\varphi(m)},a_m)$
  of $\CA\sop\CT$, with $m\le n$,
  where
  \begin{align*}
    \varphi(0) &= \max\left\{j\left|
    \forall k\in\{0,\ldots,j\},\,q_k=q_j \land \lab(s_k) \resteq{\FV(q_j)} \lab(s_j)\right.\right\} \\
\text{for~}i>0\quad \varphi(i) &= \max\left\{j\left|
    \forall k\in\{\varphi(i-1)+1,\ldots, j\},\,
    q_k=q_j \land \lab(s_k) \resteq{\FV(q_j)} \lab(s_j)\right.\right\}\\
  \end{align*}
  and $\{s_0,\ldots,s_{\varphi(0)}\}\subseteq a_0$ and
  for $i>0,\,\{s_{\varphi(i-1)+1},\ldots,s_{\varphi(i)}\}\subseteq a_i$.
\end{lemma}
\begin{proof}
  Let prove this by induction on the length of the finite path.  The
  property is true for a path of length $n=0$ by definition of
  $q^0_{\sop}$. Now assume that the lemma is true for some length $n$
  and let us consider a path
  $p_{n+1}=(q_0,s_0)\tr{f_0,ac_0}(q_1,s_1)\cdots\tr{f_{n},ac_{n}}(q_{n+1},s_{n+1})$
  of length $n+1$.  By the induction hypothesis, we know that there
  exists a finite path
  $\sigma'_m=(q_{\varphi'(0)},a_0)\tr{ac_{\varphi'(0)}}(q_{\varphi'(1)},a_1)
  \cdots\tr{ac_{\varphi'(m-1)}}(q_{\varphi'(m)},a_m)$ of $\CA\sop\CT$,
  and a function $\varphi'$ that correspond to the prefix of length
  $n$ of $\pi_{n+1}$.

  We consider the following two cases:
  \begin{enumerate}
  \item If $q_{n+1}=q_{n}$ and $\lab(s_{n+1})\resteq{\FV(q_n)} \lab(s_n)$ then
    $s_{n+1}\in a_m$ by definition of $\delta_{\sop}$.
    Therefore the path
    $\sigma_m=(q_{\varphi(0)},a_0)\tr{ac_{\varphi(0)}}(q_{\varphi(1)},a_1)
  \cdots\tr{ac_{\varphi(m-1)}}(q_{\varphi(m)},a_m\})$ and the
  function $\varphi$ defined as $\forall i<m,\,\varphi(i)=\varphi'(i)$
  and $\varphi(m)=n+1$, satisfy the lemma.
\item Otherwise, $q_{n+1}\ne q_n$ or $\lab(s_{n+1})\restneq{\FV(q_n)} \lab(s_n)$.
  In that case, according the definition of $\delta_{\sop}$, there
  exists an aggregate $a_{m+1}$ such that
  $(q_n,a_m)\tr{ac_n}(q_{n+1},a_{m+1})$ and $s_{n+1}\in a_{m+1}$.

  Since $n=\varphi'(m)$.  We can define $\varphi$ as $\forall i\le
  m,\,\varphi(i) = \varphi'(i)$ and $\varphi(m+1)=n+1$, and build
  $\sigma_{m+1}$ by extending $\sigma'_m$:
  $\sigma_{m+1}=(q_{\varphi(0)},a_0)\tr{ac_{\varphi(0)}}(q_{\varphi(1)},a_1)
  \cdots\tr{ac_{\varphi(m-1)}}(q_{\varphi(m)},a_m)
  \tr{ac_n}(q_{\varphi(m+1)}, a_{m+1})$.  This path satisfies the
  lemma.
  \end{enumerate}
\end{proof}

\begin{lemma}\label{lem:sop:right}
  Let $\CA$ and $\CT$ be defined as in Definition~\ref{def:tgbastutprod}.
  If there exists an infinite path $\pi\in\Acc(\CA\otimes\CT)$
  accepting in $\CA\otimes\CT$.  Then there exists an accepting path
  in $\CA\sop\CT$ as well.
\end{lemma}
\begin{proof}
  $\CA\otimes\CT$ has a finite number of states, so if
  $\Acc(\CA\otimes\CT)\ne\emptyset$ then it contains at least one infinite path
  $\pi\in\Acc(\CA\otimes\CT)$ that can be represented as a finite
  prefix followed by a finite cycle that is repeated infinitely often.

  Let us denote this lasso-shaped path by $\pi=(q_0,s_0)
  \tr{ac_0}\cdots(q_{k},s_{k})\tr{ac_{k}}\cdots(q_n,s_n)$ with
  $(q_n,s_n) = (q_k, s_k)$.

  Note that because $q_k,q_{k+1},\ldots,q_{n}$ is a cycle in $\CA$, we
  have $\FV(q_k)=\FV(q_{k+1})=\ldots=\FV(q_{n-1})$.

  We consider two possible cases:
  \begin{enumerate}
  \item If $\lab(s_k) \resteq{\FV(q_k)} \lab(s_{k+1}) \resteq{\FV(q_k)} \ldots
    \resteq{\FV(q_k)} \lab(s_{n-1})$, these states are homogeneous and they
    form a cycle.

    We can apply Lemma~\ref{lem:sop:finitepath} on prefix
    $(q_0,s_0)\tr{ac_0}\cdots(q_{k},s_{k})$ to build a path
    $\sigma_m=(q_{\varphi(0)},a_0)\tr{ac_{\varphi(0)}}(q_{\varphi(1)},a_1)
    \cdots\tr{ac_{\varphi(m-1)}}(q_{\varphi(m)},a_m)$ such that
    $q_k=q_{\varphi(m)}$ and $s_k\in a_m$.  Since the states
    $s_k,\ldots s_{n-1}$ are homogeneous, we also have $\{s_k,\ldots
    s_{n-1}\}\subseteq a_m$.

    Because $a_m$ contains a cycle, there exist transitions
    $(q_k,a_m)\tr{ac_k}(q_{k+1},l)\tr{ac_{k+1}}(q_{k+2},l)\cdots
    \tr{ac_{n-1}}(q_{n}=q_k,l)\tr{ac_k}(q_{k+1},l)$ according to $\delta_{\sop}$.

    The infinite sequence $\sigma_m
    \tr{ac_k}\left((q_{k+1},l)\tr{ac_{k+1}}(q_{k+2},l)\cdots
      \tr{ac_{n-1}}(q_{k},l)\tr{ac_k}\right)^\omega$ is accepting and
    satisfies the lemma.

  \item Otherwise if the states $s_k,\ldots,s_{n-1}$ are not
    homogeneous w.r.t. $\FV(q_k)$, then we can apply
    Lemma~\ref{lem:sop:finitepath} on the
    entire path $\pi$ in order to build a path
    $\sigma_m=(q_{\varphi(0)},a_0)\tr{ac_{\varphi(0)}}(q_{\varphi(1)},a_1)
    \cdots\tr{ac_{\varphi(l-1)}}(q_{\varphi(l)},a_l)
    \cdots\linebreak[1]\tr{ac_{\varphi(m-1)}}(q_{\varphi(m)},a_m)$
    such that $s_k\in a_l$ and $s_k=s_n\in a_m$.

    If $a_l=a_m$ then $\sigma_m$ is lasso-shaped and preserves the
    acceptance conditions visited by $\pi$. Hence the lemma is
    verified.

    Unfortunately it is possible that the aggregate $a_m$ and $a_k$
    are different because they were built from different predecessors.
    In that case, consider the lasso-shaped path, where the cycle has
    been unrolled $2^\Gamma$ times.  Then applying
    Lemma~\ref{lem:sop:finitepath} allows to build a path with
    $\sigma_m$ that, among other states, traverses $2^\Gamma+1$ states
    of the form $\{(q_k,a_l)\}_{l\in{0\ldots 2^\Gamma}}$ with all
    $a_l$ containing the state $s_k=s_n$.  Since an aggregate is a
    subset of $\Gamma$, at least two of these $(q_k,a_l)$ are equal,
    and therefore we can construct a lasso-shaped accepting run that
    satisfies the lemma.
  \end{enumerate}
\end{proof}

\begin{thm}\label{th:sop}
  Let \CA{} be a TGBA, and \CT{} be a Kripke structure.  The SOP of
  \CA{} and \CT{} accepts a run if and only if the synchronized
  product of these two structures accepts a run.  In other word, we
  have $\Acc(\CA\otimes\CT) \ne \emptyset \iff \Acc(\CA\sop\CT) \ne
  \emptyset$.
\end{thm}

\begin{proof}
  $\Longleftarrow$ follows from Corollary~\ref{lem:sop:left};
  $\Longrightarrow$ follows from Lemma~\ref{lem:sop:right}.
\end{proof}
% LocalWords:  l'agrégat SOG TGBA Kokkarinen al Kripke topsep pt ac bigell ss

% LocalWords:  barycentric König's

%% file: slap.tex
\section{Self-Loop Aggregation Product (SLAP)}
\label{sec:slap}

This section presents a hybrid algorithm that is not restricted to
stuttering-invariant properties. It is a specialized synchronized
product that aggregates states of the KS as long as the TGBA state
does not change, and no \emph{new} acceptance conditions are
visited.

\subsection{Definition}

The notion of self-loop aggregation is captured by $\SF(q,ac)$, the
\textbf{S}elf-loop \textbf{F}ormulas (labeling edges $q\tr{}q$) that
are weaker in terms of visited acceptance conditions than $ac$.

When synchronizing with an edge of the property TGBA bearing $ac$
leading to $q$, successive states of the Kripke will be aggregated
as long as they model $\SF(q,ac)$. More formally, for a TGBA state
$q$ and a set of accepting condition $ac \subseteq \CF$, let us
define
\[\SF(q, ac) = \bigvee_{q\tr{f, ac'}q \in \delta \text{~s.t.~} ac'
  \subseteq ac} f\]
Moreover, for $a \subseteq \Gamma$ and $f \in \bool(\AP)$, we define
$\FSucc(a, f) = \{s' \in \Gamma \mid \exists s \in a,\, s
\rightarrow s'\in\Delta \land \lambda(s) \models f\}$. That is,
first \textbf{F}ilter $a$ to only keep states satisfying $f$, then produce their
\textbf{Succ}essors.
The difference between $\SuccF$ and $\FSucc$ is whether the
filter is applied on the source or destination states.
Similarly to $\ReachF$, we denote by $\FReach(a, f)$ the least subset
of $\Gamma$ satisfying both $a \subseteq \FReach(a, f)$ and
 $\FSucc(\FReach(a, f), f) \subseteq \FReach(a, f)$.

\begin{definition}[SLAP of a TGBA and a \KS]\label{def:slap}
  Given a TGBA $\CA = \langle \AP', \CQ, \CF, \delta, q^0 \rangle$ and
  a Kripke structure $\CT = \langle \AP, \Gamma, \lambda, \Delta, s_0
  \rangle$ over $\AP \supseteq \AP'$ , the \emph{Self-Loop
    Aggregation Product} of $\CA$ and $\CT$ is the TGBA denoted $\CA
  \boxtimes \CT = \langle \emptyset, \CQ^{}_\boxtimes, \CF, \delta^{}_\boxtimes,
  q^0_\boxtimes \rangle$ where:
\begin{itemize}[topsep=0pt]
\item $\CQ^{}_\boxtimes = \CQ \times (2^\Gamma \setminus \{\emptyset\})$
\item $\displaystyle\delta^{}_\boxtimes = \left\{(q_1, a_1)\tr{\top, ac}(q_2,
  a_2) \left|
    \begin{aligned}
      \exists f \in \bool(&\AP') \text{~s.t.~} q_1 \tr{f, ac} q_2 \in \delta,\\
      & q_1 = q_2 \Rightarrow ac \neq \emptyset, \text{~and}\\
      & a_2 = \FReach(\FSucc(a_1, f), \SF(q_2, ac))\\
    \end{aligned}
    \right.\right\}$
\item $q^0_\boxtimes = (q^0, \FReach(\{s_0\}, \SF(q^0, \emptyset)))$
\end{itemize}
\end{definition}

Note that because of the way the product is built, it is not obvious
what Boolean formula should label the edges of the SLAP product. Since
in fact this label is irrelevant when checking language emptiness, we
label all arcs of the SLAP with $\top$ and simply denote
$(q_1,a_1)\fireseq{ac}(q_2,a_2)$ any transition
$(q_1,a_1)\tr{\top,ac}(q_2,a_2)$ of the SLAP.

$\CQ \times 2^\Gamma$ might seem very large but, as we will see in
section~\ref{sec:bench}, in practice the reachable states of the SLAP
is a much smaller set than that of the product $\CQ\times \Gamma$.
Furthermore the $\FReach$ operation can be efficiently implemented as
a symbolic least fix point.

Fig.~\ref{fig:slap} represents the SLAP built from our example \KS, and
the TGBA of $a \U b$. The initial state of the SLAP iteratively
aggregates successors of states verifying
  $\SF(q^0, \emptyset) = a\bar b$. Then following the edge $q^0 \tr{b,
  \emptyset} q_1$,
 states are aggregated with condition $\SF(q_1, \emptyset) =
 \bot$. Hence $q_1$ is synchronized with successors of states in
 $\{s_0, s_1, s_2, s_3, s_4\}$ satisfying $b$ (i.e., successors
 of $\{s_4\}$). Finally, when synchronizing with edge $q_1 \tr{\top,
 \bc} q_1$, we have $\SF(q_1,\{\bc\}) = \top$, hence all states of the
 cycle $\{s_4, s_5, s_6, s_7\}$ are added.

\subsection{Proof of correctness}

As for the SOP, we aim at demonstrating that, given a \KS and a TGBA,
the emptiness of the language of the corresponding SLAP is equivalent
to the emptiness of the language of the original synchronized product. This result is progressively demonstrated in the following by means of several intermediate lemmas.

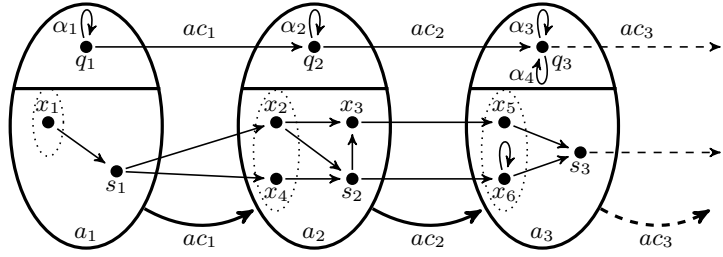
\begin{figure}[htbp]
  \centering
  \begin{tikzpicture}[automaton]
    \tikzstyle{bigell}=[ellipse,minimum width=20mm, minimum height=32.36mm]
    \begin{scope}[very thick]
    \node[draw,bigell] (S1) at (0,0) {};
    \draw (S1.27) -- (S1.153);
    \node[draw,bigell] (S2) at (3,0) {};
    \draw (S2.27) -- (S2.153);
    \node[draw,bigell] (S3) at (6,0) {};
    \draw (S3.27) -- (S3.153);
    \node[overlay,bigell] (S4) at (9,0) {};
    \path[->] (S1.-55) edge[bend right] node[below]{$ac_1$} (S2.-125)
              (S2.-55) edge[bend right] node[below]{$ac_2$} (S3.-125)
              (S3.-55) edge[bend right,dashed] node[below]{$ac_3$} (S4.-125);
    \end{scope}

    \filldraw
    (S1) +(0,10.5mm) circle(2pt) node(q1){} node[below]{\small $q_1$}
    (S2) +(0,10.5mm) circle(2pt) node(q2){} node[below]{\small $q_2$}
    (S3) +(0,10.5mm) circle(2pt) node(q3){} node[below right]{\small $q_3$}
    (S1) +(-5mm,0.5mm) circle(2pt) node(x1){} node[above]{\small $x_1$}
    (S1) +(4mm,-6mm) circle(2pt) node(ss1){} node[below]{\small $s_1$}
    (S2) +(-5mm,0.5mm) circle(2pt) node(x2){} node[above]{\small $x_2$}
    (S2) +(5mm,0.5mm) circle(2pt) node(x3){} node[above]{\small $x_3$}
    (S2) +(-5mm,-7mm) circle(2pt) node(x4){} node[below]{\small $x_4$}
    (S2) +(5mm,-7mm) circle(2pt) node(ss2){} node[below]{\small $s_2$}
    (S3) +(-5mm,0.5mm) circle(2pt) node(x5){} node[above]{\small $x_5$}
    (S3) +(-5mm,-7mm) circle(2pt) node(x6){} node[below]{\small $x_6$}
    (S3) +(5mm,-3.5mm) circle(2pt) node(ss3){} node[below]{\small $s_3$};

    \node[draw,ellipse,dotted,minimum width=4mm,minimum height=9mm] (In1) at (x1) {};
    \node[draw,ellipse,dotted,minimum width=6mm,minimum height=16mm] (In2) at (barycentric cs:x2=1,x4=1) {};
    \node[draw,ellipse,dotted,minimum width=6mm,minimum height=16mm] (In3) at (barycentric cs:x5=1,x6=1) {};

    \path[overlay] (S4) +(-5mm,10.5mm) circle(2pt) node(q4){}
                   (S4) +(-5mm,-3.5mm) circle(2pt) node(ss4){};

    \path (S1.-90) node[above] {$a_1$};
    \path (S2.-90) node[above] {$a_2$};
    \path (S3.-90) node[above] {$a_3$};

    \path[->] (q1) edge node[above]{$ac_1$} (q2)
              (q1) edge[loop above] node[left=-2.5pt,pos=.1]{$\alpha_1$} (q1)
              (q2) edge node[above]{$ac_2$} (q3)
              (q3) edge[dashed] node[above]{$ac_3$} (q4)
              (q2) edge[loop above] node[left=-2.5pt,pos=.1]{$\alpha_2$} (q2)
              (q3) edge[loop above] node[left=-2.5pt,pos=.1]{$\alpha_3$} (q3)
              (q3) edge[loop below] node[left=-2.5pt,pos=.75]{$\alpha_4$} (q3)
              (x1) edge (ss1)
              (x2) edge (x3)
              (x4) edge (ss2)
              (x2) edge (ss2)
              (ss2) edge (x3)
              (x5) edge (ss3)
              (x6) edge (ss3)
              (x6) edge[loop above] (x6)
              (ss1) edge (x4)
              (ss1) edge (x2)
              (x3) edge (x5)
              (ss2) edge (x6)
              (ss3) edge[dashed] (ss4)
              ;

  \end{tikzpicture}
  \caption{\label{fig:slapex} A prefix
    $(q_1,a_1)\fireseq{ac_1}(q_2,a_2)\fireseq{ac_2}(q_2,a_2)$ of a run
    of some SLAP $\CA\boxtimes\CT$ (with \emph{different} $\CA$ and
    $\CT$ from Fig.~\ref{fig:examples}) is shown using big ellipses
    and bended arrows.  The straight lines also shows the underlying
    connections between the states $\{q_1,q_2,q_3,\ldots\}$ of the
    automaton $\CA$ and between the states
    $\{s_1,s_2,\ldots,x_1,x_2\ldots\}$ of the Kripke structure $\CT$
    that have been aggregated as $a_1,a_2,a_3\ldots$ The acceptance
    conditions have been depicted as $ac_i$ or $\alpha_i$ and the
    labels of the transitions have been omitted for clarity.  The
    dotted ellipses show the set of input states ($In(a_1)$,
    $In(a_2)$, $In(a_3)$) as used in the proof of
    Lemma~\ref{rem:slap}.}
\end{figure}

\begin{lemma}\label{rem:slap}
  Let $\CA$ and $\CT$ be defined as in Definition~\ref{def:slap}.  Let
  $(q_1,a_1)\fireseq{ac}(q_2,a_2)\in\delta_{\boxtimes}$ be a
  transition of the SLAP $\CA\boxtimes\CT$.  For any state $s_2\in a_2$
  there exists at least one (possibly indirect) ancestor $s_1\in a_1$
  such that
  $(q_1,s_1)\fireseq{ac}(q_2,t_1)\fireseq{\alpha_1}(q_2,t_2)\fireseq{\alpha_2}\cdots(q_2,t_n)\fireseq{\alpha_n}(q_2,s_2)$
  is a sequence of the synchronized product $\CA\otimes\CT$ with
  $\forall i,\,t_i\in a_2$, and $\forall i,\,\alpha_i\subseteq ac$.
\end{lemma}

\noindent
For example consider transition $(q_1,a_1)\fireseq{ac}(q_2,a_2)$ on
Fig.~\ref{fig:slapex}, and some state in $a_2$, say $s_2$.  Then
$s_1\in a_1$ is an indirect ancestor of $s_2$ s.t.
$(q_1,s_1)\fireseq{ac}(q_2,x_2)\fireseq{\alpha_2}(q_2,s_2)$.

\begin{proof}
  Let us define the set of input states of the aggregate $a_2$ as
  $In(a_2)=\{ s'\in a_2\mid \exists s\in a_1, s\fireseq{}s' \in \Delta
  \}$.  This set cannot be empty since
  $(q_1,a_1)\fireseq{ac}(q_2,a_2)$.

  Consider a state $s_2\in a_2$.  By construction of $a_2$, $s_2$ is
  reachable from some state in $t_1\in In(a_2)$, so there exists a
  path $t_1\fireseq{}t_2\fireseq{}\cdots\fireseq{}s_2$ in the Kripke
  structure.

  By definition of $\delta_\boxtimes$, if $t_1,t_2,\ldots,s_2$ belong
  to $a_2$, the transitions between these states of $\CT$ have been
  synchronized with self-loops $q_2\tr{\alpha_i}q_2$ of $\CA$ with
  $\alpha_i\subseteq ac$.
  Therefore the sequence
  $(q_2,t_1)\fireseq{\alpha_1}(q_2,t_2)\fireseq{\alpha_2}\cdots(q_2,t_n)\fireseq{\alpha_n}(q_2,s_2)$
  is a sequence of the synchronized product $\CA\otimes\CT$.

  Moreover, since $t_1\in In(a_2)$, there exists a state $s_1$ in
  $a_1$ such that $(q_1,s_1)\fireseq{ac}(q_2,t_1)$.

  Consequently, the path
  $(q_1,s_1)\fireseq{ac}(q_2,t_1)\fireseq{\alpha_1}(q_2,t_2)\fireseq{\alpha_2}\cdots(q_2,t_n)\fireseq{\alpha_n}(q_2,s_2)$
  satisfies the lemma.
\end{proof}

\begin{lemma}\label{lem:left}
  If there exists $\sigma\in\Acc(\CA\boxtimes\CT)$ an infinite run
  accepted by the SLAP, then there exists an accepting run
  $\pi\in\Acc(\CA\otimes\CT)$ in the classical product.
\end{lemma}
\begin{proof}
  Let us denote
  $\sigma=(q_1,a_1)\fireseq{ac_1}(q_2,a_2)\fireseq{ac_2}(q_3,a_3)\fireseq{ac_3}\cdots$
  an accepting run of $\CA\boxtimes\CT$.
  Let us build an infinite tree in which all nodes (except the root) are
  states of $\CA\otimes\CT$.  Let us call $\top$ the root, at depth
  $0$.  The set of nodes at depth $n>0$ is exactly the finite set
  of pairs $\{(q_n,s)\mid s\in a_n\}\subseteq \CQ\times\Gamma$.

  The parent of any node at level $1$ is $\top$.  For any $i>0$, the
  parent of a node $(q_{i+1},s')$ with $s'\in a_{i+1}$ is the node
  $(q_i,s)$ for is any state $s\in a_i$ such that $(q_{i},s)$ is a
  (possibly indirect) ancestor of $(q_{i+1},s')$ such that we observe
  $ac_i$ on the path between these two states.  We know such a state
  $(q_i,s)$ exists because of Lemma~\ref{rem:slap}.  As a consequence
  of this parenting relation, every edge in this tree, except those
  leaving the root, correspond to a path between two states of
  $\CA\otimes\CT$.

  Because the set of nodes at depth $n>0$ is finite, this infinite
  tree has finite branching.  By König's Lemma it therefore contains
  an infinite branch.  By following this branch and ignoring the first
  edge, we can construct a path of $\CA\otimes\CT$ that starts in
  $(q_1,s_1)$ for some $s_1\in a_1$, and that visits at least all the
  acceptance conditions $ac_i$ of $\sigma$ in the same order (and
  maybe more).  To prove that this accepting path we have constructed
  actually occurs in a run of $\CA\otimes\CT$, it remains to show that
  $(q_1,s_1)$ is a state that is accessible from the initial state of
  $\CA\otimes\CT$.

  Obviously $q_1=q^0$ because $(q_1,a_1)=q^0_\boxtimes$ is the initial
  state of $\CA\boxtimes\CT$.  Furthermore we have $s_1\in a_1$, so by
  definition of $q^0_\boxtimes$, $(q^0,s_1)$ must be reachable from
  (or equal to) $(q^0,s_0)$ in $\CA\otimes\CT$.
\end{proof}
\begin{lemma}\label{lem:finitepath}
  For a given $n$ and a finite path
  $\pi_n=(q_0,s_0)\tr{f_0,ac_0}(q_1,s_1)\cdots\tr{f_{n-1},ac_{n-1}}(q_n,s_n)$
  of $\CA\otimes\CT$, there exists a finite path
  $\sigma_n=(q'_0,a_0)\tr{ac_{\varphi(0)}}(q'_1,a_1)\cdots\tr{ac_{\varphi(m-1)}}(q'_m,a_m)$
  of $\CA\boxtimes\CT$, with $m\le n$, $q_n=q'_m$, $s_n\in a_m$ and
  $\varphi_n:\{0,\ldots,m-1\}\to\{0,\ldots,n-1\}$ is a strictly
  increasing function such that $\forall j\,(\exists i,
  \varphi_n(i)=j\iff ac_i\ne \emptyset)$.
\end{lemma}
\begin{proof}
  Let us prove this lemma by induction on $n$.
  It is true if $n=0$: Given $\pi_0=(q_0,s_0)$, the path
  $\sigma_0=(q'0,a_0)=q^0_\boxtimes=(q_0, \FReach(\{s_0\},
  \{\lambda(s_0)\} \cap \lambda(q_0, \emptyset))$ satisfies the
  conditions (with $\varphi$ being a null function).

  Let us now assume that the lemma is true for $n+1$ assuming it is
  true for $n$.  Given a path
  $\pi_{n+1}=\pi_n\tr{f_{n},ac_{n}}(q_{n+1},s_{n+1})$, we know by
  hypothesis that we have a matching $\sigma_n$ for $\pi_n$.  Let us
  consider how to extend $\sigma_n$ into $\sigma_{n+1}$ to handle the
  new transition $(q_n,s_n)\tr{f_{n},ac_{n}}(q_{n+1},s_{n+1})$ of
  $\pi_{n+1}$.

  There are two cases to consider:
  \begin{enumerate}[topsep=0pt]
  \item If $q_n=q_{n+1}$ and $acc_n=\emptyset$ and $\lambda(s_{n+1})\models\SF(q_n,ac)$,
    then by definition of $\FSucc$ and $\SF$ the last state of $\sigma_n$, $(q'_m,a_m)$ is
    such that $s_{n+1}\in a_m$ and $q'_m=q_n=q_{n+1}$.  In that case $\sigma_{n+1}=\sigma_n$,
    and $\varphi_{n+1}=\varphi_n$.
  \item If $q_n\ne q_{n+1}$ or $acc_n\ne\emptyset$ or $\lambda(s_{n+1})\not\models\SF(q_n,ac)$,
    then because $\lambda(s_n)\models f_n$ and $s_n\tr{}s_{n+1}$, by definition of
    $\delta_\boxtimes$ there exists $(q'_m,a_m)\tr{acc_n}(q'_{m+1},a_{m+1})$ such that
    $s_{n+1}\in a_{m+1}$ and $q'_{m+1}=q_{n+1}$.  In this case, we can define
    $\sigma_{n+1}=\sigma_{n}\tr{acc_n}(q'_{m+1},a_{m+1})$ with
    $\forall i<n,\,\varphi_{n+1}(i)=\varphi_n(i)$ and $\varphi_{n+1}(n)=n$.
  \end{enumerate}
  So by induction this lemma is true for all $n\in\nat$.
\end{proof}
\begin{lemma}\label{lem:right}
  If there exists an infinite path $\pi\in\Acc(\CA\otimes\CT)$
  accepting in $\CA\otimes\CT$.  Then there exists an accepting path
  in $\CA\boxtimes\CT$ as well.
\end{lemma}
\begin{proof}
  $\CA\otimes\CT$ has a finite number of states, so if
  $\Acc(\CA\otimes\CT)\ne\emptyset$ then it contains at least one infinite path
  $\pi\in\Acc(\CA\otimes\CT)$ that can be represented as a finite
  prefix followed by a finite cycle that is repeated infinitely often.

  Lemma~\ref{lem:finitepath} tells us that any prefix $\pi_n$ of $\pi$
  corresponds to some prefix $\sigma_n$ of a path in $\CA\boxtimes\CT$
  in which the acceptance conditions of $\pi_n$ occur in the same
  order.  We have $|\sigma_n|\le|\pi_n|=n$ but because $\pi$ will
  visit all acceptance conditions infinitely often, and these
  transitions will all appear in $\sigma_n$ (only transition without
  acceptance conditions can be omitted from $\delta_\boxtimes$), we
  can find some value of $n$ for which $|\sigma_n|$ is arbitrary
  large.  Because $|\sigma_n|$ can be made larger than the size of the
  SLAP, at some point this finite sequence will have to loop in a way
  that visits the acceptance conditions exactly in the same order as
  they appear in the cycle part of $\pi$.  By repeating this cycle
  part of $\sigma_n$ we can therefore construct an infinite path
  $\sigma$ that is accepted by $\CA\boxtimes\CT$.
\end{proof}
\begin{thm}
  Let \CA{} be a TGBA, and \CT{} be a Kripke structure.  We have
  \[
     \Acc(\CA\otimes\CT) \ne \emptyset \iff \Acc(\CA\boxtimes\CT) \ne \emptyset
  \]
  In other words, the SLAP of \CA{} and \CT{} accepts a run if and
  only if the synchronized product of these two structures accepts a
  run.
\end{thm}
\begin{proof}
  $\Longleftarrow$ follows from Lemma~\ref{lem:left};
  $\Longrightarrow$ follows from Lemma~\ref{lem:right}.
\end{proof}

\subsection{Mixing SLAP and Fully Symbolic Approaches}
\label{sec:slap-fst}

This section informally presents a variation on the SLAP algorithm, to
use a fully symbolic algorithm in cases where the automaton state will
no longer evolve.

The principle is the following: when the product has reached a state
where the formula automaton state is terminal (i.e., it has itself as
only successor), we proceed to use a fully symbolic search for an
accepted path in the states of the current aggregate. This variant is
called SLAP-FST, standing for Fully Symbolic search in Terminal
states. Note that we suppose here that such a terminal state allows
accepting runs, otherwise semantic simplifications would have removed
the state from the TGBA.

In this variant, if $q_1$ is a terminal state, i.e., $\nexists q_1
\tr{f, ac} q_2 \in \delta$, with $q_1 \neq q_2$, a state $(q_1,a_1)$
of the product has itself as sole successor through an arc labeled
$(\top, \CF)$ if and only if $a_1$ admits a solution computed using a
fully symbolic algorithm, or has no successors otherwise.

The fully symbolic search uses the self-loop arcs on the formula TGBA
state to compute the appropriate transition relation(s), and takes
into account possibly multiple acceptance conditions.

The rationale is that discovering this behavior when the aggregate is
large, and particularly if there are long prefixes before reaching the
SCC that bears all acceptance conditions, tends to create large SLAP
structures in explicit size.  The counterpart is that when no such
solution exists, the fully symbolic SCC hull search may be quite
costly.

In practice this variation on the SLAP was proposed after manually
examining cases where SLAP performance was disappointing, typically
because the SLAP was much larger in explicit size than the SOP. As
discussed in the performance section, this variation is on average
more effective than the basic SLAP algorithm.

% LocalWords:  TGBA Kripke pt ac acc FST SCC

%% file: perfs.tex
\section{Experimentations}
\label{sec:perf}
\label{sec:bench}

In this section we present experimental results comparing several algorithms for hybrid or fully symbolic
LTL model-checking.
We use two different benchmark sets, one based on Petri net models using randomly generated LTL formulas, and one based on BEEM (Benchmarks for Explicit Model checkers~\cite{BEEMspin07}) models using meaningful LTL properties.
We first present the context of these experiments, then focus on the results for Petri nets before detailing results for BEEM.

\input{implement.tex}

\subsection{Benchmark description}

We use here classic scalable Petri net examples taken from Ciardo's
benchmark set~\cite{ciardo03saturation}: slotted ring, Kanban,
flexible manufacturing system, and dining philosophers.
Table~\ref{tab:size} gives the size of each model.

\begin{table}[tbp]
  \centering
\begin{tabular}{lrclr}
model & state space &\hspace*{1cm}& model & state space\\
\cline{1-2}\cline{4-5}
fms5 & $2.9\times 10^6$ && kanban5 & $2.5\times 10^6$\\
fms10 & $2.5\times 10^9$ && kanban10 & $1\times 10^9$\\
philo10 & $4.6\times 10^6$ && ring6 & $5.8\times 10^5$\\
philo50 & $2.3\times 10^{33}$ && ring7 & $6.2\times 10^6$\\
philo100 & $5.1\times 10^{66}$ && ring10 & $8.3\times 10^9$\\
\end{tabular}
\caption{Number of reachable states in the selected models.\label{tab:size}}
\end{table}

The formulas considered include a selection of random LTL formulas,
which were filtered to have a (basic TGBA/Kripke) product size of at
least 1000 states.  We also chose to have as many verified formulas
(empty products) as violated formulas (non-empty products) to avoid
favoring on-the-fly algorithms too much.  To produce TGBA with several
acceptance conditions, this benchmark includes 200 formulas for each
model built from fairness assumptions of the form: $(\G\F p_1 \land
\G\F p_2 \ldots) \implies \varphi$.

We also used 100 random formulas that use the next operator, and hence
are not stuttering invariant (these where not used for SOG that does
not support them).

We killed any process that exceeded 120 seconds of runtime, and set
the garbage collection threshold at 1.3GB.  Cases where all considered
methods performed under 0.1s were filtered out from the results
presented here: theses trivial cases represent only 4.2\% of the
entire benchmark, and were too fast too be allow any pertinent
comparison.

\subsection{General comparison}

Table~\ref{tab:winner} gives a synthetic overview of the results
presented hereafter.  SLAP or SLAP-FST are the fastest methods in over
half of all cases, and they are rarely the slowest.  Furthermore, they
have the least failure rate.  This table also shows that BCZ has the
highest failure rate and that the fully symbolic algorithms (OWCTY,
EL) have trouble with non-empty products.

\begin{table}[tbp]
\centering
\input tablewin-tr
\caption{On all experiments (grouped with respect to the existence of
  a counterexample and the use of a $\X$ operator in the LTL formula),
  we count the number of cases a specific method has (Win) the best
  time or (Lose) it has either run out of time or it has the worst
  time amongst successful methods.  The Fail line shows how much of the
  Lost cases were timeouts.  The sum of a line may exceed 100\% if
  several methods are equally placed.\label{tab:winner}}
\end{table}

\begin{figure}[t]
\centering
\begin{tabular}{cc}
\epsfig{file=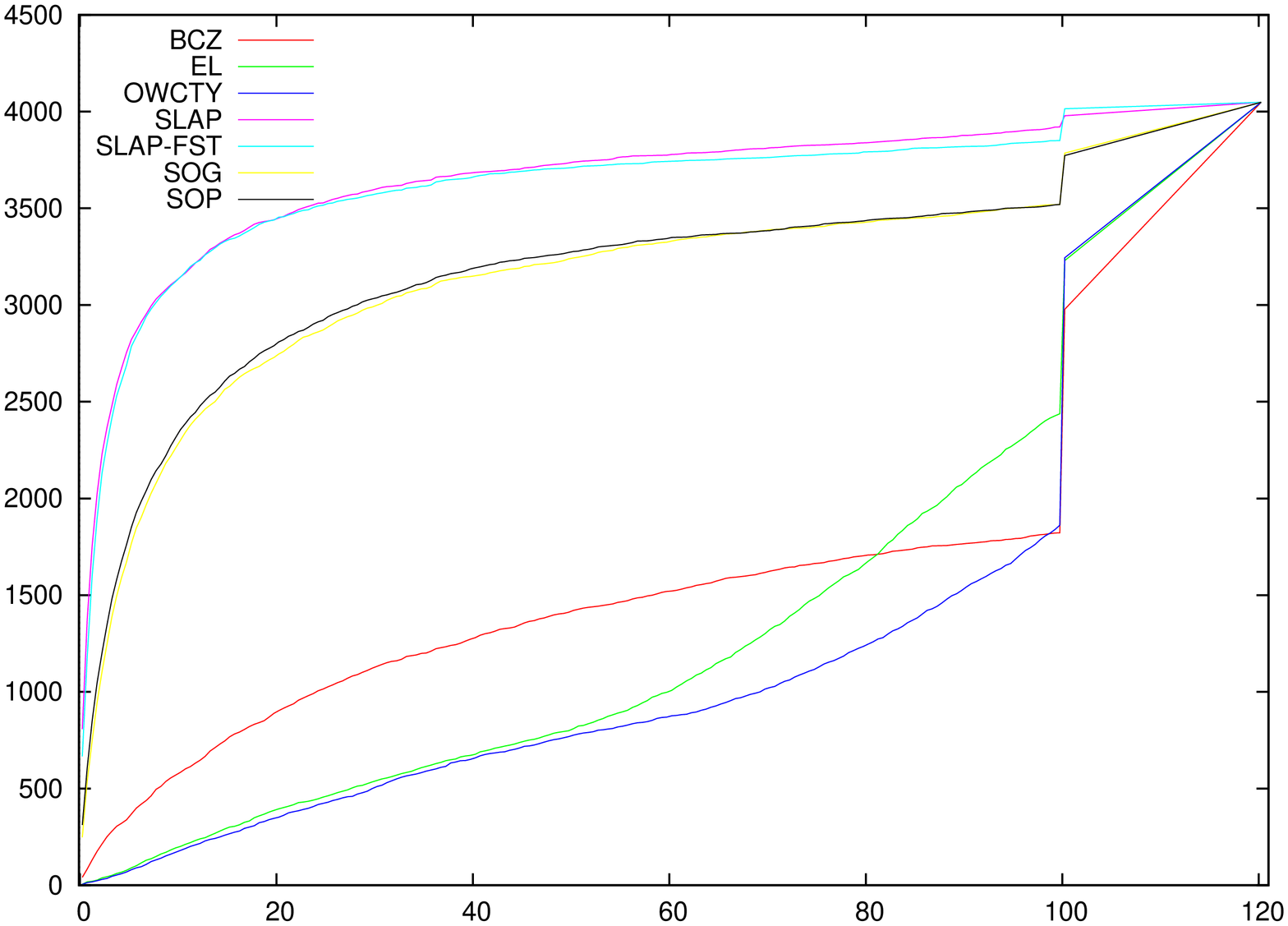,angle=-90,width=0.5\linewidth,clip=} &
\epsfig{file=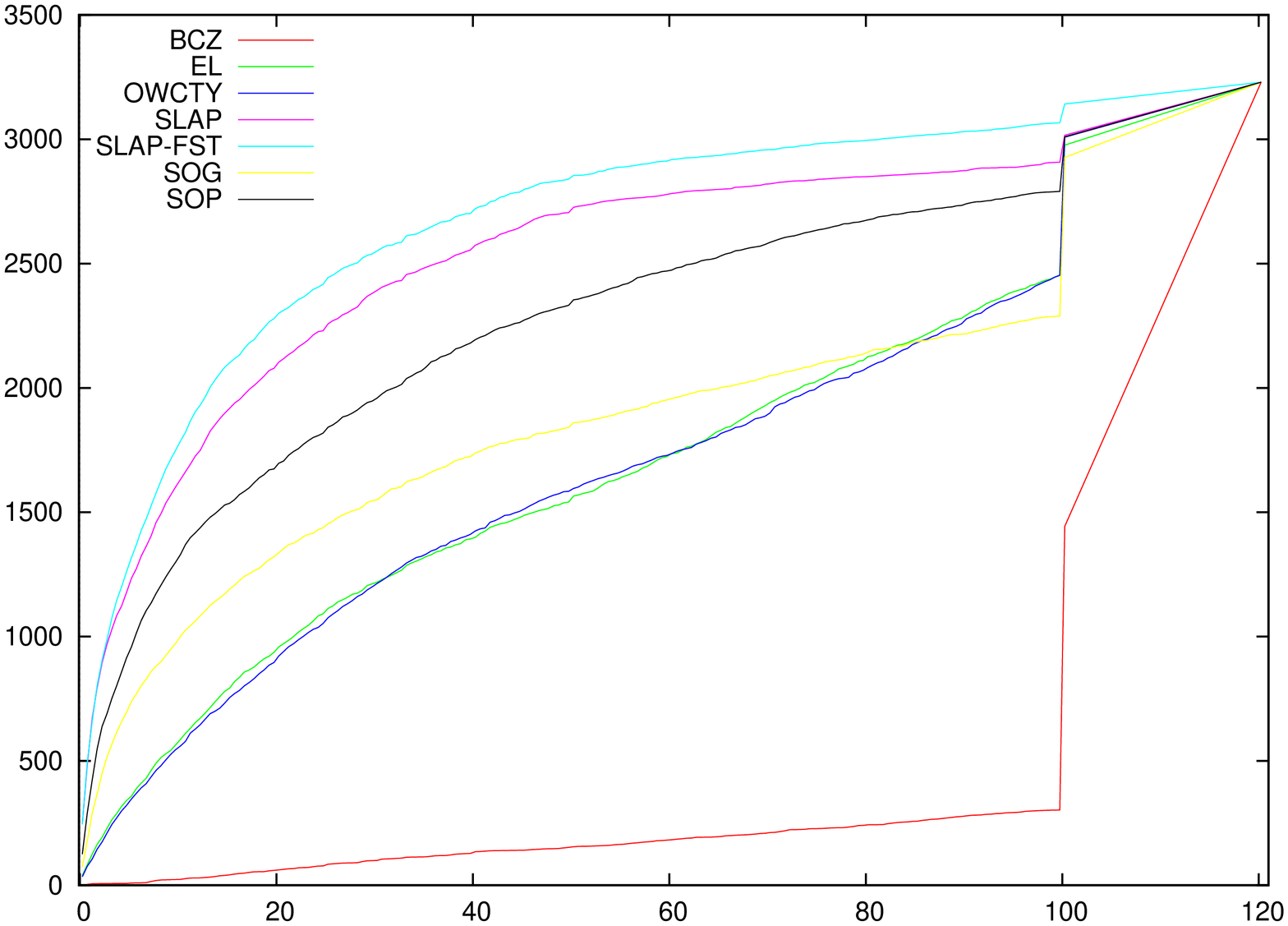,angle=-90,width=0.5\linewidth,clip=} \\
\end{tabular}
\caption{Cumulative plots comparing the time of all methods on stuttering invariant properties only. Non-empty
  products are shown on the left, and empty products on the
  right.\label{fig:cumul}}
\end{figure}

\begin{figure}[t]
\centering
\begin{tabular}{cc}
\epsfig{file=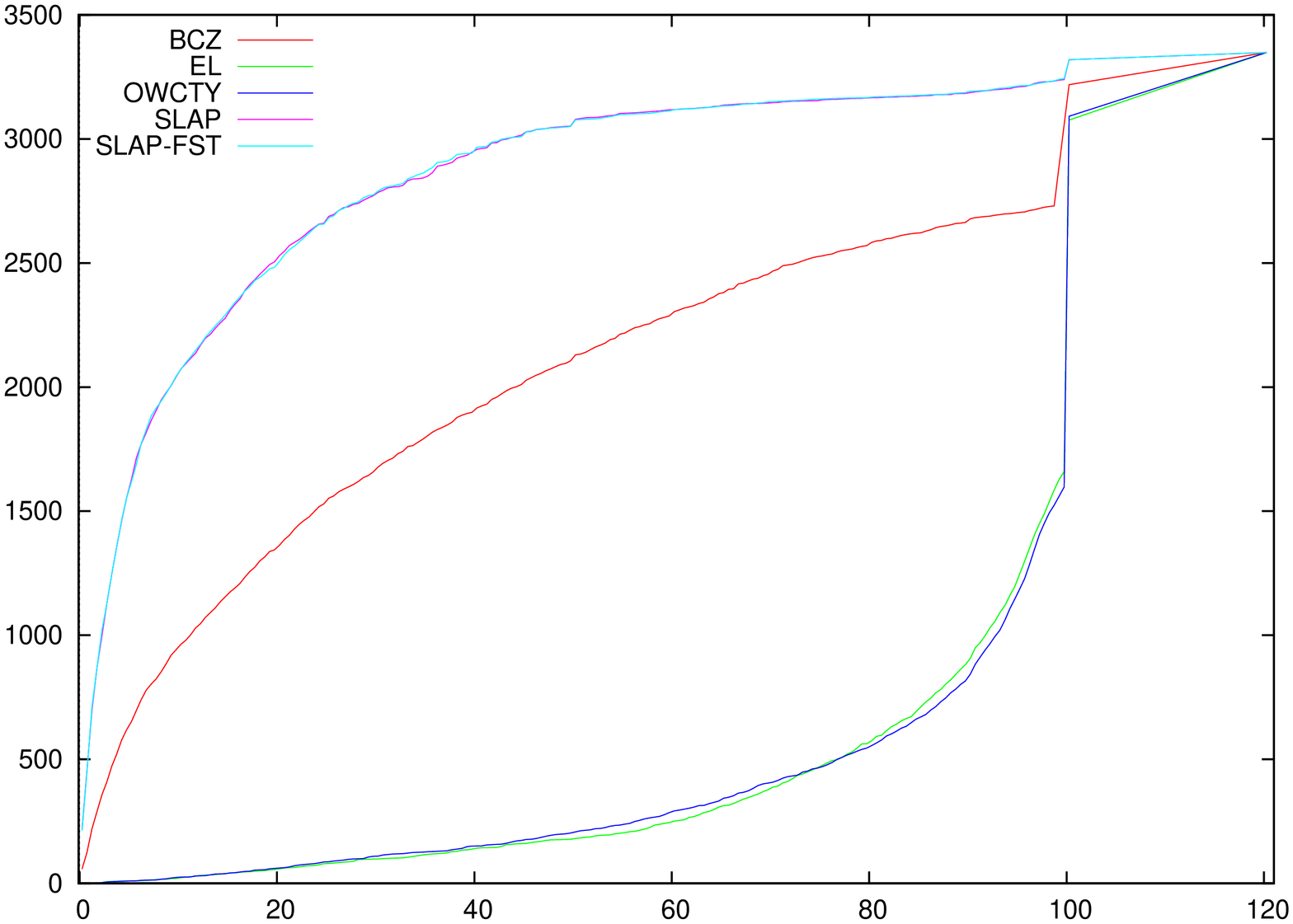,angle=-90,width=0.5\linewidth,clip=} &
\epsfig{file=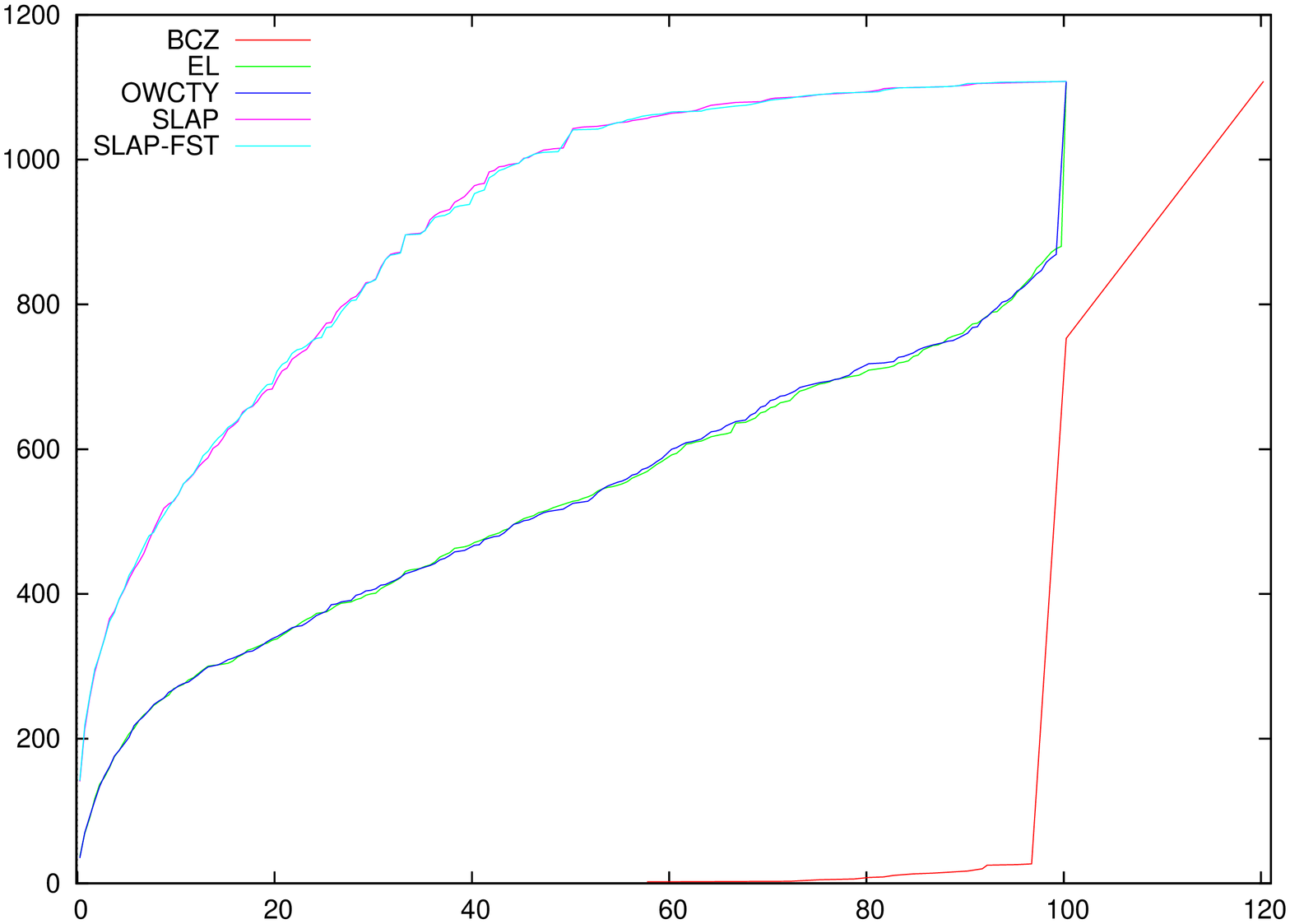,angle=-90,width=0.5\linewidth,clip=} \\
\end{tabular}
\caption{Cumulative plots comparing the time of all methods (except SOG and SOP) for non stuttering invariant properties. Non-empty
  products are shown on the left, and empty products on the right.\label{fig:cumul-x}}
\end{figure}

Table~\ref{tab:winner} presents only the best and the worst methods.
While Fig.~\ref{fig:cumul} and ~\ref{fig:cumul-x} allow to compare the different methods in
a finer manner.

For each experiment (model/formula pair) we first collect the maximum
time reached by a technique that did not fail, then compute for the
other approaches what percentage of this maximum was used. The
vertical segments visible at 100\% thus show the number of runs for
which this technique was the worst of those that did not fail.  Any
failures are plotted arbitrarily at 120\%.  This gives us a set of
values between 0\% and 120\% for which we plot the cumulative
distribution function.  For instance, if a curve goes through the
(20\%,2000) point, it means that for this technique, 2000 experiments
took at most 20\% of the time taken by the worst technique for the
same experiments.

The behavior at 120\% represents the ``Fail'' line of previous table,
while the behavior at 100\% represents the difference between the
``Slow'' and ``Fail'' lines (``Slow'' methods include methods that
failed).

The left plot of Fig~\ref{fig:cumul} for the non-empty cases shows that the on-the-fly
mechanism allows all hybrid algorithms (SLAP, SLAP-FST, SOG, SOP, BCZ) to
outperform the symbolic ones (OWCTY, EL).  However as seen previously,
BCZ still fails more often than other methods.  The SLAP and SLAP-FST
method take less than 10\% of the time of the slowest method in 80\%
of the cases. On left of Fig~\ref{fig:cumul-x}, the same effect is visible, although
BCZ actually has less failures than the fully symbolic algorithms.

The right plots for the empty cases show that fully symbolic algorithm
behave relatively far better (all methods have to explore the full
product anyway).  BCZ spends too much time exploring enormous
products, and timeouts.

For stuttering-invariant properties SLAP-FST and SLAP have similar
performance, with a slight edge for SLAP-FST when the product is
empty; however on Fig~\ref{fig:cumul-x} SLAP and SLAP-FST are
not significantly different.

EL appears slightly superior to OWCTY in the non-empty case, while
they have similar performances in the empty case. This is mostly
discernible on stuttering-invariant properties.

SOG and SOP show good results when there is a counterexample, and they
perform better than BCZ in most cases.  However SOG and SOP only
support stuttering-invariant properties. As shown in
Fig~\ref{fig:cumul-x} BCZ is a good alternative to fully symbolic
algorithms in presence of a counter-example; it is however
systematically outperformed by the new algorithms we propose in this
paper.

\subsection{SLAP versus SLAP-FST}

\begin{figure}[tbp]
\centering
\begin{tabular}{cc}
\epsfig{file=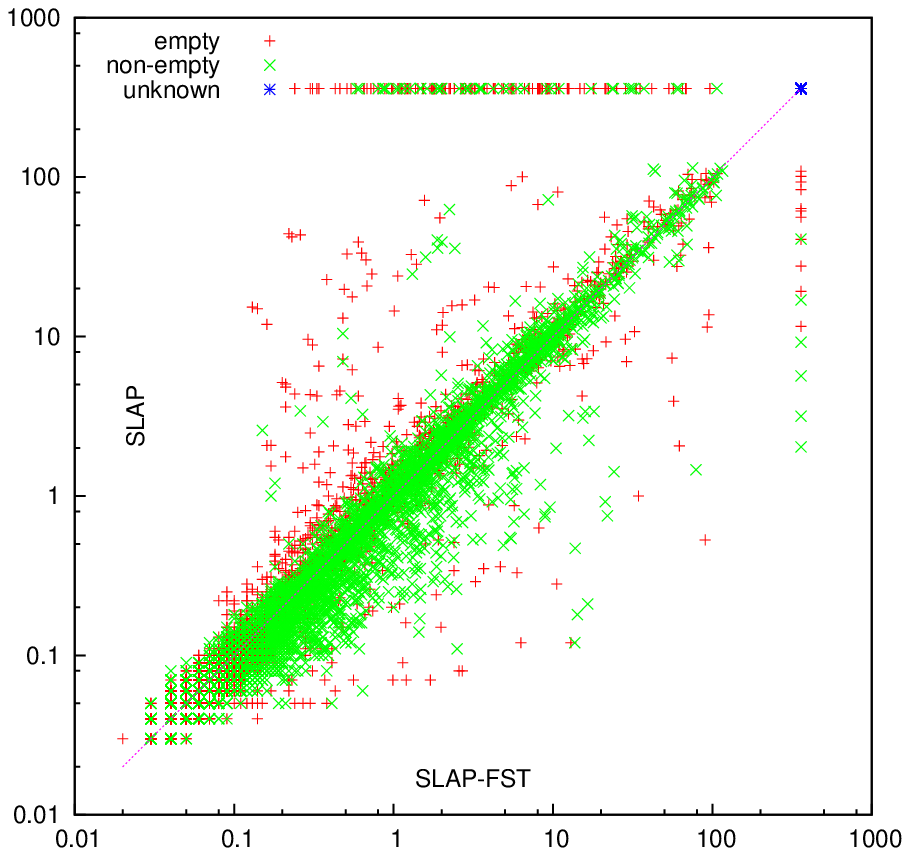,width=0.5\linewidth,clip=}&
\epsfig{file=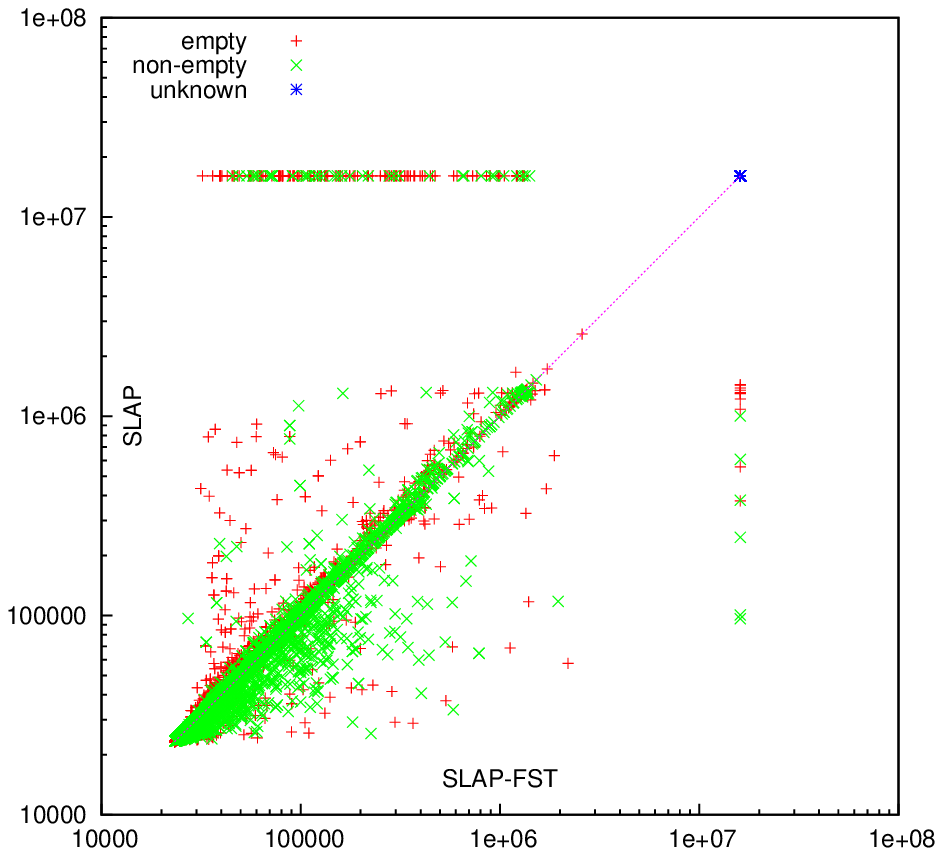,width=0.5\linewidth,clip=}\\
\end{tabular}
\epsfig{file=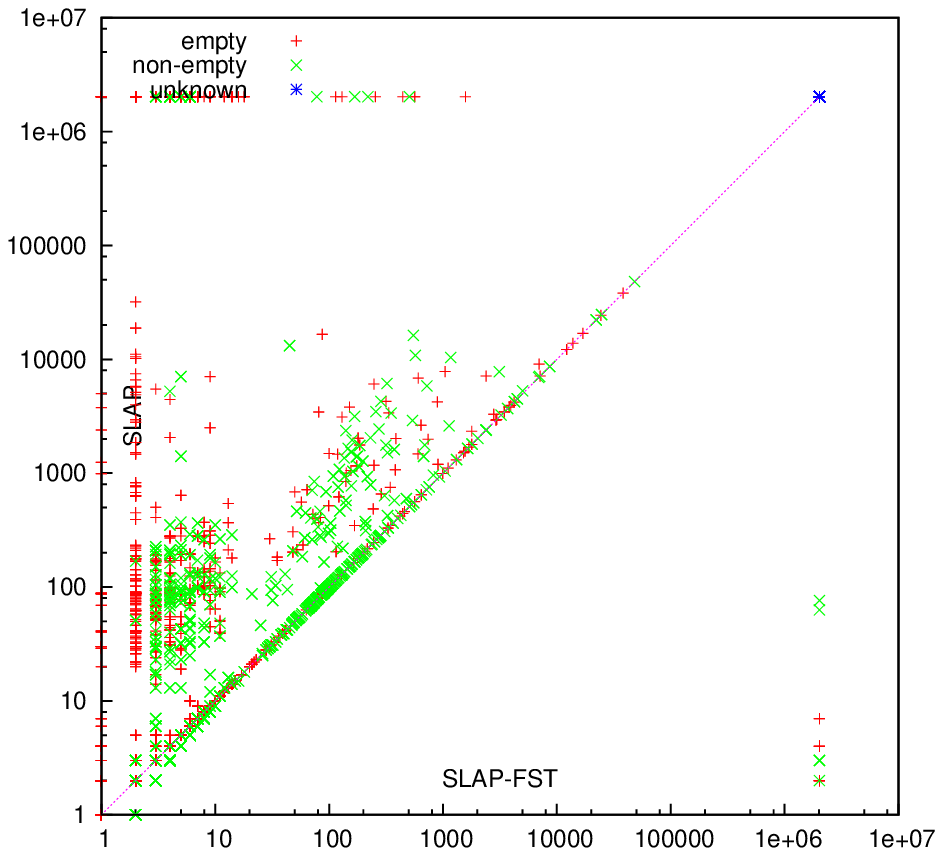,width=0.5\linewidth,clip=}
\caption{Comparison of SLAP-FST against SLAP. Top left: time (in seconds); top right: memory (in kilobytes); bottom: product size (in states).\label{fig:slap-fst-slap}}
\end{figure}

To study the differences between SLAP and SLAP-FST consider the
scatter plots from Fig.~\ref{fig:slap-fst-slap}.  The performances are
presented using a logarithmic scale.  Each point represents an
experiment, i.e., a model and formula pair.  We plot experiments that
failed (due to timeout) as if they had taken 360 seconds, so they are
clearly separated from experiments that didn't fail (by the wide white
band).

In these plots we have 11733 experiments, of which 132 proved too hard
to solve for either algorithm within the time limit.  Overall SLAP
algorithm solved 17 problems that SLAP-FST did not, and SLAP-FST
solved 179 instances that SLAP did not.  SLAP was at least a hundred
times slower than SLAP-FST in 5 cases, ten times slower in 50 cases,
and twice as slow in 164 cases.  SLAP-FST was one hundred times slower
in 3 cases, ten times slower in 34 cases, and twice as slow in 396
cases.

SLAP is on the average faster and consume less memory than SLAP-FST
for non-empty products, but fails more often.  SLAP-FST is better
overall for empty products. Indeed the explicit product size of
SLAP-FST is always smaller than that of SLAP, and often by several
orders of magnitude.  In some cases the SLAP degenerates to a
state-space proportional to size of the explicit product while the
SLAP-FST is able to keep the symbolic advantage.

\begin{figure}[tbp]
\centering
\begin{tabular}{cc}
\epsfig{file=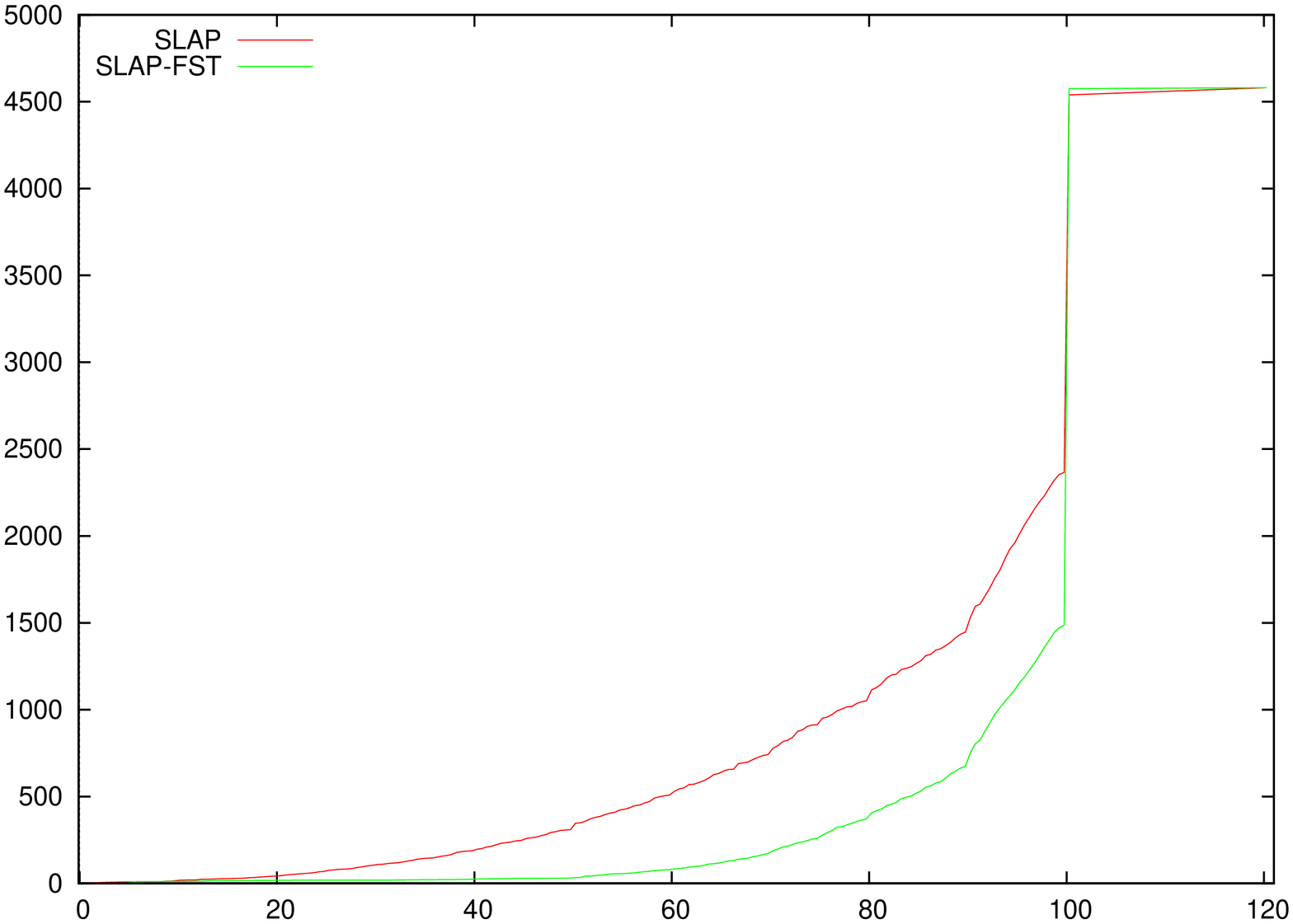,angle=-90,width=0.5\linewidth,clip=} &
\epsfig{file=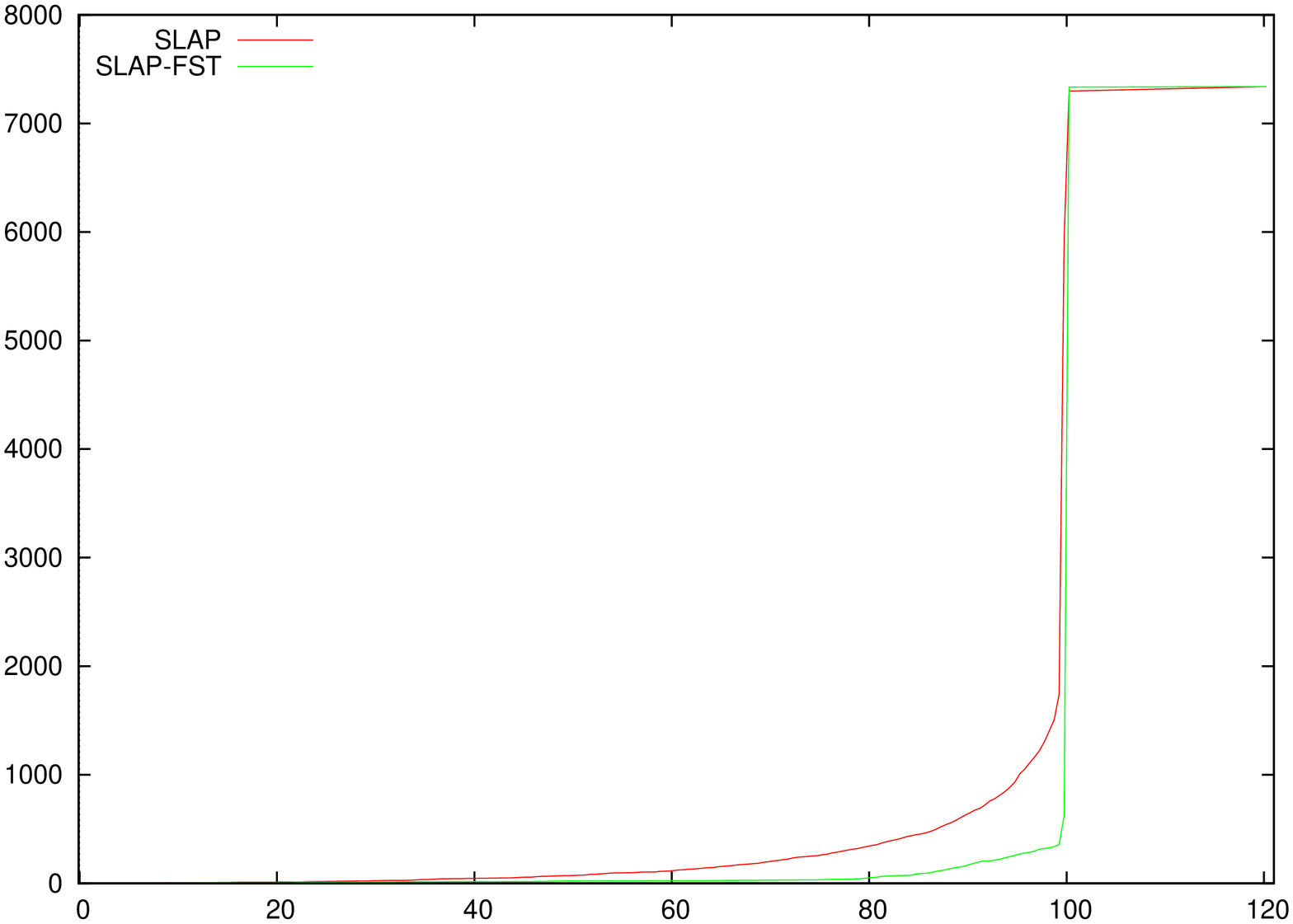,angle=-90,width=0.5\linewidth,clip=} \\
\epsfig{file=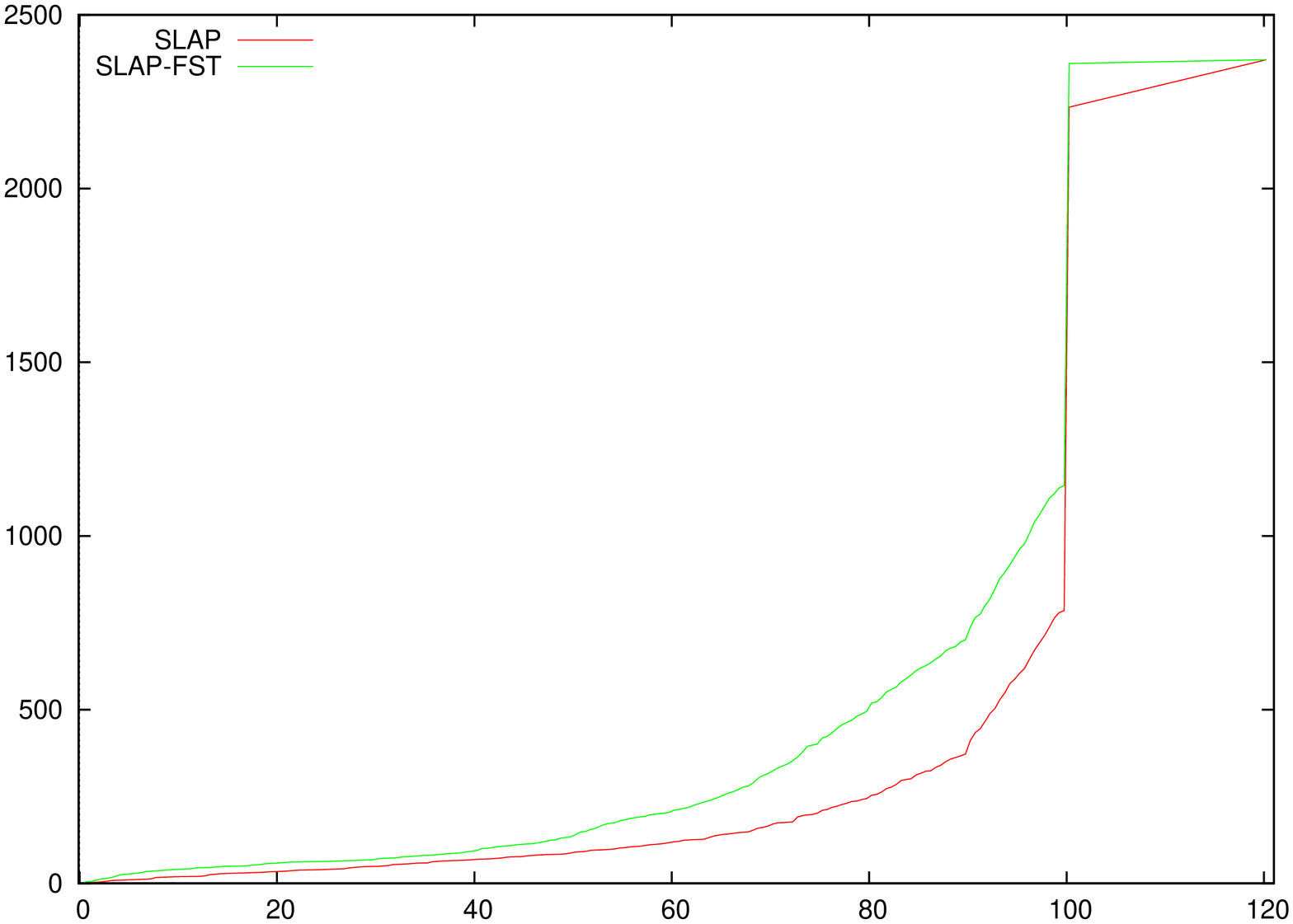,angle=-90,width=0.5\linewidth,clip=} &
\epsfig{file=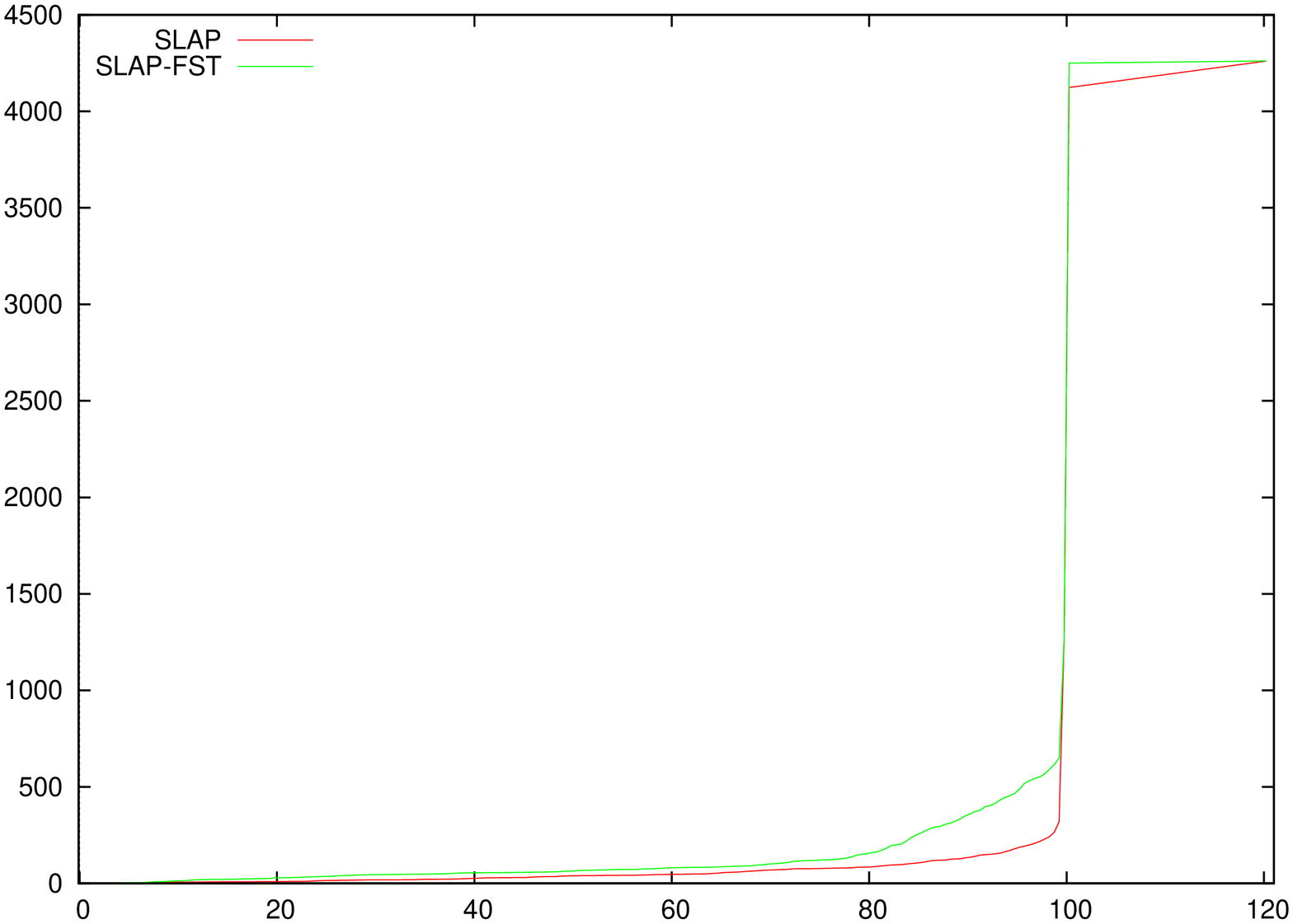,angle=-90,width=0.5\linewidth,clip=} \\
\end{tabular}
\caption{Compared cumulative performances in time (left) and memory (right) of SLAP variants. Top for non-empty products, bottom for empty ones.}
\label{fig:slaps-cumul}
\end{figure}

The cumulative plots of Fig.~\ref{fig:slaps-cumul} make this advantage
even more visible.
Indeed at the cost of slight memory overhead, and
a more significant overall time overhead (when counter-examples are
present), SLAP-FST produces much smaller explicit structures (hence
wins significantly for empty products).

\subsection{SLAP-FST versus other techniques}

\begin{figure}[tbp]
\centering
\begin{tabular}{cc}
\epsfig{file=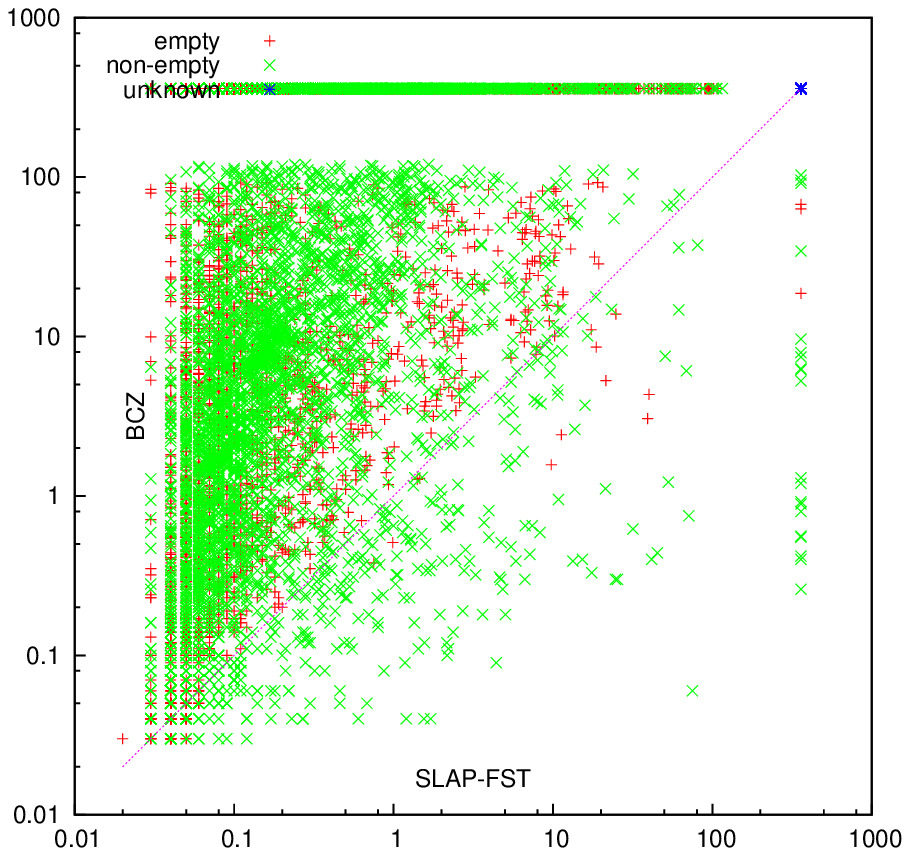,width=0.5\linewidth,clip=} &
\epsfig{file=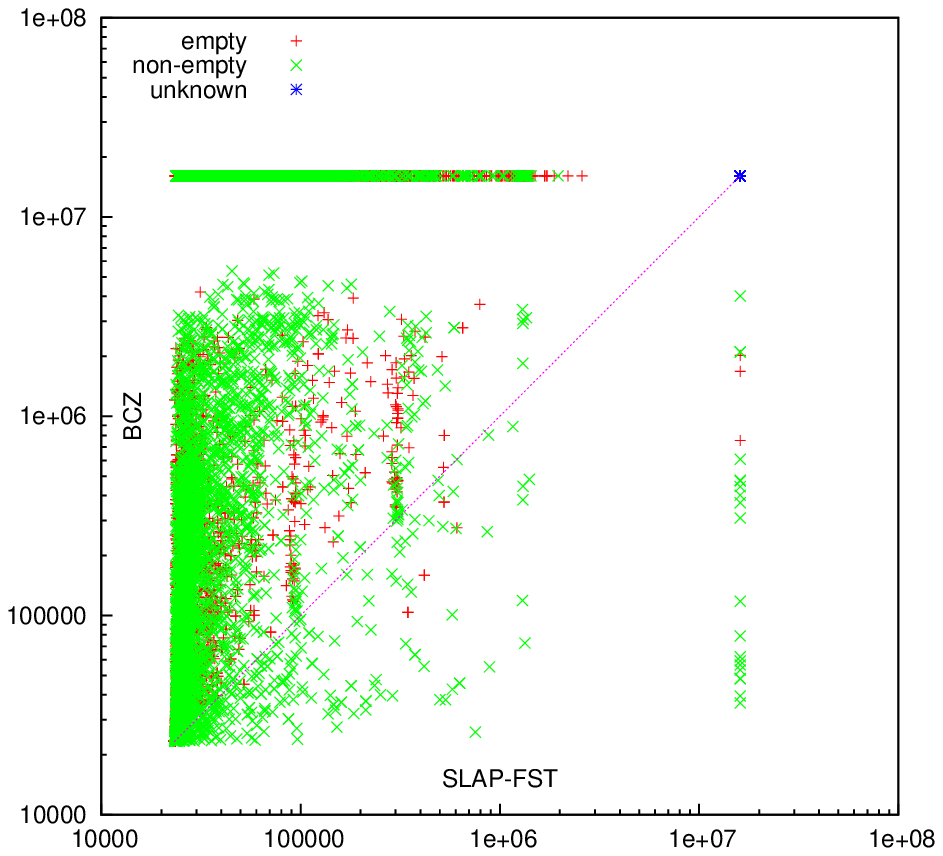,width=0.5\linewidth,clip=} \\
\epsfig{file=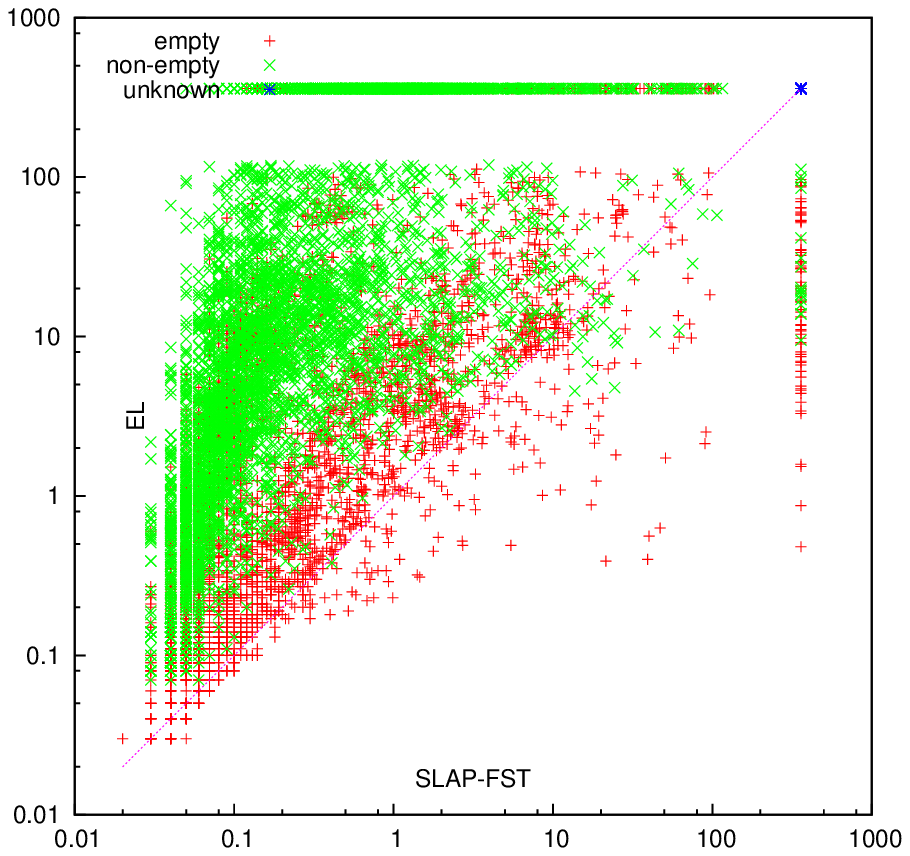,width=0.5\linewidth,clip=} &
\epsfig{file=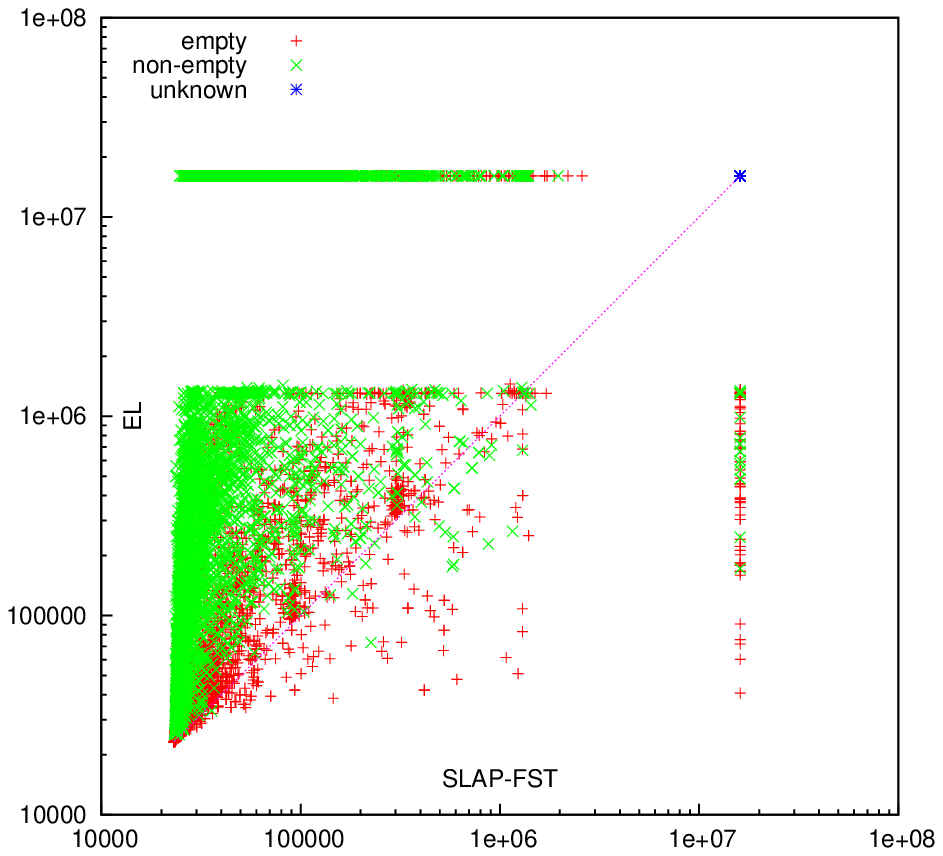,width=0.5\linewidth,clip=} \\
\epsfig{file=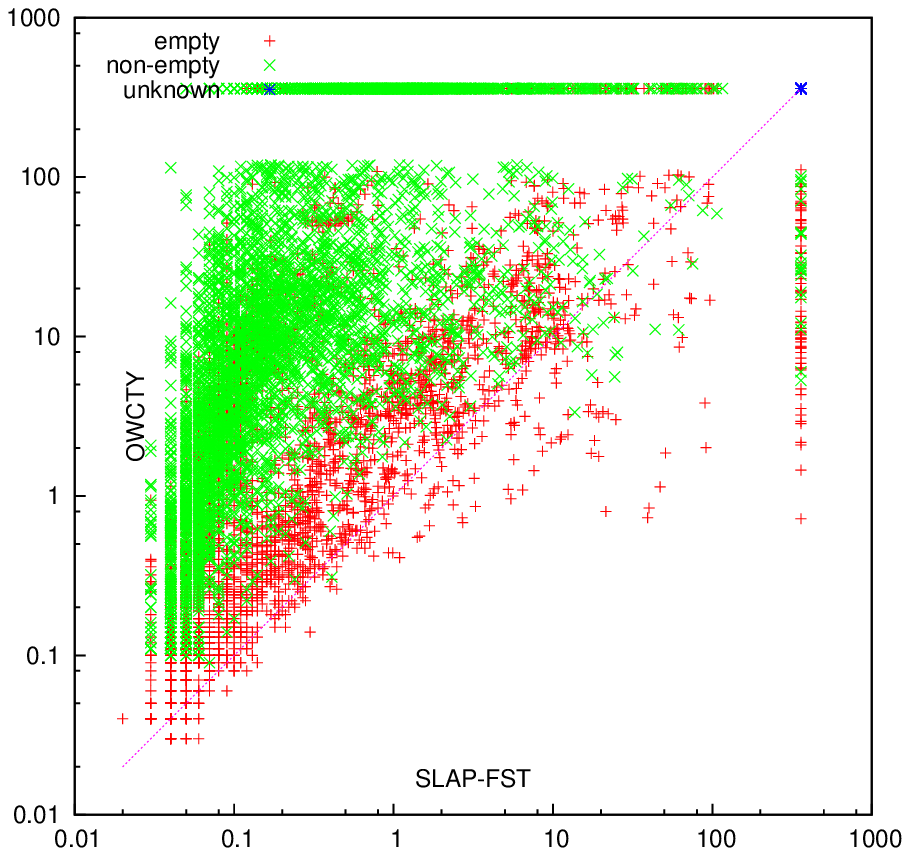,width=0.5\linewidth,clip=} &
\epsfig{file=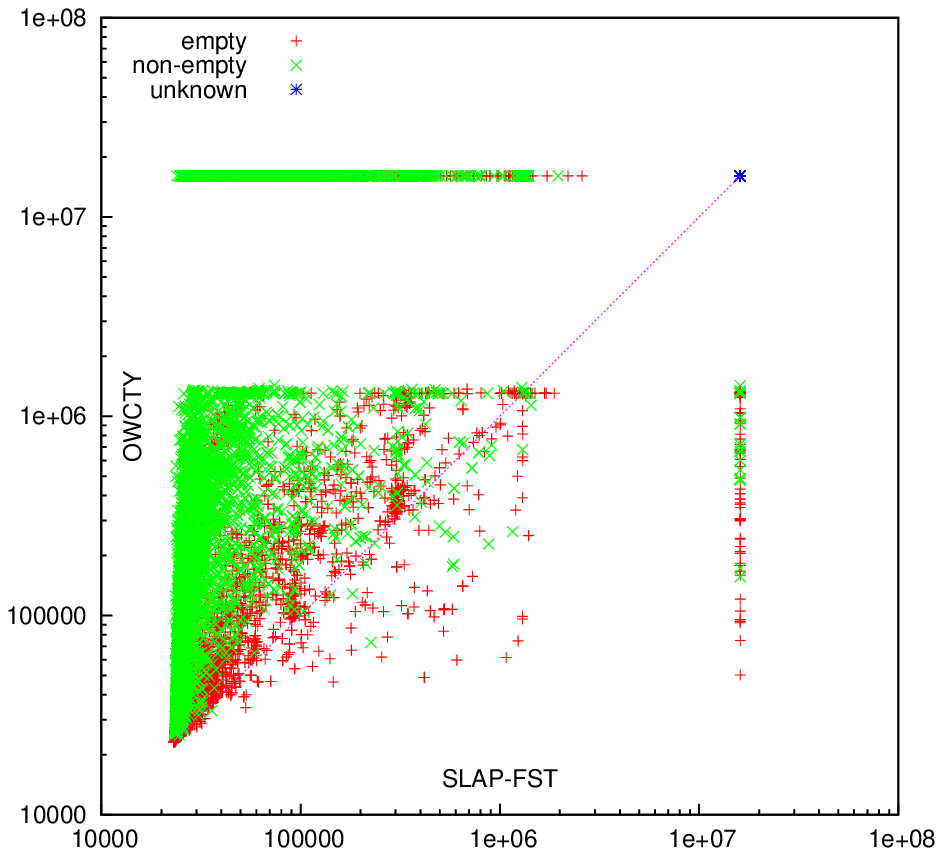,width=0.5\linewidth,clip=} \\
\end{tabular}
\caption{Comparison of SLAP-FST against BCZ, EL, and OWCTY in time (left) and memory (right).\label{fig:slap-fst-all}}
\end{figure}

\begin{figure}[tbp]
\centering
\begin{tabular}{cc}
\epsfig{file=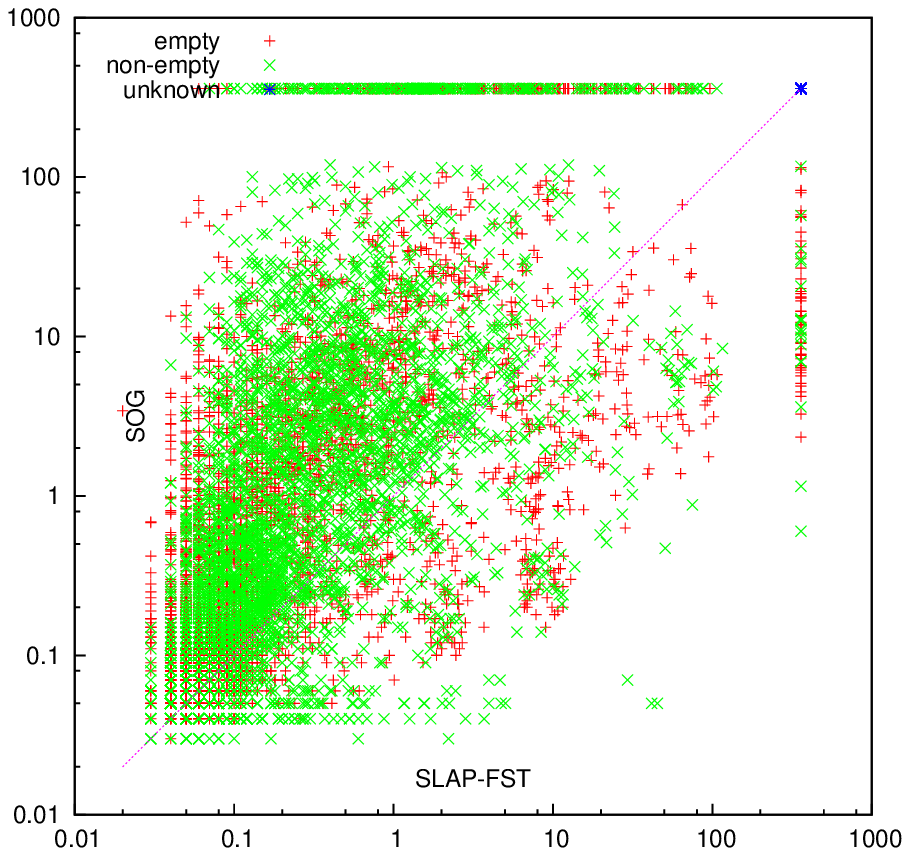,width=0.5\linewidth,clip=} &
\epsfig{file=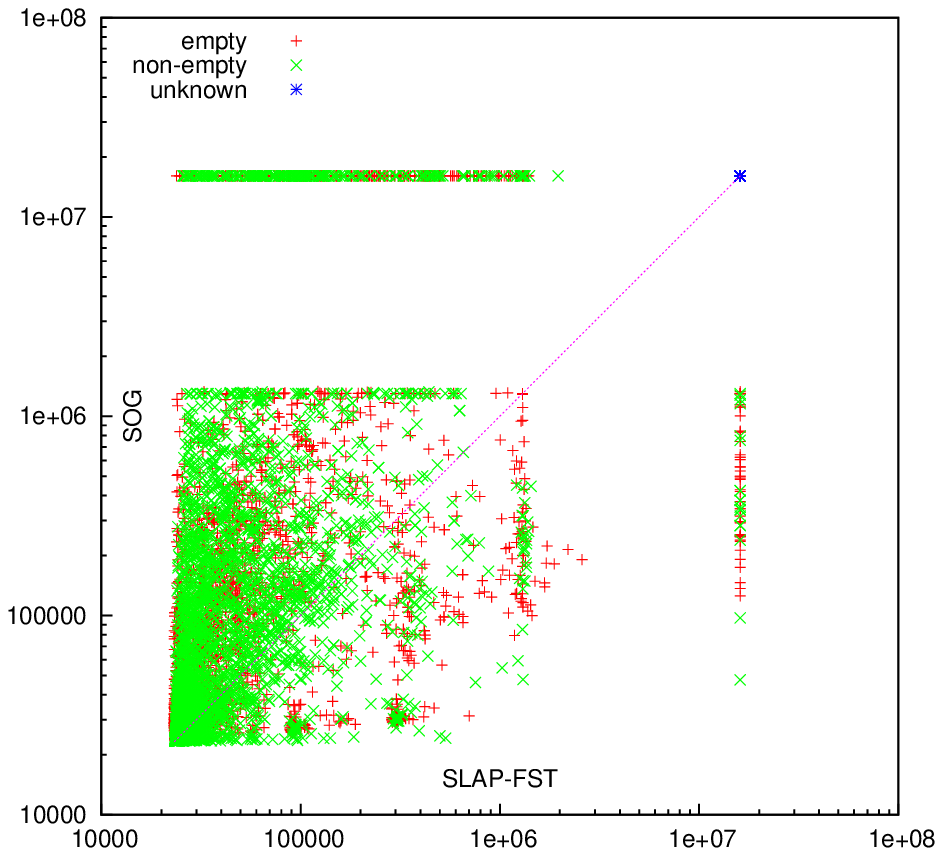,width=0.5\linewidth,clip=} \\
\epsfig{file=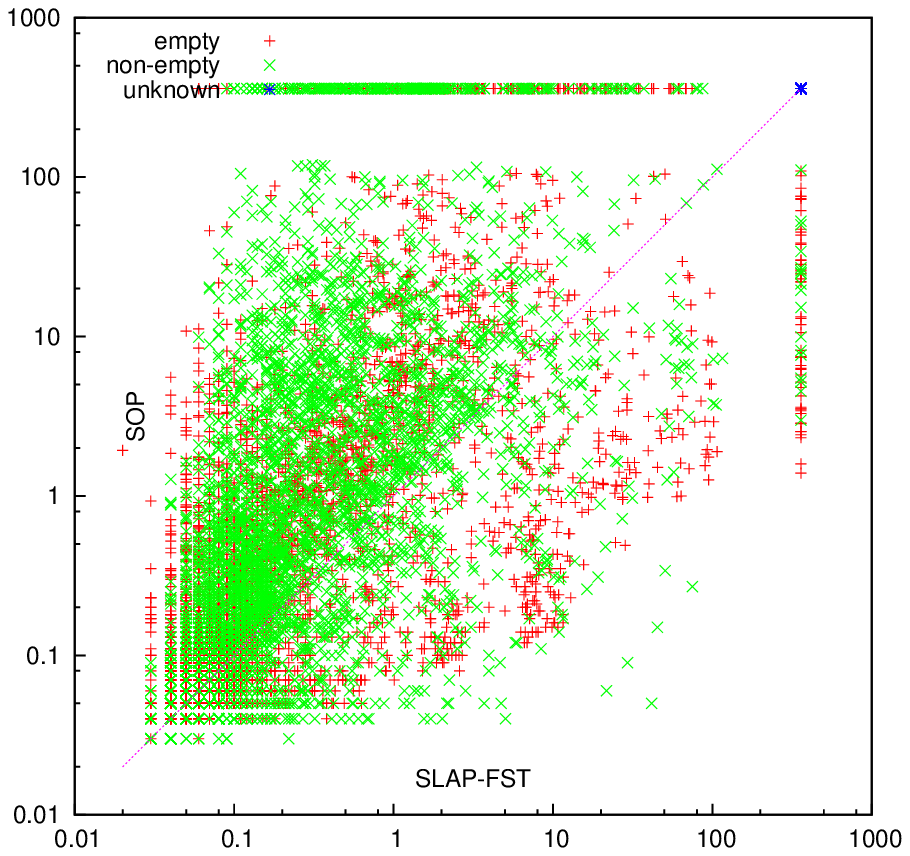,width=0.5\linewidth,clip=} &
\epsfig{file=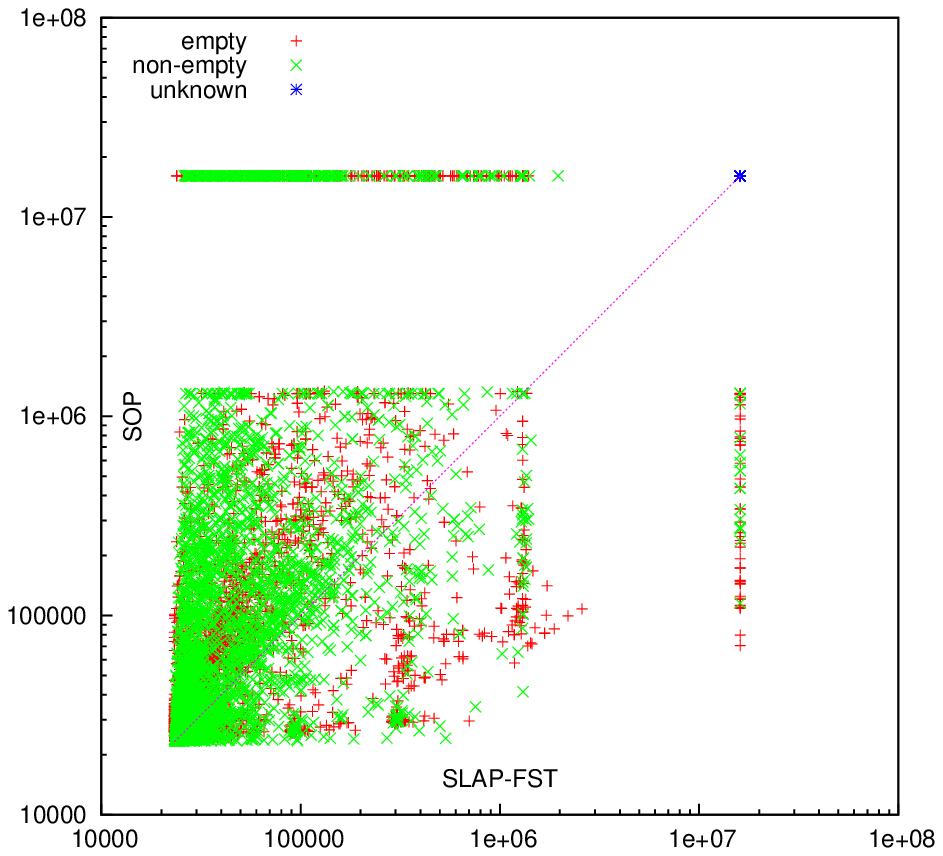,width=0.5\linewidth,clip=} \\
\end{tabular}
\caption{Comparison of SLAP-FST against SOG and SOP in time (left) and
  memory (right).\label{fig:slap-fst-all2}}
\end{figure}

In Fig.~\ref{fig:slap-fst-all} and~\ref{fig:slap-fst-all2} we compare
SLAP-FST against the other methods, using the same kind of logarithmic
scatter plots.  These plots only use stuttering invariant properties
so they can be more easily compared. Unsurprisingly, the only methods that appear
competitive are SOG and SOP; but to the advantage of SLAP-FST, SOG and SOP are not
able to handle non stuttering-invariant properties.

\subsection{Fully symbolic algorithms: EL vs. OWCTY}

These two algorithms are worth comparing because they differ only is
in the way that the acceptance conditions are alternated throughout
the fixpoint computation.

\begin{figure}[ht]
\centering
\begin{tabular}{cc}
\epsfig{file=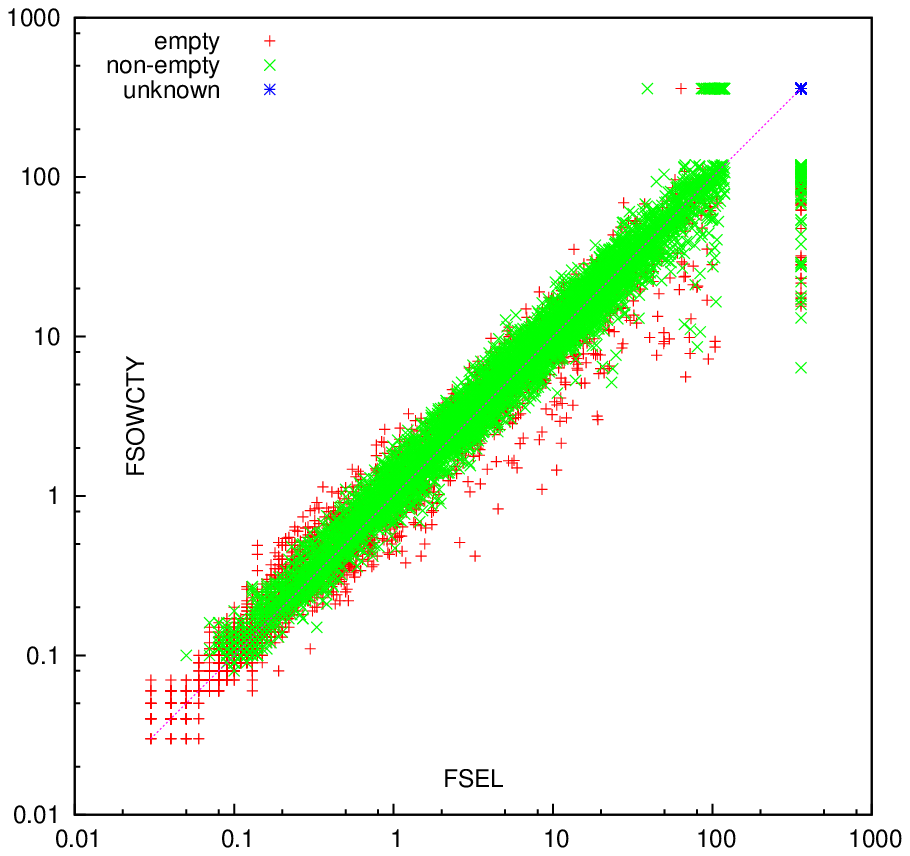,width=0.5\linewidth,clip=} &
\epsfig{file=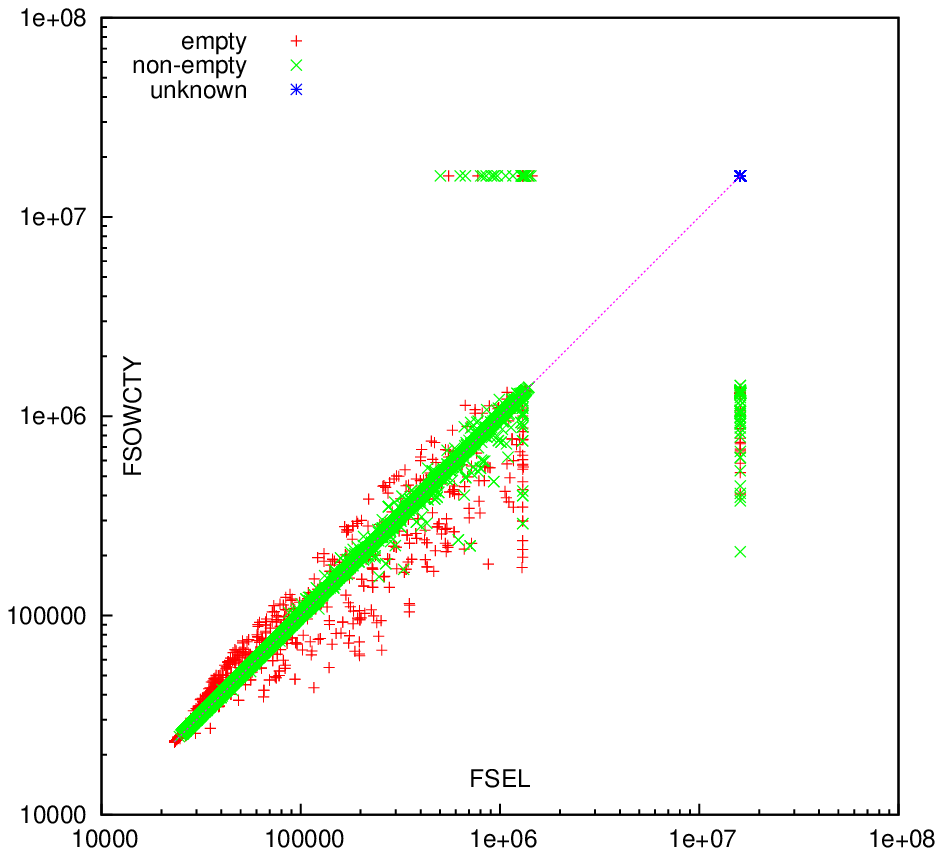,width=0.5\linewidth,clip=} \\
\end{tabular}
\caption{Performances in time (left, in seconds) and memory (right, in kilobytes) of fully symbolic algorithms for 11738 experiments.}
\label{fig:fsalgos-scatter}
\end{figure}

In Fig.~\ref{fig:fsalgos-scatter}, we have 11738 experiments, of which
1225 proved too hard to solve for either algorithm within the time
limit.  Overall EL algorithm solved 58 problems that OWCTY did not,
and OWCTY solved 119 instances that EL did not.  EL was at least ten
times slower than OWCTY in 4 cases, and at least twice as slow in 141
cases, whereas OWCTY was twice as slow in 91 cases.

Overall these plots show very little perceptible difference for
non-empty products and seem to slightly favor OWCTY for empty
products.  Given the overall aspect of these plots that do not stray
much from the diagonal, we can state that both algorithms have
comparable empirical complexities on this benchmark.

\begin{figure}[ht]
\centering
\begin{tabular}{cc}
\epsfig{file=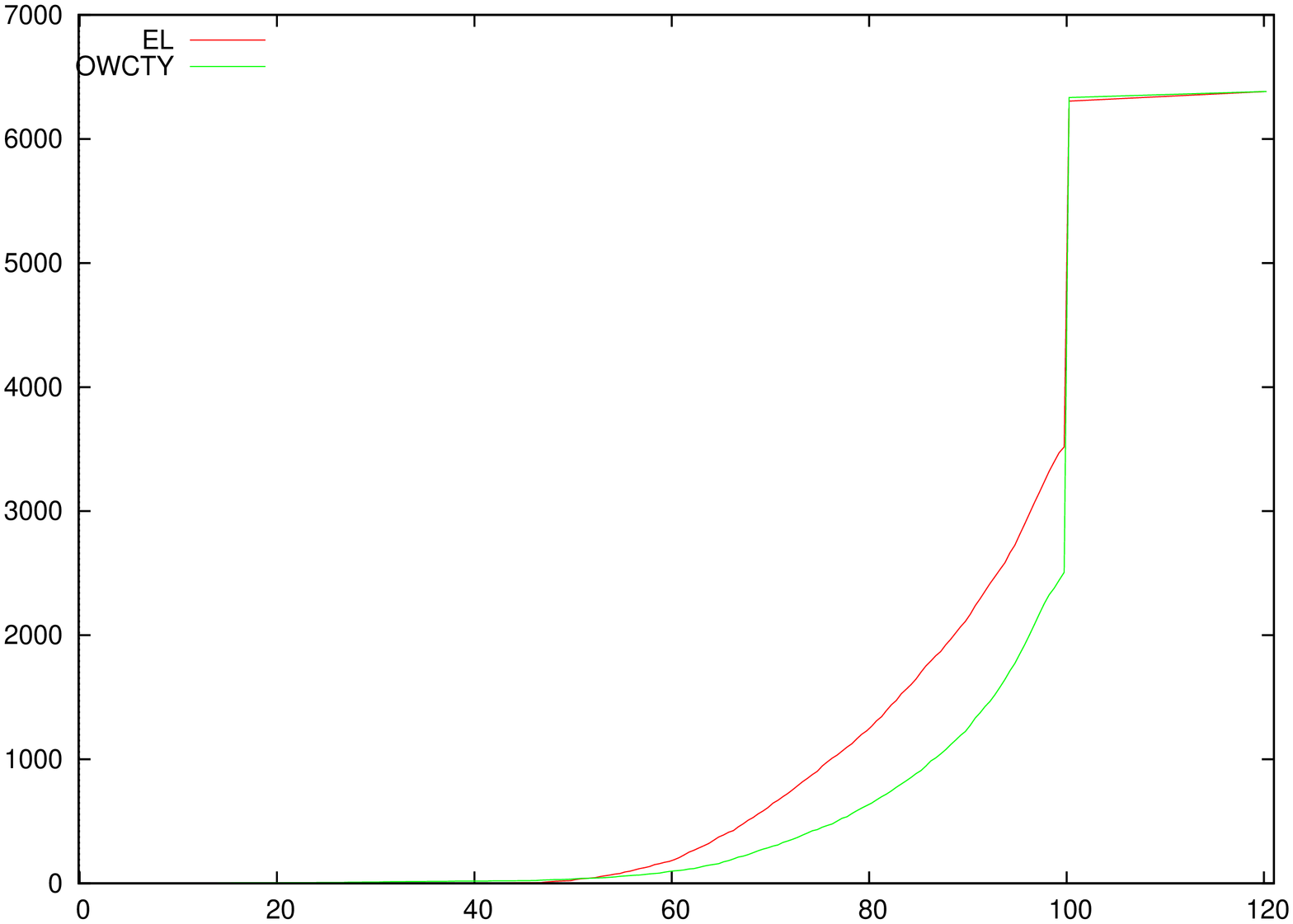,angle=-90,width=0.5\linewidth,clip=} &
\epsfig{file=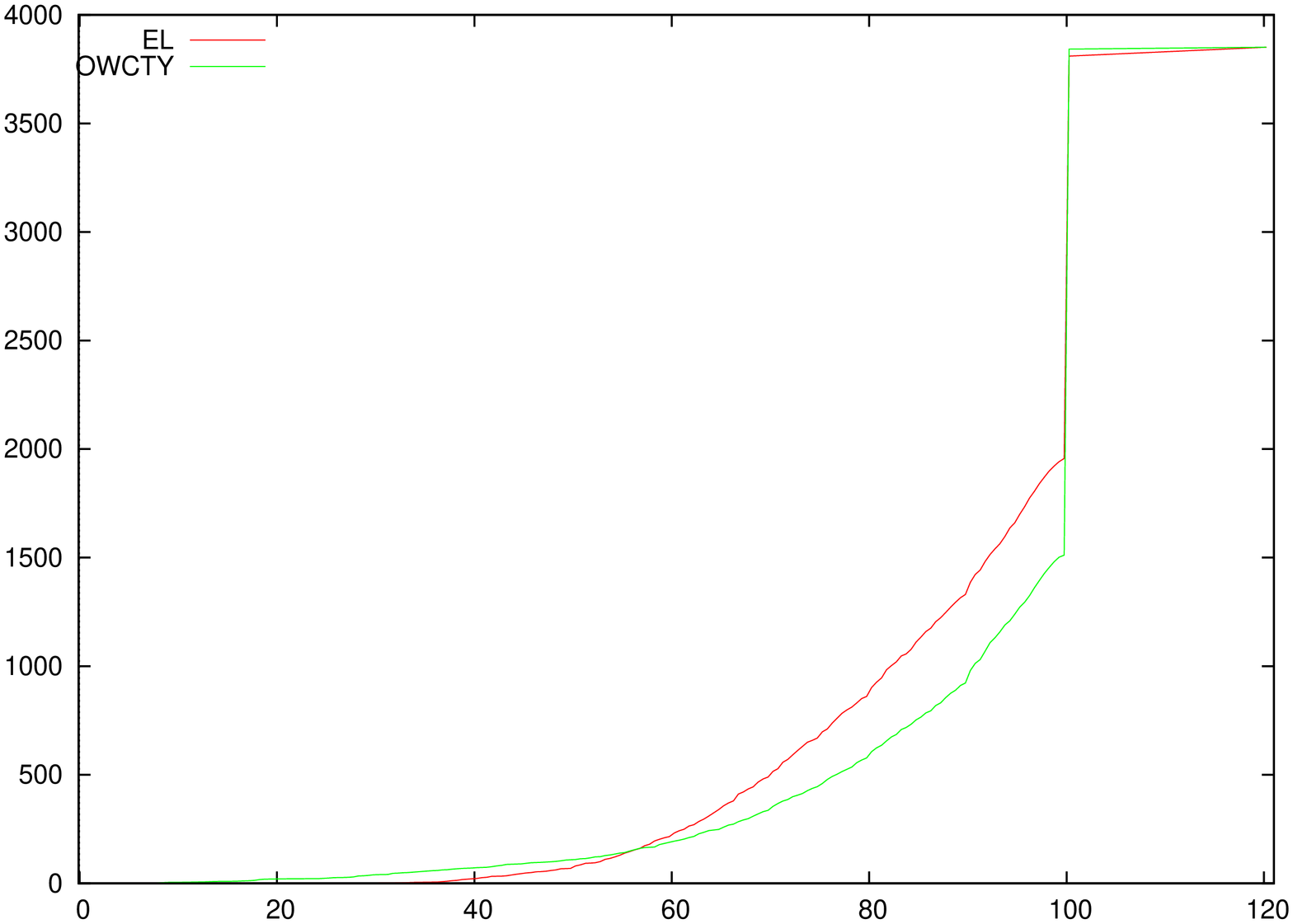,angle=-90,width=0.5\linewidth,clip=} \\
\end{tabular}
\caption{Compared cumulative performances in time of fully symbolic algorithms with (left) or without  counterexamples (right).}
\label{fig:fsalgos-cumul}
\end{figure}

Although the scatter plot does not highlight this fact very blatantly,
the density of experiments where EL outperformed OWCTY is actually
quite high.  The cumulative plots from Fig.~\ref{fig:fsalgos-cumul}
make this more visible. They also show that the difference between the
two algorithms only very rarely exceed a ratio of 2.

\subsection{Hybrid stuttering invariant algorithms: SOG vs. SOP}

We compare here the SOP and SOG algorithms.

\begin{figure}[ht]
\centering
\begin{tabular}{cc}
\epsfig{file=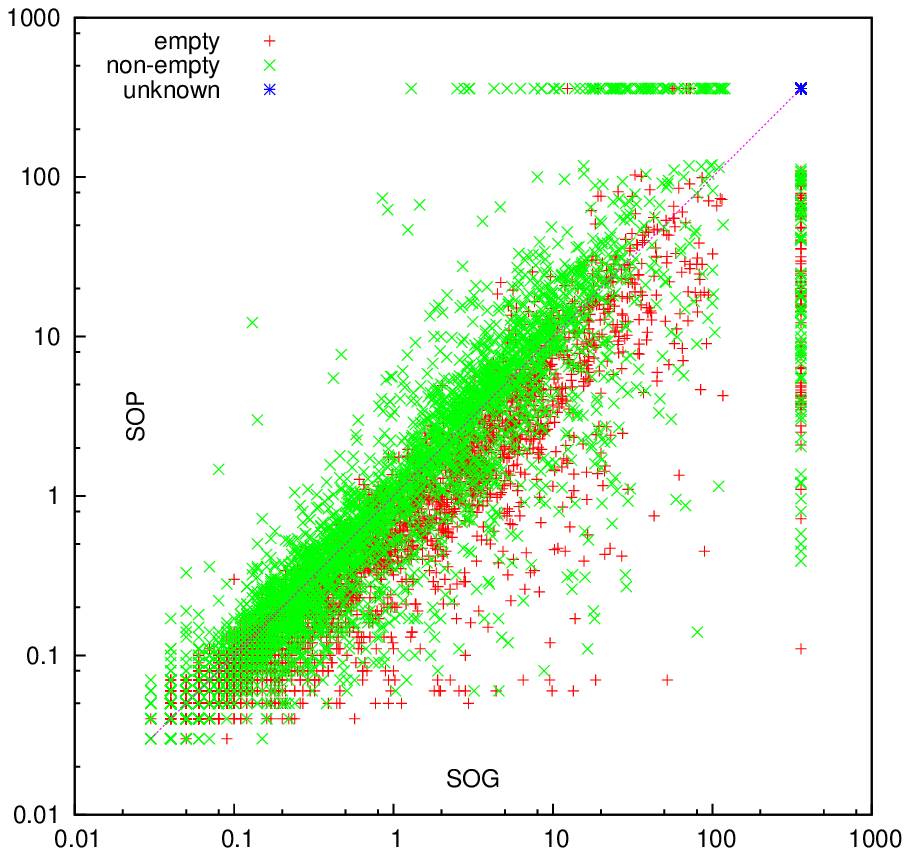,width=0.5\linewidth,clip=} &
\epsfig{file=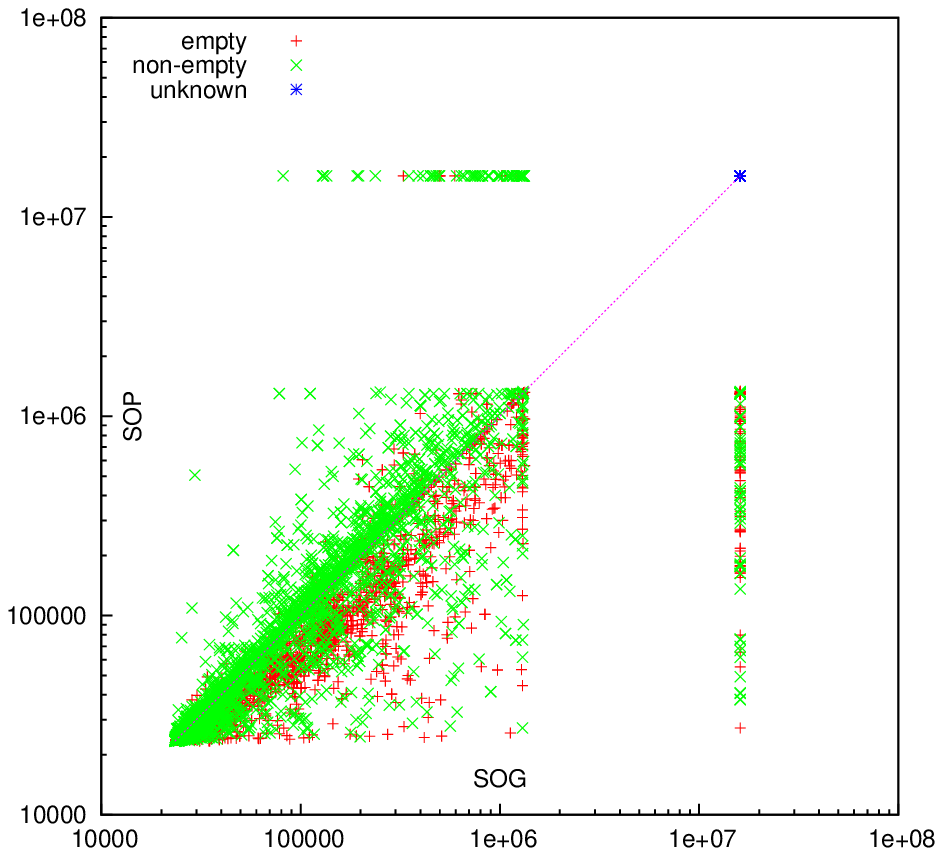,width=0.5\linewidth,clip=} \\
\end{tabular}
\epsfig{file=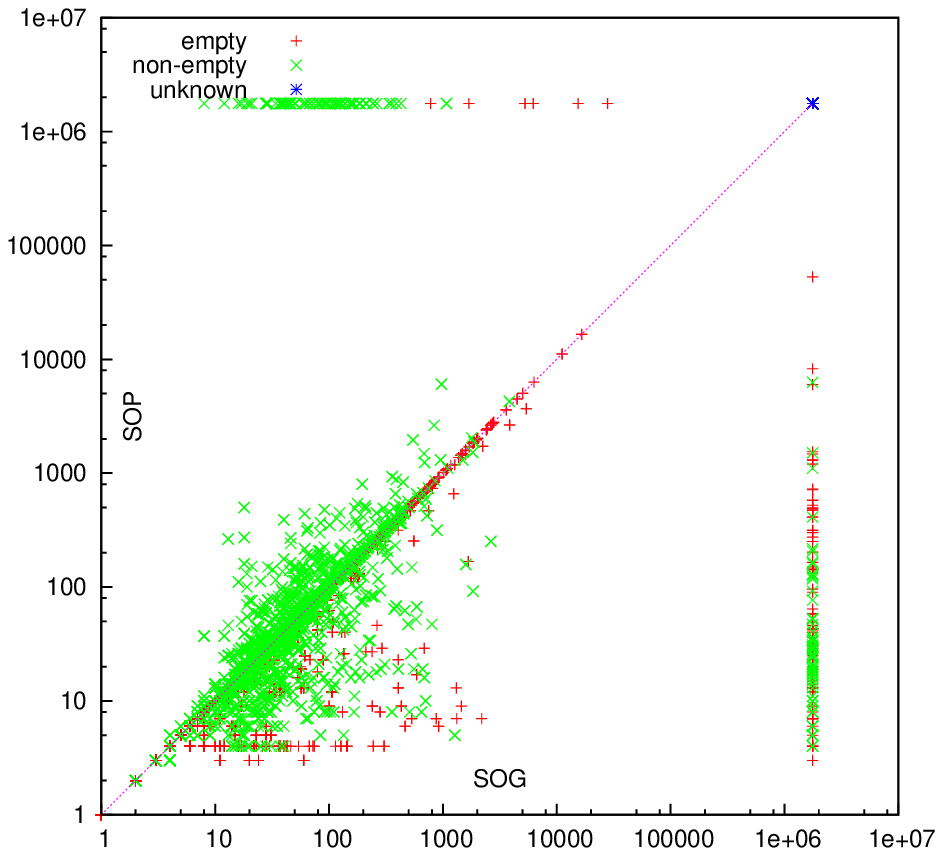,width=0.5\linewidth,clip=}
\caption{Performances in time (top-left, in seconds) and memory
  (top-right, in kilobytes) of observation graph algorithms for 7277
  experiments, and product size (bottom). The product size show the
  number of states of the SOP against the number of state of the
  product between the Kripke structure and the SOG.}
\label{fig:sopsog-scatter}
\end{figure}

In these plots (Fig.~\ref{fig:sopsog-scatter}), we have 7277 experiments, of
which 399 proved too hard to solve for either algorithm within the time limit.
Overall SOG algorithm solved 95 problems that SOP did not, and SOP solved 166
instances that SOG did not. SOG was at least a hundred times slower than SOP
in 11 cases, ten times slower in 132 cases, and twice as slow in 1326 cases.
A contrario, SOP was ten times slower than SOG in 13 cases, and twice as slow
in 280 cases.

Overall these plots show that SOP significantly outperforms SOG in
many problem instances, particularly when the full product needs to be
built (i.e., it is empty so on-the-fly mechanism does not come into
play).

\begin{figure}[ht]
\centering
\begin{tabular}{cc}
\epsfig{file=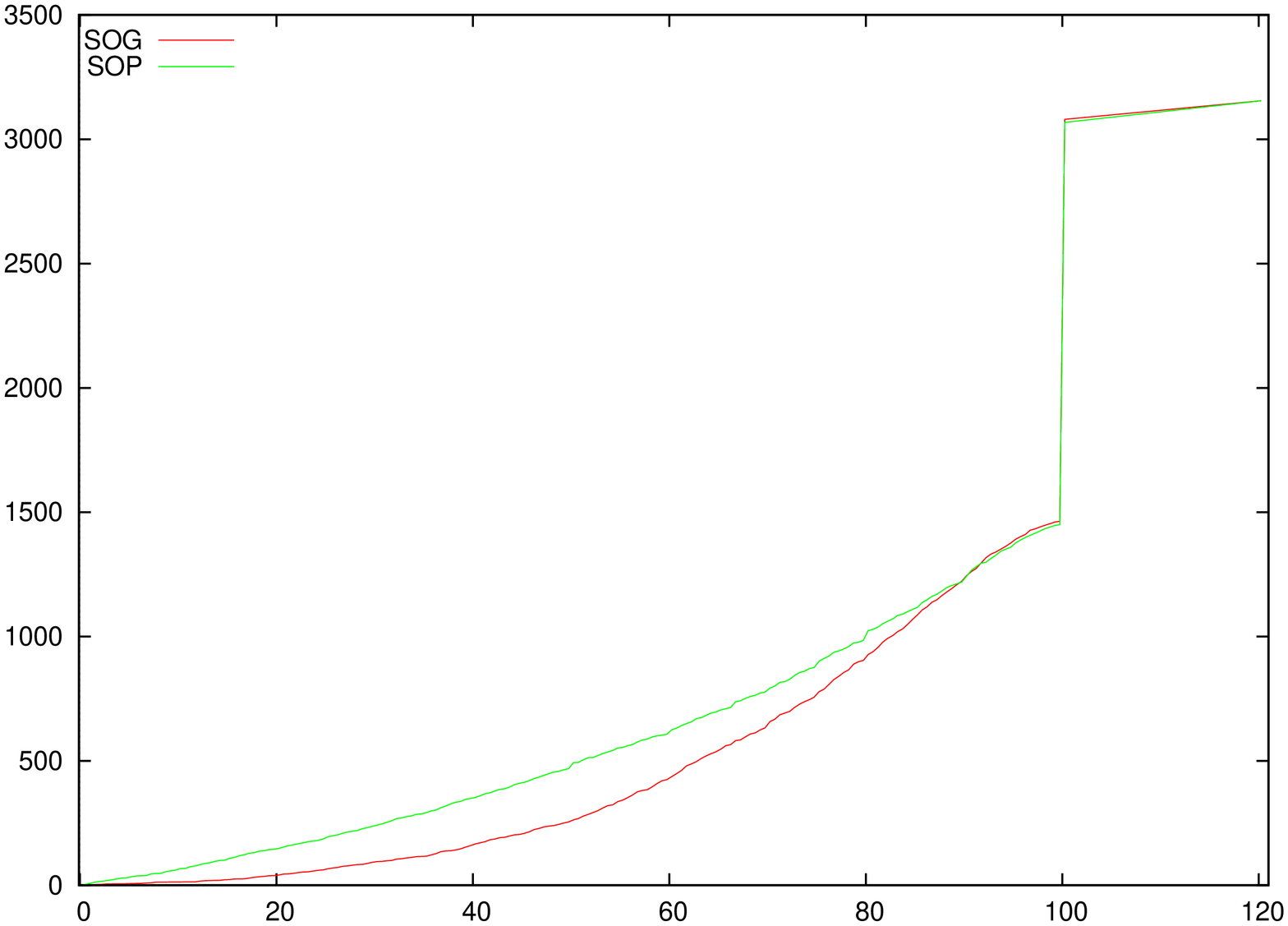,angle=-90,width=0.5\linewidth,clip=} &
\epsfig{file=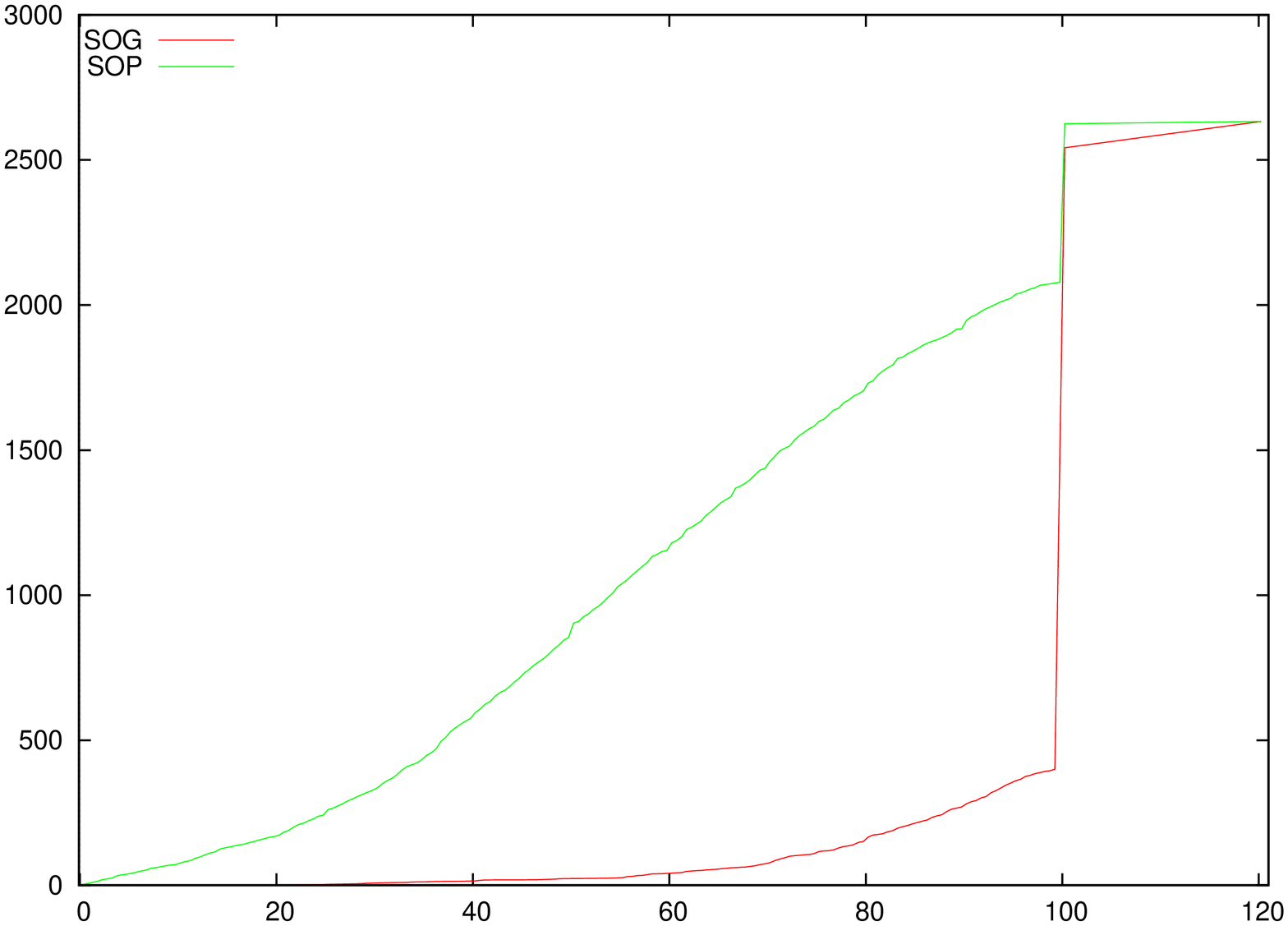,angle=-90,width=0.5\linewidth,clip=} \\
\end{tabular}
\caption{Compared cumulative performances in time of hybrid stuttering-invariant algorithms with (4048 experiments, left) or without  counterexamples (3229 experiments, right).}
\label{fig:sogsop-cumul}
\end{figure}

The following cumulative plots (Fig.~\ref{fig:sogsop-cumul}) make this
more visible.  It also shows that SOP does not necessarily outperform
SOG, particularly when the product is non-empty.  In fact the SOP may in
some cases be much larger in explicit size than the SOG.  This is
shown in the bottom graph of figure~\ref{fig:sopsog-scatter}.

\subsection{BEEM models}

We performed extensive experimentations using the BEEM (Benchmarks for Explicit Model checkers~\cite{BEEMspin07})
models and LTL formulas.

The BEEM database contains a large set of examples modeling various network
protocols, mutual exclusion or consensus problems.  We used all the
examples for which LTL formulas are provided, and ran the verification
for both the formula and its negation, to increase the number of
formulas.

Surprisingly enough, all LTL formulas provided by BEEM are stuttering
invariant.  Thus we also generated a few random formulas which are not
stuttering invariant.
This benchmark is interesting as it shows some concurrent software
oriented examples,
 and real formulas. However, the number of formulas is quite limited
 with respect to the previous benchmark. These formulas are also simpler,
 hence less able to discriminate the various algorithms that depend on the
formula automaton. The number of reachable states in these models is also
much lower than in the Petri net examples; this makes it more
difficult to measure the impact of large Kripke structures on the
explicit size of hybrid algorithms. The transition relation, which is
built by interacting with the LTSmin package is also less efficient
than that of the Petri net benchmark, and takes less advantage of the
automatic saturation features \cite{atpn08saturation} of the SDD
library that are heavily sollicited in the SOG, SOP and SLAP algorithms.

A total of 729 formula/model pairs were computed for each
algorithm, of which 292 are not stutterin invariant and not used for
SOG or SOP. We filtered out model/formula pairs that took less than
0.1 seconds to solve for all methods.

\begin{figure}[tbp]
\centering
\begin{tabular}{cc}
\epsfig{file=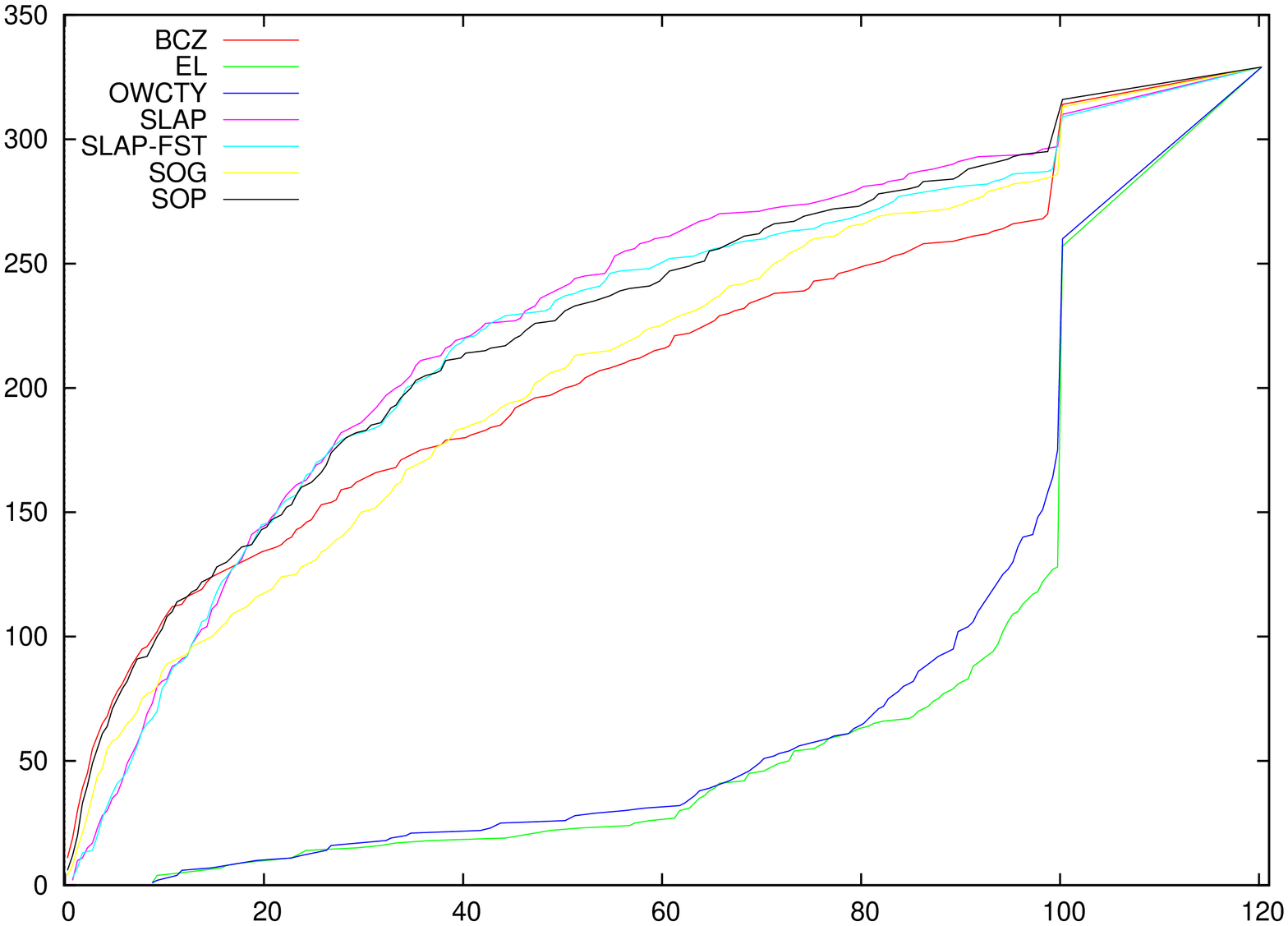,angle=-90,width=0.5\linewidth,clip=} &
\epsfig{file=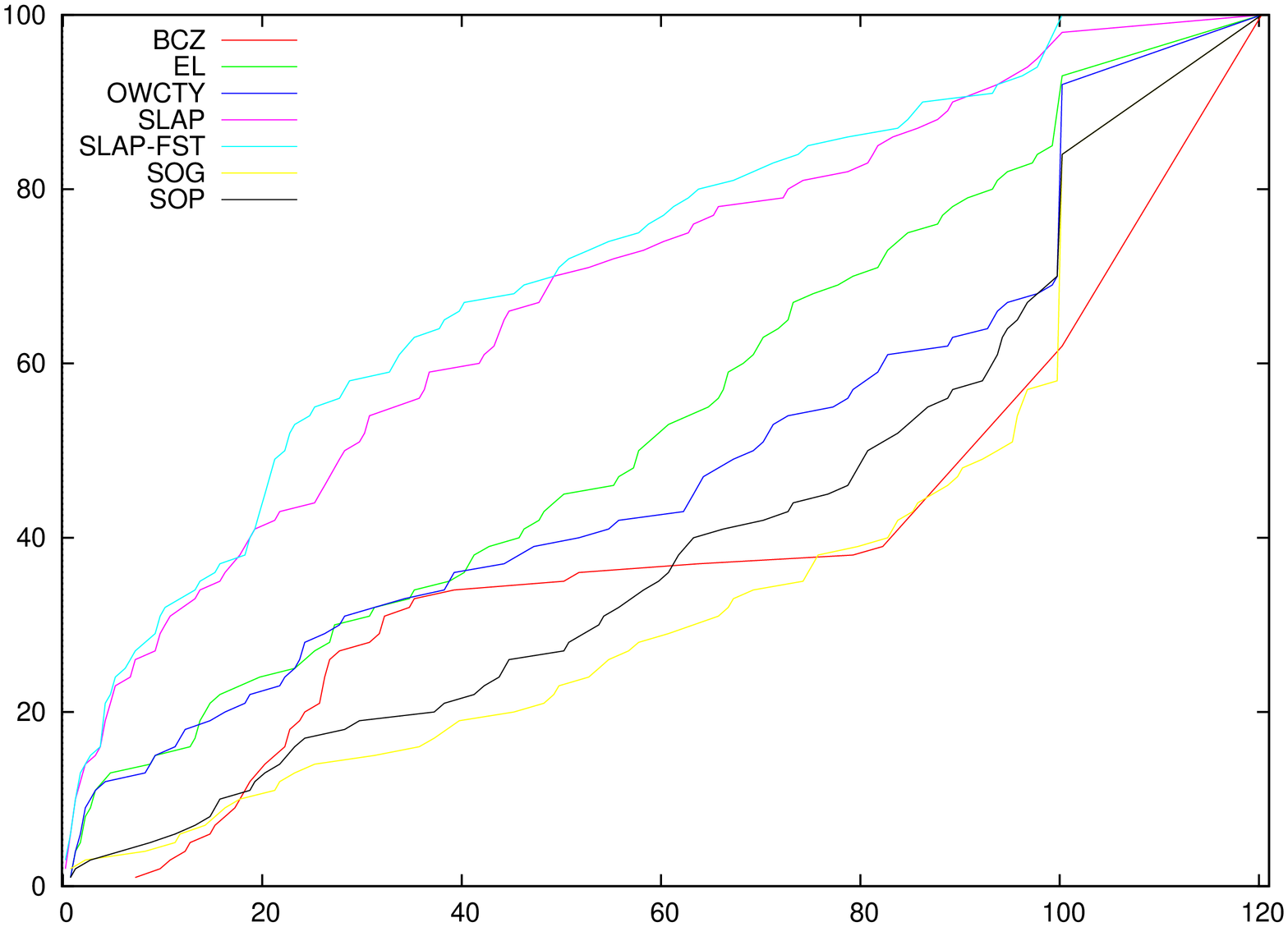,angle=-90,width=0.5\linewidth,clip=} \\
\epsfig{file=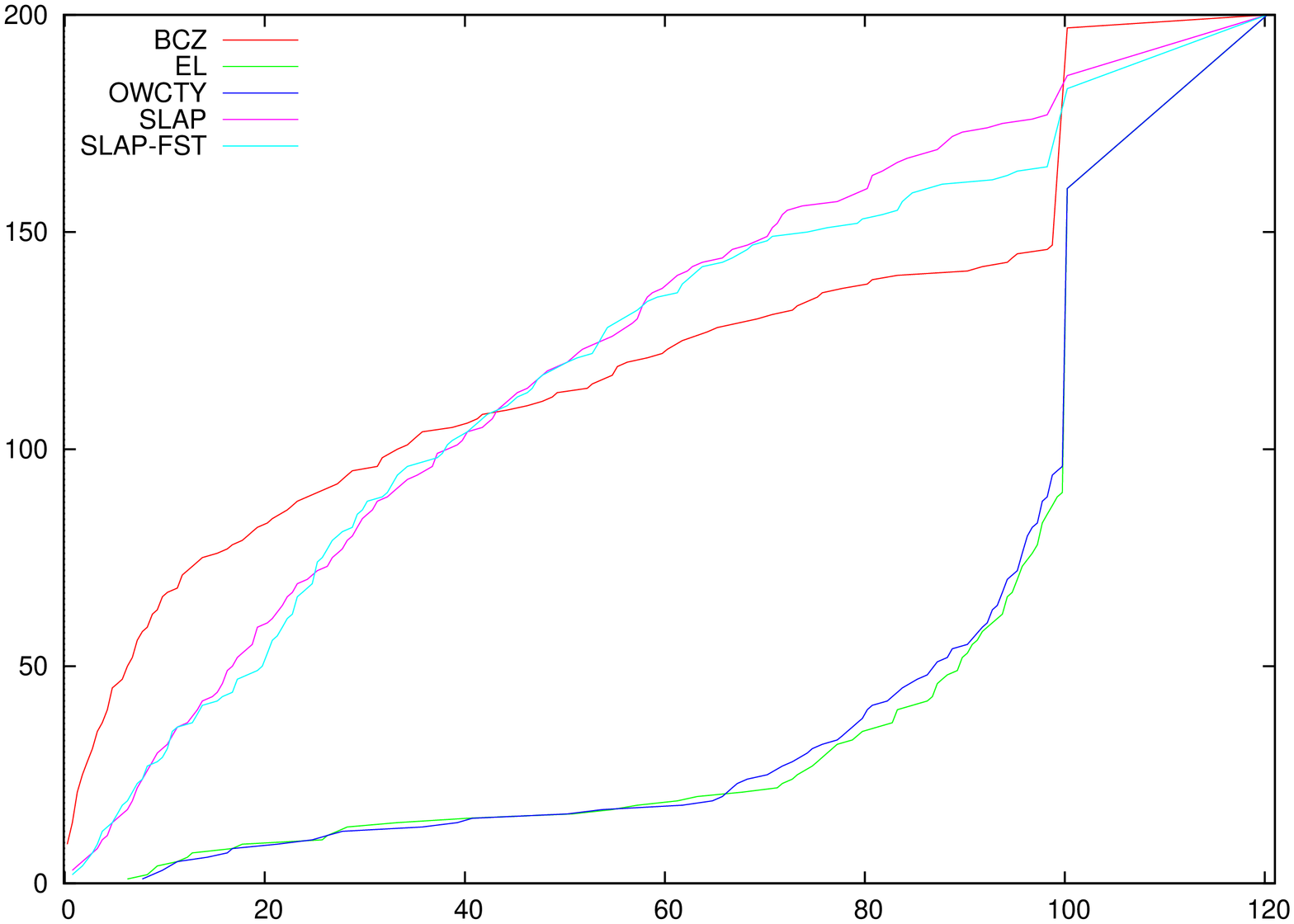,angle=-90,width=0.5\linewidth,clip=} &
\epsfig{file=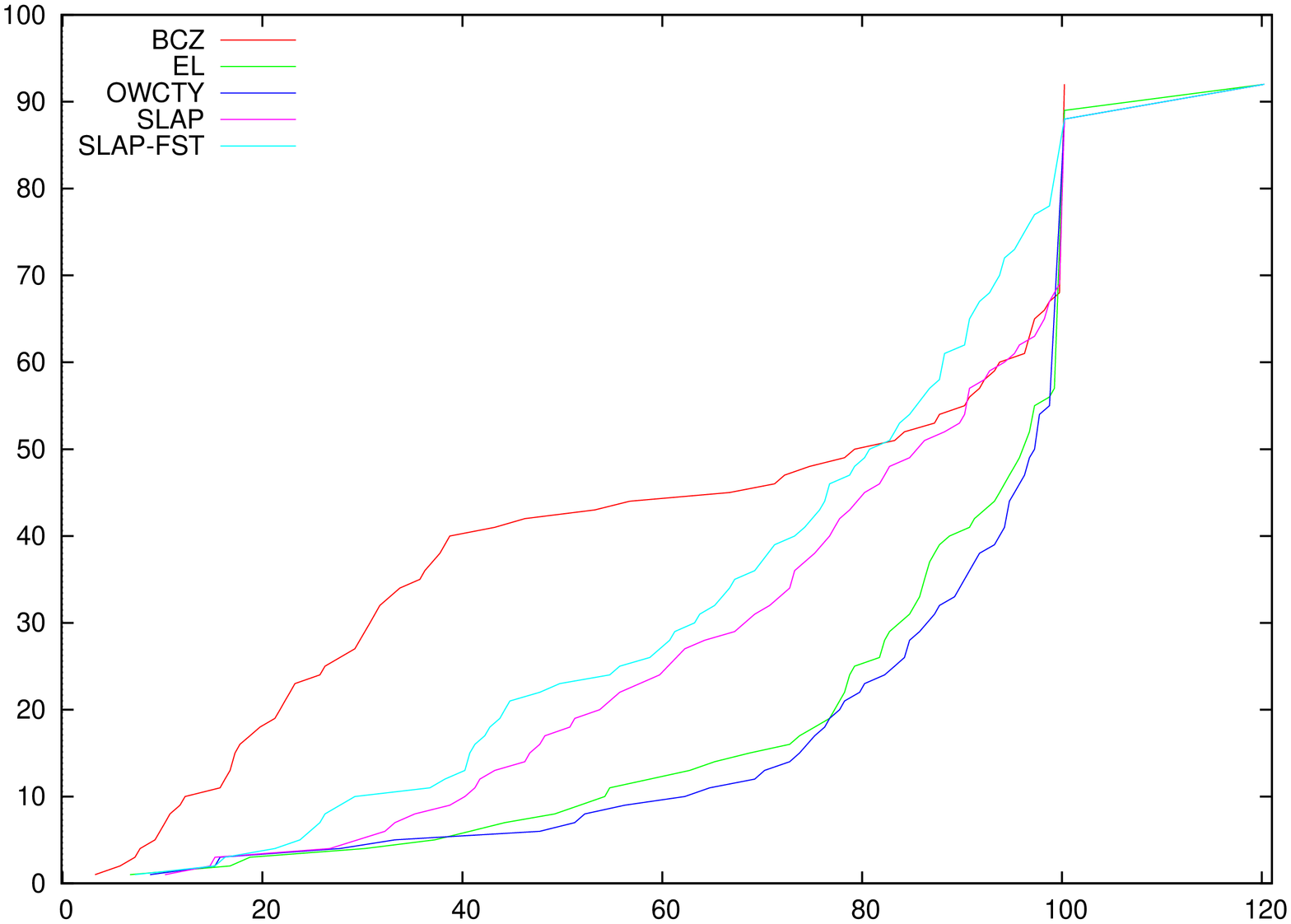,angle=-90,width=0.5\linewidth,clip=} \\
\end{tabular}
\caption{Cumulative plots comparing the time of all methods on the
  BEEM models. Non-empty products are shown on the left, and empty
  products on the right. Top for stuttering invariant properties and bottom for LTL formulae with the $\X$ operator.
\label{fig:cumul-beem}}
\end{figure}

Table~\ref{tab:beeemwinner} gives a synthetic overview of the results
presented hereafter and Fig.~\ref{fig:cumul-beem} details these
measures with a cumulative distribution function plot.

For stuttering-invariant examples, when the product is empty
 SLAP or SLAP-FST are the fastest methods in over half of all cases,
 and they are rarely the slowest. Furthermore, they have the least
 failure rate, whether the product is empty or not.  This table also
 shows that BCZ has the highest failure rate when the product is
 empty, although its is the fastest method in one third of cases
 whether the product is empty or not. SOG and SOP perform honorably
 when the product is non-empty (hence the on the fly mechanism comes
 into play), but behave quite porrly in the empty product case. The
 fully symbolic algorithms (OWCTY, EL) have trouble with non-empty
 products, but have a low failure rate when the product is empty
 though they rarely win. On this benchmark set (like on the Petri net measures)
EL seems to perform slightly better than OWCTY.

For non stuttering-invariant properties (which were randomly
generated), BCZ behaves very well whether the product is empty or
not. In non-empty case, as in our other measures, fully symbolic
algorithms EL and OWCTY behave poorly with a lot of failures and
slowest runtimes. SLAP and SLAP-FST are outperformed by BCZ on this
benchmark set, but remain competitive.  As shown in the cumulative
distribution plot of Fig.~\ref{fig:cumul-beem}(bottom),
 SLAP and SLAP-FST overtake the BCZ curve around 50%. BCZ for a
 significant number of problem instances has very good performance (as
 shown by the steep start for BCZ curve), but then flattens out, while
 the slope of SLAP and SLAP-FST is more regular.  SLAP and SLAP-FST
 seem to perform more poorly on this benchmark set than in our other
 measures, however the number of experiments is much more limited here.

\begin{table}[tbp]
\centering
\input tablebeem-tr
\caption{On all experiments (grouped with respect to the existence of
  a counterexample and the use of a $\X$ operator in the LTL formula),
  we count the number of cases a specific method has (Win) the best
  time or (Lose) it has either run out of time or it has the worst
  time amongst successful methods.  The Fail line shows how much of the
  Lost cases were timeouts.  The sum of a line may exceed 100\% if
  several methods are equally placed.\label{tab:beeemwinner}}
\end{table}

The performances are presented as scatter plots using logarithmic
scale.  Each point represents an experiment, i.e., a model and formula
pair.  We killed any process that exceeded 800
seconds of runtime; hence for some formulas we were not able to
compute the answer.  We plot experiments that failed (due to timeout)
as if they had taken 2400 seconds, so they are clearly separated from
experiments that didn't fail (by the wide white band).

\begin{figure}[tbp]
\centering
\begin{tabular}{cc}
\epsfig{file=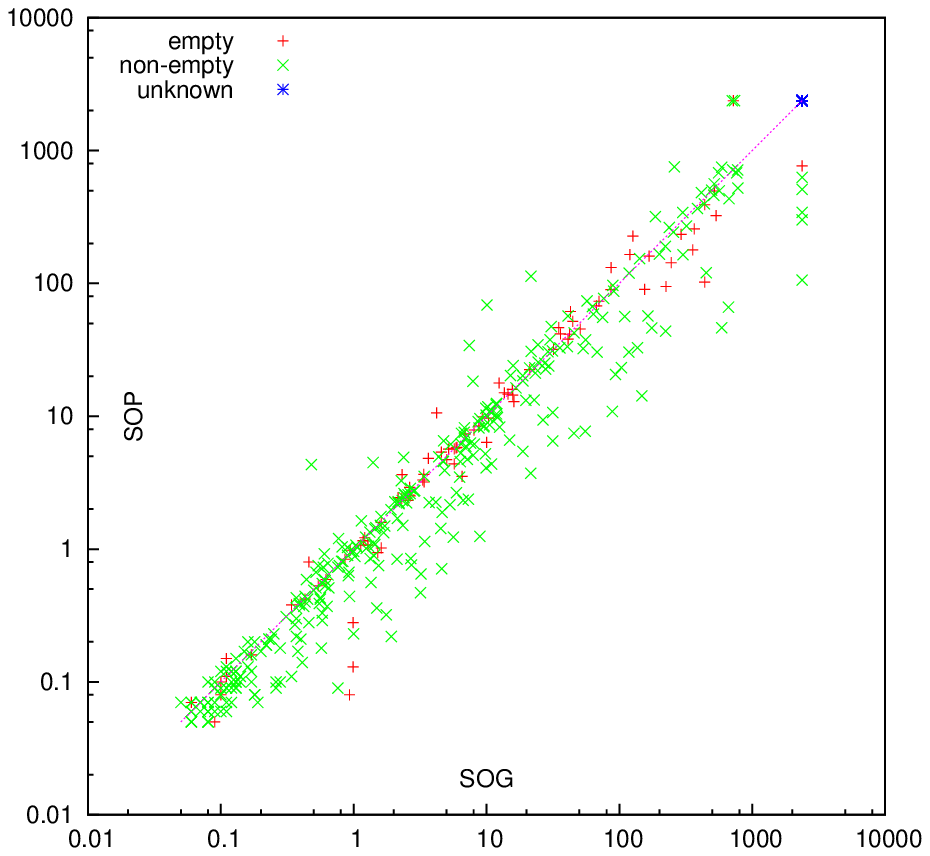,width=0.5\linewidth,clip=} &
\epsfig{file=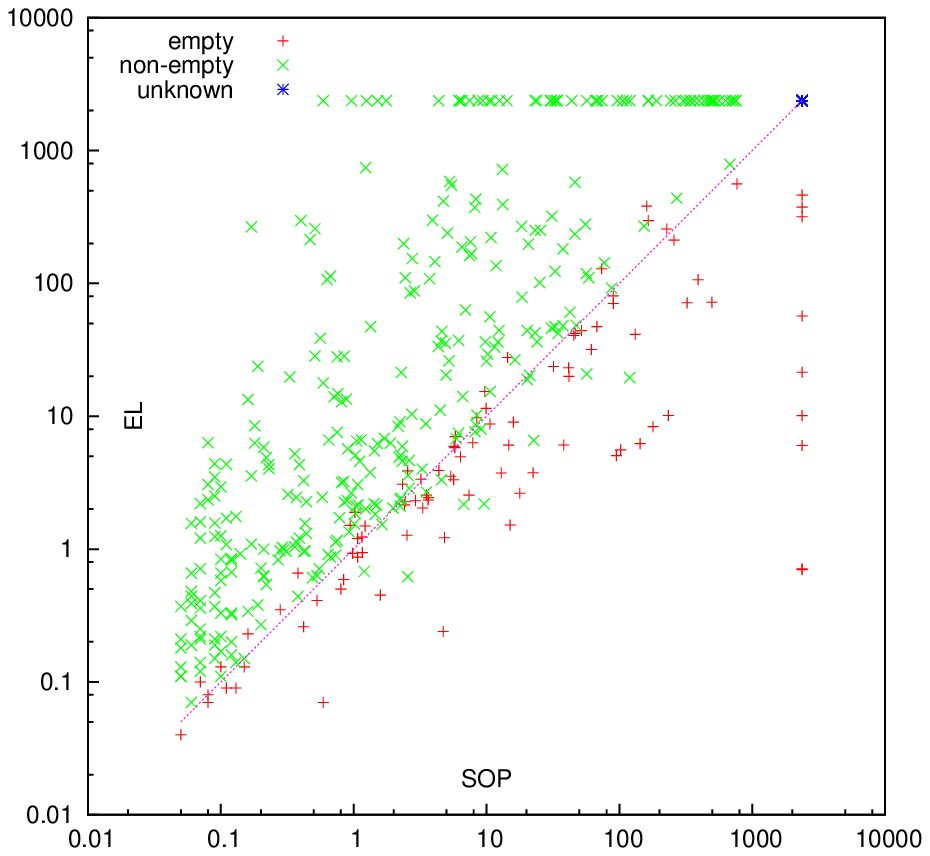,width=0.5\linewidth,clip=} \\
\epsfig{file=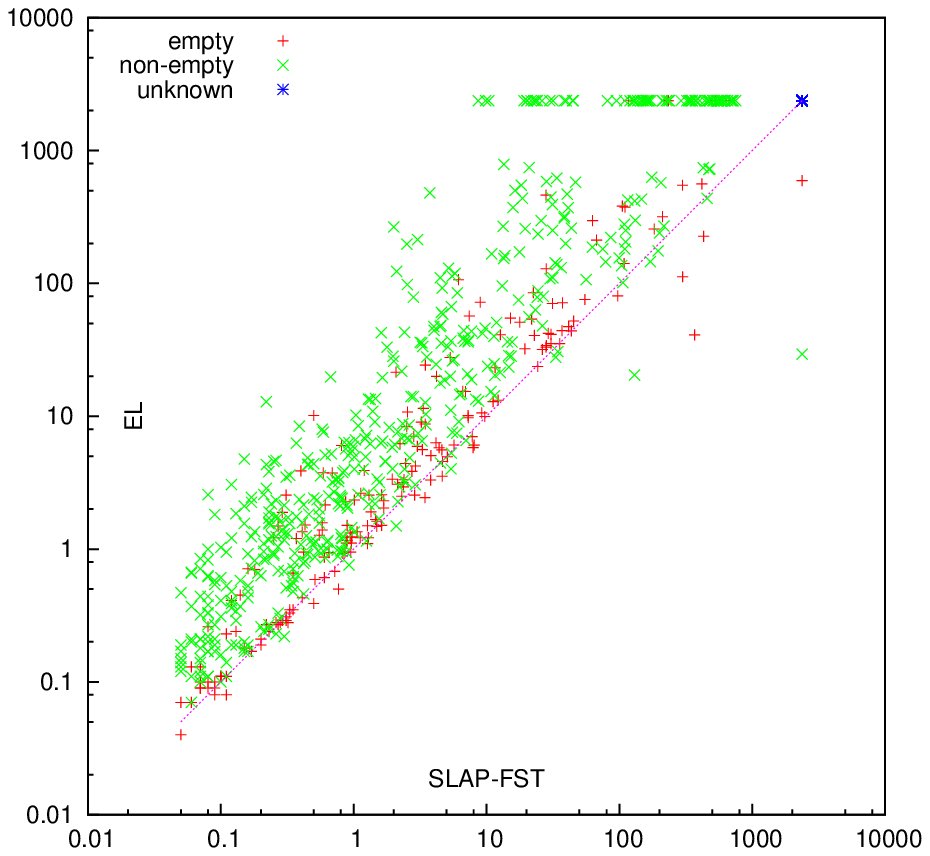,width=0.5\linewidth,clip=} &
\epsfig{file=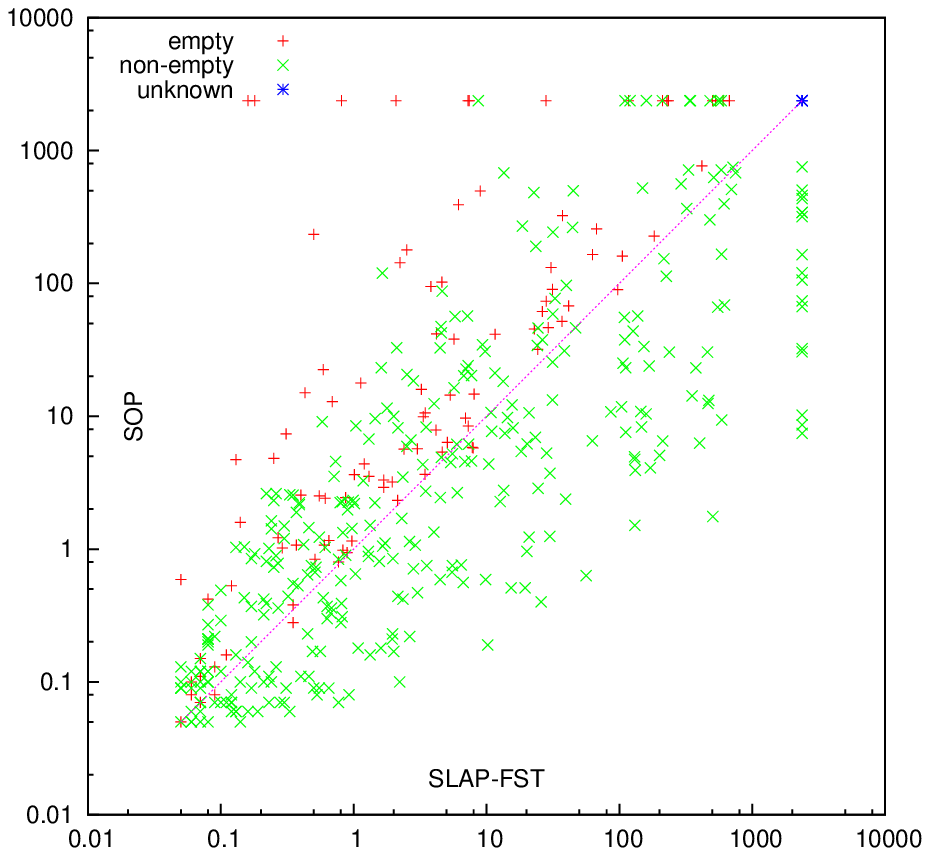,width=0.5\linewidth,clip=} \\
\epsfig{file=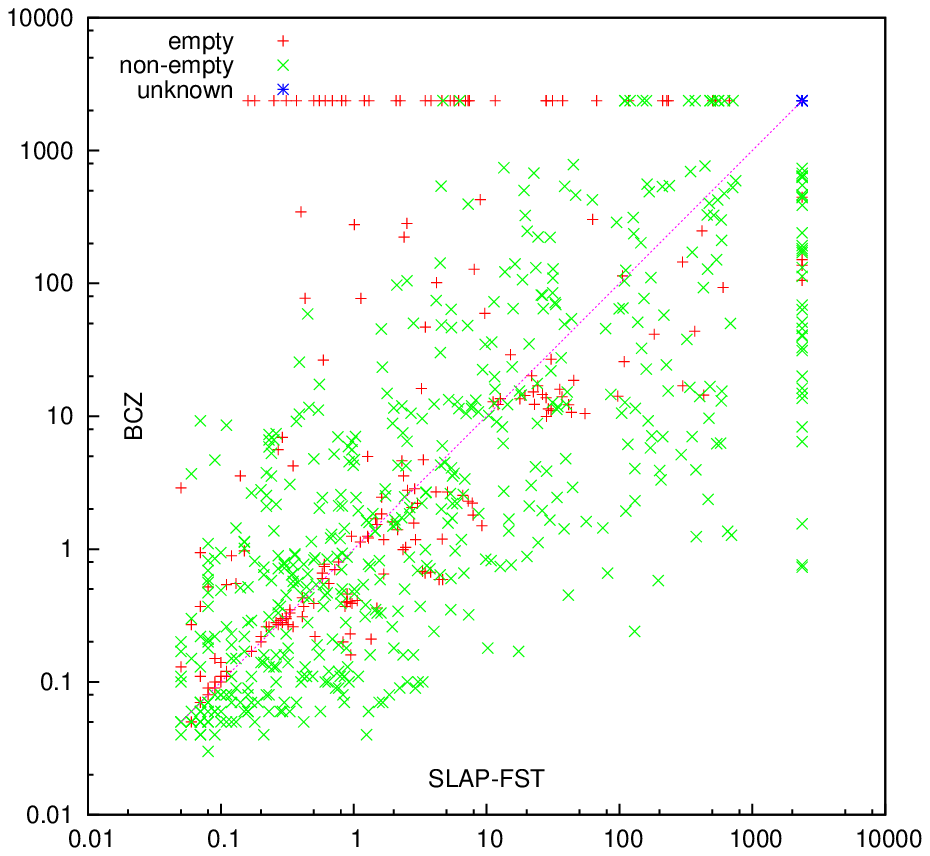,width=0.5\linewidth,clip=} &
\epsfig{file=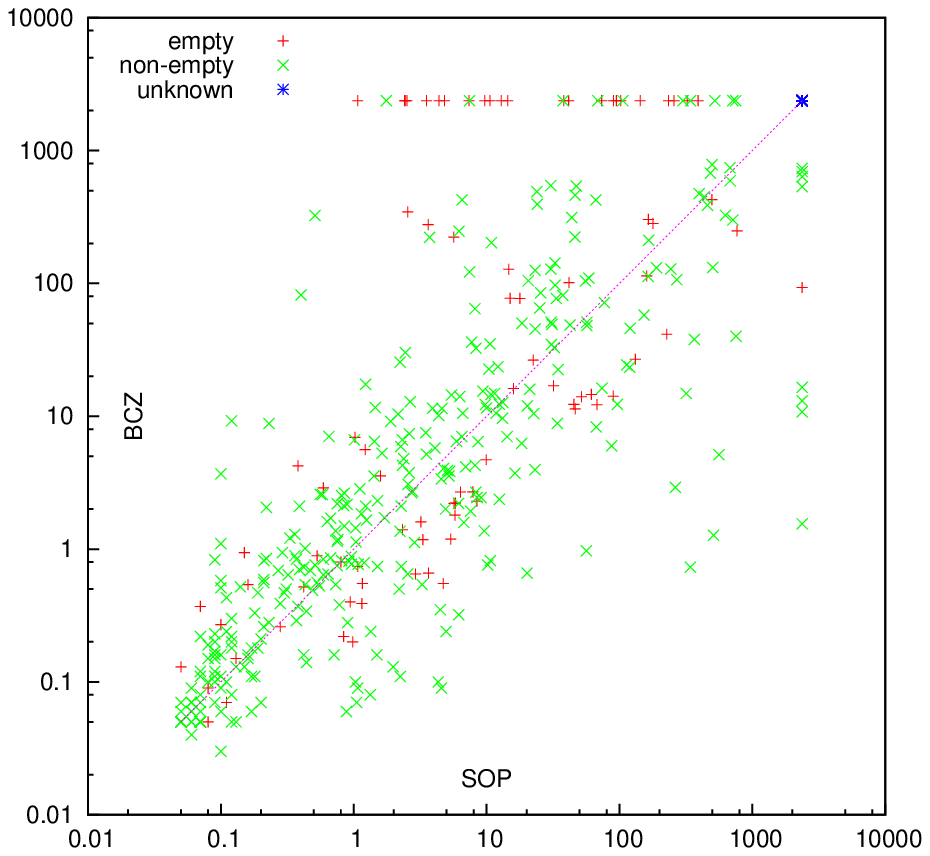,width=0.5\linewidth,clip=} \\
\end{tabular}
\caption{Performances in time for 721 experiments using the BEEM models. Plots with SOG or SOP only contain the 429 stuttering-invariant experiments. }
\end{figure}

The comparison between SOP and SOG slightly favors SOP for these
examples.  The two methods often have very similar complexity, as
shown by the numerous points on the diagonal.  These correspond to
cases where the alphabet was not significantly reduced during the
verification.  Overall SOP outperforms SOG in practically all problem
instances.
Since SOP is similar but overall better than SOG, we compare other
methods to SOP rather than SOG in the other plots.

The comparison between SOP and EL favors SOP for non empty products
(unsatisfied formulas) and EL for empty products (satisfied
formulas).  This means that the on-the-fly mechanism of SOP often
allows to answer quite fast (when an accepting cycle exists), but when
the full product needs to be explored EL is actually more effective
than SOP.

The comparison between EL and SLAP-FST clearly favors SLAP-FST in
practically all experiments, with a significant portion of experiments
that failed with EL but not with SLAP-FST.  SLAP-FST like SOP benefits from
the on-the-fly verification that often allows to answer without
building the full product. But because the SLAP-FST is often quite small
in explicit size, it performs quite well whether the product is empty
or not.

The plot comparing SOP to SLAP-FST is much more ambiguous.  SLAP
clearly outperforms SOP when the product is empty, but there are many
problem instances where SOP find an accepting cycle faster.  This
could be due to the DFS order chosen during the on-the-fly
verification, which makes these results difficult to interpret, as
either of the algorithms could be lucky. However, a significant number
of problem instances were solved by SOP and not by SLAP-FST; hence overall
both algorithms can be useful in practice.

Comparison of SLAP-FST to BCZ, and of SOP to BCZ show that these three
methods can be complementary. BCZ unfortunately has a large number of
failures for empty products on stuttering invariant examples, but can
perform quite well in a significant number of problem instances.

Comparison to the LTL model-checker provided by DiVine was attempted
but when running in non compiled mode the run times are prohibitive,
and when using the compiled mode the results are wrong on a
significant number of problem instances, hence we have little confidence in
 the current distribution of DiVine.

% LocalWords:  Ciardo's LTL TGBA Kripke SOG OWCTY fixpoint FST BEEM DFS DiVine
% LocalWords:  NuSMV LTSmin ETF FMS Kanban

%% file: implement.tex
\subsection{Implementation}
\label{sec:implem}

We have implemented these new techniques (SOP, SLAP and SLAP-FST), the
SOG~\cite{HaIlKa04,KlaiP08} as well as the classical fully symbolic
algorithms (OWCTY~\cite{KestenPR98} and EL~\cite{EL86}) and the hybrid
approach of Biere et al.~\cite{BiereCZ99} (noted BCZ in the following)
to allow comparisons. The software, available from \url{ddd.lip6.fr},
builds upon three existing components: Spot, SDD/ITS, and LTSmin.

Spot (\url{http://spot.lip6.fr}) is a model checking
library~\cite{duret.04.mascots}: it provides bricks to build your own
model checker based on the automata theoretic approach using TGBAs. It
has been evaluated as "one of the best explicit LTL
model-checkers"~\cite{rozier.07.spin}. Spot provides translation
algorithms from LTL to TGBA, an implementation of a product between a
Kripke structure and a TGBA (def.~\ref{def:tgbaprod}), and various
emptiness-check algorithms to decide if the language of a TGBA is
empty (among other things). The library uses abstract interfaces, so
any object that can be wrapped to conform to the Kripke or TGBA
interfaces can interoperate with the algorithms supplied by Spot.

SDD/ITS (\url{http://ddd.lip6.fr}) is a library representing
Instantiable Transition Systems efficiently using Hierarchical Set
Decision Diagrams~\cite{Thierry-MiegPHK09}. ITS are essentially an
abstract interface for (a variant of) labeled transition systems, and
several input formalisms are supported (discrete time Petri nets,
automata, and compositions thereof). SDD are a particular type of
decision diagram that a) allow hierarchy in the state encoding,
yielding smaller representations, b) support rewriting rules that
allow the library to automatically~\cite{atpn08saturation} apply the
symbolic saturation algorithm~\cite{ciardo03saturation}.  These
features allow the SDD/ITS package to offer very competitive
performance.

LTSmin\footnote{\url{http://fmt.cs.utwente.nl/tools/ltsmin}}~\cite{ltsmin.cav2010}
is a tool allows to build a symbolic representation of the transition
relation of a system using an explicit firing engine in background.
The tool supports a wide range of input formalisms and explicit
engines.  For our experiments, we used LTSmin to build ETF files
representing the transition relation. These files are then our input
model, they were wrapped to conform to the ITS interface, thus
allowing to apply our algorithms to any of the formalisms accepted by
LTSmin or by ITS. We used the DVE variant of LTSmin to process the
models from the BEEM benchmark. As noted by
Blom~et~al.~\cite{ltsmin.cav2010} this benchmark is not particularly
favorable to symbolic approaches.

The fully symbolic OWCTY algorithm is implemented directly on top of
the ITS interface; it uses an ITS representing the TGBA derived from
the LTL formula by Spot composed (at the ITS formalism level) with the
ITS representing the system. The resulting ITS is then analyzed using
OWCTY with the forward transition relation.

The SOG is implemented as an object conforming to Spot's Kripke
interface.  It loads an ITS model, then builds the SOG on the fly, as
required by the emptiness check of the product with the formula
automaton.

Both SOP and SLAP are implemented as objects conforming to Spot's
product interface. The SOP and the SLAP classes both take an ITS model
and a TGBA (the formula automaton) as input parameters, and build
their specialized product on the fly, driven by the emptiness-check
algorithm.

%For the SOG approach, we built an object that reads an ITS then build its
%SOG on-the-fly using the Kripke interface from Spot.  We could then
%make the product between the SOG and an automaton representing the LTL
%formula, and check the result for emptiness.
%
%Implementing the SOP and SLAP approaches required us to define new
%classes behaving like a product.  The SOP and the SLAP classes both
%take an ITS model and a TGBA (the formula automaton) as input
%parameters, and they then build the SOP or SLAP on demand.  Because
%they honor the TGBA interface from Spot, we can directly plug them
%into the emptiness check algorithms.
%
%The SOG-ITS package, containing the implementation of the SOG, SOP,
%and SLAP approach for ITS, can be downloaded from \textbf{XXX}.

%% file: tablewin-tr.tex
\begin{tabular}{cccrlrlrlrlrlrlrl}
&&& \multicolumn{2}{c}{OWCTY} & \multicolumn{2}{c}{EL} & \multicolumn{2}{c}{BCZ} & \multicolumn{2}{c}{SOG} & \multicolumn{2}{c}{SOP} & \multicolumn{2}{c}{SLAP} & \multicolumn{2}{c}{SLAP-FST} \\
\hline\multirow{6}{*}{\begin{sideways}without $\X$\end{sideways}}
& empty      & Win   &    103 &  (3\%) &    173 &  (5\%) &     53 &  (1\%) &    161 &  (4\%) &    735 & (22\%) &   1256 & (38\%) &   1703 & (52\%) \\
&(3229 cases)& Lose  &    260 &  (8\%) &    272 &  (8\%) &   2909 & (90\%) &    481 & (14\%) &    256 &  (7\%) &    246 &  (7\%) &     94 &  (2\%) \\
&            & Fail  &    221 &  (6\%) &    253 &  (7\%) &   1786 & (55\%) &    302 &  (9\%) &    219 &  (6\%) &    213 &  (6\%) &     87 &  (2\%) \\
\cline{2-17}
& non empty  & Win   &      2 &  (0\%) &     10 &  (0\%) &    196 &  (4\%) &    513 & (12\%) &    645 & (15\%) &   2393 & (59\%) &   1293 & (31\%) \\
&(4048 cases)& Lose  &   1846 & (45\%) &   1378 & (34\%) &   1924 & (47\%) &    305 &  (7\%) &    318 &  (7\%) &     70 &  (1\%) &     40 &  (0\%) \\
&            & Fail  &    804 & (19\%) &    818 & (20\%) &   1070 & (26\%) &    263 &  (6\%) &    275 &  (6\%) &     69 &  (1\%) &     33 &  (0\%) \\
\hline\multirow{6}{*}{\begin{sideways}with $\X$\end{sideways}}
& empty      & Win   &     14 &  (1\%) &     13 &  (1\%) &      2 &  (0\%) &        &        &        &        &    810 & (73\%) &    823 & (74\%) \\
&(1108 cases)& Lose  &     21 &  (1\%) &     21 &  (1\%) &   1081 & (97\%) &        &        &        &        &      1 &  (0\%) &      1 &  (0\%) \\
&            & Fail  &      0 &  (0\%) &      0 &  (0\%) &    355 & (32\%) &        &        &        &        &      0 &  (0\%) &      0 &  (0\%) \\
\cline{2-17}
& non empty  & Win   &     12 &  (0\%) &      7 &  (0\%) &    778 & (23\%) &        &        &        &        &   1912 & (57\%) &   1872 & (55\%) \\
&(3348 cases)& Lose  &   1697 & (50\%) &   1627 & (48\%) &    470 & (14\%) &        &        &        &        &     54 &  (1\%) &     48 &  (1\%) \\
&            & Fail  &    257 &  (7\%) &    272 &  (8\%) &    129 &  (3\%) &        &        &        &        &     29 &  (0\%) &     29 &  (0\%) \\
\end{tabular}

%% file: tablebeem-tr.tex
\begin{tabular}{cccrlrlrlrlrlrlrl}
&&& \multicolumn{2}{c}{OWCTY} & \multicolumn{2}{c}{EL} & \multicolumn{2}{c}{BCZ} & \multicolumn{2}{c}{SOG} & \multicolumn{2}{c}{SOP} & \multicolumn{2}{c}{SLAP} & \multicolumn{2}{c}{SLAP-FST} \\
\hline\multirow{6}{*}{\begin{sideways}without $\X$\end{sideways}}
& empty      & Win   &      2 &  (2\%) &      3 &  (3\%) &     31 & (31\%) &      3 &  (3\%) &      1 &  (1\%) &     30 & (30\%) &     40 & (40\%) \\
& (100 cases)& Lose  &     22 & (22\%) &     10 & (10\%) &     61 & (61\%) &     26 & (26\%) &     25 & (25\%) &      3 &  (3\%) &      2 &  (2\%) \\
&            & Fail  &      8 &  (8\%) &      7 &  (7\%) &     38 & (38\%) &     16 & (16\%) &     16 & (16\%) &      2 &  (2\%) &      0 &  (0\%) \\
\cline{2-17}
& non empty  & Win   &      0 &  (0\%) &      0 &  (0\%) &    106 & (32\%) &     24 &  (7\%) &    101 & (30\%) &     70 & (21\%) &     57 & (17\%) \\
& (329 cases)& Lose  &    152 & (46\%) &    199 & (60\%) &     48 & (14\%) &     28 &  (8\%) &     21 &  (6\%) &     19 &  (5\%) &     23 &  (6\%) \\
&            & Fail  &     69 & (20\%) &     72 & (21\%) &     15 &  (4\%) &     16 &  (4\%) &     13 &  (3\%) &     19 &  (5\%) &     20 &  (6\%) \\
\hline\multirow{6}{*}{\begin{sideways}with $\X$\end{sideways}}
& empty      & Win   &     10 & (10\%) &     10 & (10\%) &     47 & (51\%) &        &        &        &        &     18 & (19\%) &     30 & (32\%) \\
&  (92 cases)& Lose  &     37 & (40\%) &     33 & (35\%) &     21 & (22\%) &        &        &        &        &     23 & (25\%) &     14 & (15\%) \\
&            & Fail  &      4 &  (4\%) &      3 &  (3\%) &      0 &  (0\%) &        &        &        &        &      4 &  (4\%) &      4 &  (4\%) \\
\cline{2-17}
& non empty  & Win   &      1 &  (0\%) &      2 &  (1\%) &    125 & (62\%) &        &        &        &        &     39 & (19\%) &     43 & (21\%) \\
& (200 cases)& Lose  &    102 & (51\%) &    108 & (54\%) &     33 & (16\%) &        &        &        &        &     15 &  (7\%) &     24 & (12\%) \\
&            & Fail  &     40 & (20\%) &     40 & (20\%) &      3 &  (1\%) &        &        &        &        &     14 &  (7\%) &     17 &  (8\%) \\
\end{tabular}

%% file: conclusion.tex
\section{Conclusion and Perspectives}

We have presented two new hybrid techniques: the \textit{symbolic
  observation product} (SOP), is a generalization of the
\textit{symbolic observation graph} (SOG) that diminishes the set of
observed atomic propositions as we progress in the product, and the
\textit{Self-Loop Aggregation Product} (SLAP) that exploits
the self-loops of the property automaton even if it does not express
a stuttering formula.

During our evaluation, we have found that SOP improves SOG, and
outperform fully symbolic algorithms EL and OWCTY in the presence of counterexamples.  
SLAP surpasses both SOG (always), OWCTY (always), and SOP (when the product is
empty). BCZ performs better than EL or OWCTy when the product is not empty  
and more poorly otherwise; the only set of experiments where it shows favorable results is the
BEEM models with randomly generated formulas using the next LTL operator. 
When the product is not empty, SOP and SLAP techniques seem
complementary. SLAP-FST provides the overall best performance and
lowest failure rate of all the methods we compared.

This work opens several perspectives.

Firstly, the above two techniques replace the product used in the
traditional automata-theoretic approach to model-checking in order to
reduce the product graph while preserving the result of the
emptiness-check.  Another technique with the same goal is the
\textit{Symbolic Synchronized Product} (SSP)~\cite{baarir.07.acsd}.
The SSP studies the symmetries of the model with respect to the
current state of the property automaton, to aggregate symmetrically
equivalent states.  A classical emptiness-check of the SSP is
possible, but Baarir and Duret-Lutz~\cite{baarir.07.acsd} also
suggested two emptiness checks variants taking advantage of the
inclusion between the aggregates.  It would be interesting to see if a
similar inclusion-aware emptiness checks could be used with SOG, SOP,
and SLAP.

Secondly, representing a stuttering property as a \textit{testing
  automata}~\cite{hansen.02.fmics} is another way to take advantage of
stuttering transitions in the model.  In the product between a KS and
a testing automaton, the latter does not move when the KS is
stuttering.  A possible perspective would be to adapt our
stuttering-based techniques (SOG and SOP) to aggregate all states from
the KS corresponding to one state of the testing automaton.

Finally, since the SOG is a KS, and the SLAP is built upon a KS, it is
possible to construct the SLAP of SOG.  This is something we did not
implement due to technical issues: in this case the aggregates are
sets of sets of states.

% LocalWords:  SOG OWCTY SSP Baarir Duret Lutz